\documentclass[%
 reprint,
superscriptaddress,
nofootinbib,
 amsmath,amssymb,
 amsfonts,
 aps,
]{revtex4-2}

\usepackage{graphicx}
\usepackage{enumitem}
\usepackage{dcolumn}
\usepackage{bm}
\usepackage{hyperref}
\hypersetup{
    colorlinks=true,
    linkcolor=blue,
    filecolor=blue,      
    urlcolor=blue,
    citecolor=blue, 
    pdftitle={GFL paper},
    pdfpagemode=FullScreen,
    }
\usepackage{appendix}
\usepackage{aas_macros}
\usepackage[mathlines]{lineno}

\usepackage[normalem]{ulem}
\usepackage{tabularx}
\newcolumntype{Y}{>{\centering\arraybackslash}X}

\usepackage[usenames,dvipsnames]{color}

\newcommand{\appropto}{\mathrel{\vcenter{
  \offinterlineskip\halign{\hfil$##$\cr
    \propto\cr\noalign{\kern2pt}\sim\cr\noalign{\kern-2pt}}}}}

\begin{document}

\preprint{APS/123-QED}

\title{The Need For Speed: Rapid Refitting Techniques for Bayesian Spectral Characterization of the Gravitational Wave Background Using PTAs}

\author{William G. Lamb}
\email{william.g.lamb@vanderbilt.edu}
 \affiliation{Department of Physics \& Astronomy, Vanderbilt University, 2301 Vanderbilt Place, Nashville, TN 37235, USA}
\author{Stephen R. Taylor}
 \email{stephen.r.taylor@vanderbilt.edu}
 \affiliation{Department of Physics \& Astronomy, Vanderbilt University, 2301 Vanderbilt Place, Nashville, TN 37235, USA}
\author{Rutger van Haasteren}
\email{rutger.v.haasteren@aei.mpg.de}
\affiliation{Max-Planck-Institut f{\"u}r Gravitationsphysik (Albert-Einstein-Institut), Callinstrasse 38, D-30167, Hannover, Germany}

\date{\today}

\begin{abstract}
Pulsar timing arrays (PTAs) have recently found evidence for a nanohertz-frequency stochastic gravitational-wave background (SGWB). Constraining its spectral characteristics will reveal its origins. In order to achieve this, we must understand how data and modeling conditions in each pulsar influence the precision and accuracy of SGWB spectral recovery. These goals typically require many Bayesian analyses on real data sets and large-scale simulations that are slow and computationally taxing. To combat this, we have developed several new rapid approaches that instead operate on intermediate SGWB analysis products. These techniques refit SGWB spectral models to previously-computed Bayesian posterior estimates of the timing power spectra. We test our new techniques on simulated PTA data sets and the NANOGrav $12.5$-year data set, where in the latter our refit posterior achieves a Hellinger distance---bounded between $0$ for identical distributions and $1$ for zero overlap---from the current full production-level pipeline that is $\lesssim 0.1$. Our techniques are $\sim10^2$--$10^4$ times faster than the production-level likelihood, and scale much more favorably (sub-linearly) as a PTA is expanded with new pulsars or observations. Our techniques also allow us to demonstrate conclusively that SGWB spectral characterization in PTA data sets is driven by the longest-timed pulsars and the best-measured power spectral densities, which is not necessarily the case for SGWB detection that is predicated on correlating many pulsars. Indeed, the common-process spectral properties found in the NANOGrav $12.5$-year data set are given by analyzing only the $\sim14$ longest-timed pulsars out of the full $45$ pulsar array, and we find that the ``shallowing'' of the common-process power-law model occurs when gravitational-wave frequencies higher than $\sim 50$~nanohertz are included. The implementation of our techniques is openly available as a software suite to allow fast and flexible PTA SGWB spectral characterization and model selection.

\end{abstract}

\maketitle

\section{\label{sec:intro}Introduction}
Pulsar Timing Array (PTA) experiments \citep{1990ApJ...361..300F} across the world have now reported compelling evidence for a nanohertz-frequency stochastic gravitational wave background (SGWB) \citep{2023ApJ...951L...8A,2023arXiv230616214A,2023ApJ...951L...6R,2023RAA....23g5024X}. This new insight into the gravitational-wave (GW) spectrum was achieved by measuring small deviations between the expected and observed radio-pulse times-of-arrival (TOAs) from a set of Galactic millisecond pulsars, wherein the distinctive imprint of an SGWB is inferred through a quasi-quadrupolar correlation signature imparted between pulsars in the PTA. This Hellings \& Downs correlation pattern \citep{hellings1983upper} has now been inferred with varying levels of significance by most regional PTA collaborations, with the promise of higher significance and enhanced scientific returns when these are synthesized into an updated International Pulsar Timing Array data set \citep{iptadr2}. 

While PTA detection statistics are centered around the cross-correlation of distinct pulsars, it is an interesting consequence of the PTA data model that spectral characterization of the SGWB is dominated by pulsar auto-correlations \citep{Romano2020-mq, astro4cast}. In fact, multiple PTA collaborations saw the first hints of the SGWB through emerging common spectral behavior in many pulsars, which was modeled as a common uncorrelated red noise (CURN) signal \citep{Arzoumanian2020-br,Goncharov_2021,10.1093/mnras/stab2833}. Even now, with evidence of GW-induced cross-correlations, SGWB spectral characteristics derived from a CURN data model provide an excellent approximation to a full Hellings \& Downs--correlated model (HD), yet at a fraction of the computational cost. Adequately modeled, the shape of the inferred SGWB spectrum encodes information about the emitting source, e.g., the dynamics and demographics of a black-hole binary population, or the details of early-Universe processes. Assuming that the source is a population of circular supermassive black-hole binaries (SMBHBs) evolving via GW radiation reaction, the SGWB's characteristic strain $h_\mathrm{c}$ as a function of frequency follows a power-law, $h_\mathrm{c}(f) = A\left(f / f_\mathrm{yr}\right)^{\alpha}$,
where $\alpha=-2/3$ \citep{Phinney_2001}. The amplitude $A$ is the characteristic strain referenced to a frequency of $f_\mathrm{yr}=1/\mathrm{year}$, and determined by the demographics of the binary population, e.g., the number density of emitting systems per redshift, primary mass, and mass ratio \citep[e.g.,][and references therein]{Burke-Spolaor2018-io}.

While measuring this amplitude parameter $A$ can provide interesting constraints on the SMBHB population, the inward migratory dynamics of supermassive black holes after a galaxy merger are likely much more complicated \citep[][and references therein]{2023ApJ...952L..37A}. Binary orbital eccentricity and interactions of binaries with gas and stars (particularly at wider orbital separations) will attenuate the expected SGWB characteristic strain at lower frequencies \citep{Sesana_2013, Burke-Spolaor2018-io}. This causes a deviation from a power-law \citep{Sampson2015-qj} and as such, the SGWB may carry information about the dynamical interactions of the SMBHB population within the final parsec of orbital evolution \citep{2017PhRvL.118r1102T}. Even finiteness of the emitting population under the most simplified conditions above may cause spectral deviations from $f^{-2/3}$ \citep{sesana_vecchio_colacino, Roebber_2016, kelley_17}, rendering fixed-$\alpha$ studies of limited utility for astrophysical inference. 

Beyond SMBHBs, searches are underway for relic signatures of processes in the early Universe \citep{2023ApJ...951L..11A}, e.g., cosmic strings \citep{1986coco}, primordial gravitational waves \citep{Lasky_2016}, and cosmological phase transitions \citep{cpt2015}. While it is thought that these signals are likely to be an additional, weaker contribution to the SMBHB signal, current searches can not yet arbitrate on the dominant contributing source of the SGWB. \citet{Kaiser2022-su} investigated the separability of a SGWB signal into its component sources: a circular-SMBHB-population signal, and a background from primordial gravitational waves. Using simulated data sets developed by \citet{astro4cast}, they found that they could begin to distinguish two injected power-law spectra with 45 pulsars and 17 years of data, while after 20 years they could begin to characterize the sub-dominant power-law GWB signal. However, their sub-dominant injected GWB spectrum had a cosmological energy density that was still rather strong, comparable to upper limits on primordial gravitational waves \citep[e.g.][]{arzoumanian2018nanograv, Lasky_2016}.

The goal for SGWB spectral characterization is to be a bridge between pulsar timing data and the physics of these sources. Our guiding principle is for spectral characterization to be scalable and modular; testing a new spectral model should not need the analysis to be started from scratch back at the level of timing residuals, nor should adding a new pulsar to the PTA require us to ignore that the analysis has already been successfully performed on the other pulsars. With this being said, the current production-level pipelines do indeed start from scratch whenever a new model is tested or a pulsar is added. The current PTA data model is formulated in the time domain. Uneven observational sampling of the pulsars, and concerns over the potential for spectral leakage from windowing, renders fast searches directly in the Fourier domain impractical. The computational bottleneck of this time-domain Gaussian likelihood is the required inversion of the data covariance matrix containing the SGWB signal contributions. Elegant accelerations can be achieved simply by modeling low-frequency processes (like the SGWB or per-pulsar red noise) with only a small number of Fourier basis functions \citep{2014PhRvD..90j4012V,2015MNRAS.446.1170V}. However, even with these accelerations and optimized sparse linear-algebra routines, Bayesian SGWB parameter estimation with the PTA likelihood via Markov Chain Monte Carlo (MCMC) sampling can require several days to weeks of computation. This is the status quo, and will worsen as more pulsars are added, and further observations of existing pulsars are incorporated into data sets. 

There is a tremendous need for robust, efficient, and flexible analysis methods for PTAs that follow our previously mentioned guiding principles of scalability and modularity. For example, high-cadence timing campaigns from telescopes such as CHIME \citep{2021ApJS..255....5C} generate large data volumes that will slow current pipelines. More pressing is that the synthesis of all current regional PTA data sets will result in a combined IPTA data set with more than 100 pulsars, which will tax existing pipelines. Significant acceleration of parameter estimation was achieved by \citet{Taylor2022-nq}, who modeled the SGWB as a CURN, which thereby allows the PTA likelihood to be factorized into parallelized per-pulsar analyses \citep[see e.g.][]{Arzoumanian2020-br, iptadr2, Johnson_2022, Goncharov_2021}. This Factorized Likelihood (FL) method shows excellent agreement with the full production-level PTA likelihood. Unfortunately, the FL method assumes a power-law model with a fixed spectral index, which limits its usefulness for spectral model selection and source inference. A more general approach would maintain the likelihood computational speed-up, parallelization over pulsars, and the intended modularity of this FL technique while permitting analyses of SGWB models with arbitrary spectral parameterizations.

In this paper, we introduce the aforementioned generalization of the FL approach, allowing for rapid SGWB spectral characterization under arbitrary parametrized models, rather than just a fixed-index power-law. This is made possible by condensing the pulsar timing data down to what we call \textit{Bayesian periodograms}: probability density reconstructions of the pulsar timing-residual power spectral density at each frequency. Models are then re-fit to combinations of these Bayesian periodograms. In \autoref{sec:method}, we discuss current analysis methods before introducing our new analysis techniques. We present the results of our comparison tests between the current and new methods on simulated and real data in \autoref{sec:results}, before sharing our conclusions and goals for further developments in \autoref{sec:discuss}.

\section{\label{sec:method}Methods}
Here we describe current PTA data-analysis techniques as they pertain to SGWB spectral characterization, and discuss expected future limitations as PTA data sets expand. We then introduce the factorized likelihood (FL) approach \citep{Taylor2022-nq}, and the concept of refitting spectral models to Bayesian periodograms of PTA timing residuals.

    \subsection{\label{sec:current}Current spectral characterization methods}
    
    PTA analyses model pulsar timing residuals $\vec{\delta t}$ as the sum of a deterministic pulsar timing model and stochastic red and white noise components:
    \begin{equation} \label{eq:residuals}
        \vec{\delta t} = \mathbf{M}\vec{\epsilon} + \mathbf{F}\vec{a} + \vec{n}.
    \end{equation}
    The $(N_\mathrm{TOA}\times m)$-shaped \textit{design matrix} $\mathbf{M}$ is a matrix of partial derivatives of the TOAs with respect to $m$ timing-ephemeris parameters evaluated at an initial fitting solution, with a vector of linear offsets from the initial fit $\vec{\epsilon}$. Red-noise processes, such as the common gravitational wave signal and red noise intrinsic to each pulsar, are modeled as a Fourier sum over $N_f$ sampling frequencies such that, for the $i$-th timing residual observed at time $t_i$,
\begin{equation} \label{eq:redfourier}
    \left[\mathbf{F}\vec{a}\right]_i = \sum_{k=1}^{N_f} \left\{a_{s,k}\sin\left(\frac{2\pi k t_i}{T}\right) + a_{c,k}\cos\left(\frac{2\pi k t_i}{T}\right)\right\},
\end{equation}
    where $T$ is the timing baseline (typically the total timespan of the data set being analyzed). As such, $\mathbf{F}$ is a $N_\mathrm{TOA}\times2N_f$ matrix of sines and cosines evaluated at observation times, and $\vec{a} = \left(a_{s,1}, a_{c,1}, a_{s,2}, a_{c,2}, ..., a_{s,N_f}, a_{c,N_f}\right)^\mathrm{T}$ is a vector of Fourier coefficients. We model intrinsic red noise (IRN) as independent between pulsars, and the SGWB as a common signal to all pulsars. For a single pulsar $p$, its total red noise is the sum, $(\mathbf{F}\vec{a})_p=(\mathbf{F}\vec{a})_p^\mathrm{IRN}+(\mathbf{F}\vec{a})^\mathrm{SGWB}$. Finally, $\vec{n}$ is uncorrelated white noise due to radiometer noise, instrumental effects, and pulsar phase jitter. We rearrange \autoref{eq:residuals} to model residual, unmodeled noise as $\vec{r}$:
    \begin{equation} \label{eq:WNresiduals}
        \vec{r} = \vec{\delta t} - \mathbf{M}\vec{\epsilon} - \mathbf{F}\vec{a} = \vec{\delta t} - \mathbf{T}\vec{b},
    \end{equation}
    where $\mathbf{T}=[\mathbf{M}\ \mathbf{F}]$ and $\vec{b}=[\vec{\epsilon}\ \vec{a}]^\mathrm{T}$. Other contributions to the timing residuals include correlated white noise between TOAs within the same timing epoch, and radio frequency-dependent red noise due to time-dependent variation in dispersion from the interstellar medium\citep[see e.g.][]{cordes2010measurement, keith2013measurement}.
    
    We place a zero-mean Gaussian prior on $\vec{b}$ such that, for model hyper-parameters $\vec{\eta}$,
    \begin{equation}\label{eq:b-prior}
        p(\vec{b}|\vec{\eta}) = \frac{\exp{\left(-\frac{1}{2}\vec{b}^\mathrm{T}\mathbf{B}^{-1}\vec{b}\right)}}{\sqrt{\det{\left(2\pi \mathbf{B}\right)}}},
    \end{equation}
    where $\mathbf{B}\equiv\mathbf{B}(\vec{\eta})=\langle\vec{b}\vec{b}^\mathrm{T}\rangle=\mathrm{diag}(\mathbf{\infty},\mathbf{\phi})$. The matrix $\mathbf{\phi}$ is the Fourier-domain covariance on red-noise processes, while the $\infty$-block effectively converts the Gaussian prior into a improper uniform prior on the timing model. Given that we will eventually only deal with the inverse of $\mathbf{B}$, we need not worry about the practicalities of treating infinities.

    The full hierarchical likelihood of the timing residuals given the model hyper-parameters and $b$-coefficients is given by $p(\vec{\delta t}|\vec{b},\vec{\eta}) = p(\vec{\delta t}|\vec{b}) \times p(\vec{b}|\vec{\eta})$. However, we are only interested in the model hyper-parameters $\vec{\eta}$ that describe the statistical properties of various stochastic processes; thus we marginalize over the Gaussian $b$-coefficients to recover a more concise likelihood:
    \begin{equation} \label{eq:PTA like}
        p(\vec{\delta t}|\vec{\eta}) = \frac{\exp{\left(-\frac{1}{2}\vec{\delta t}^\mathrm{T}\mathbf{C}^{-1}\vec{\delta t}\right)}}{\sqrt{\det{\left(2\pi \mathbf{C}\right)}}}.
    \end{equation}
    Here, $\mathbf{C}=\mathbf{N} + \mathbf{TBT}^\mathrm{T}$ is the full time-domain covariance matrix of the data model, with white-noise covariance $\mathbf{N}$, where
    \begin{align} \label{eq:PTA cov}
        [\mathbf{C}]_{(pi),(qj)} = [\mathbf{N}]_{p,(ij)}\delta_{pq}\delta_{ij} &+ [\mathbf{C}^\mathrm{IRN}]_{p,(ij)}\delta_{pq} \nonumber \\
        &+ \Gamma_{pq}[\mathbf{C}^\mathrm{SGWB}]_{(ij)}.
    \end{align}
    \autoref{eq:PTA cov} indexes over pulsars ($p,q$) and TOAs ($i,j$). $[\mathbf{N}]_{p,(ij)}$ and $[\mathbf{C}^\mathrm{IRN}]_{p,(ij)}$ are the white noise and intrinsic red noise covariance matrix components respectively for pulsar $p$ and $i$-th TOA, while $[\mathbf{C}^\mathrm{SGWB}]_{(ij)}$ is the covariance matrix components for the SGWB between the $i$-th and $j$-th TOAs. The expected GW-induced cross-correlation in timing residuals between pulsars is given by the overlap reduction function (ORF) coefficient $\Gamma_{pq}$, which, for an isotropic SGWB is the aforementioned Hellings \& Downs (HD) curve \citep{hellings1983upper}.

    All current spectral characterization techniques involve computing the PTA likelihood in \autoref{eq:PTA like} under different models or assumptions \citep{coles2011pulsar, Arzoumanian2020-br}. When cross-correlations between pulsars are modeled (hereafter referred to as inter-pulsar correlations), inverting $\mathbf{C}$ should scale as $\mathcal{O}(N_p^3N_b^3)$. As more TOAs and more pulsars are added to the array, evaluation of this likelihood will slow down significantly because of this scaling. The autocorrelation blocks in the PTA data covariance matrix contain white noise, pulsar-intrinsic red noise, and the SGWB, while the inter-pulsar blocks only feature the SGWB. However, we now know that spectral characterization of an SGWB is dominated by PTA autocorrelation information \citep{Romano2020-mq,astro4cast}. Therefore, for the class of techniques below where the PTA likelihood is factorized over pulsars, we assume no inter-pulsar correlations (i.e., a CURN model) such that $\Gamma_{pq}=\delta_{pq}$.
    
    Modeling only the diagonal blocks of the PTA data covariance matrix reduces the likelihood evaluation scaling to $\mathcal{O}(N_pN_b^3)$. Physically speaking, this significant acceleration arises because the PTA likelihood is factorized as a product over pulsars:
    \begin{align} \label{eq:model2a}
        p(\{\vec{\delta t}\} | \vec\eta) &= \frac{\exp{\left(-\frac{1}{2}\sum_{p=1}^{N_p}\vec{\delta t}_{p}^\mathrm{T}C_{pp}^{-1}\vec{\delta t}_{p}\right)}}{\sqrt{\det{\left(2\pi C\right)}}} \nonumber \\
        &= \prod_{p=1}^{N_p} \frac{\exp{\left(-\frac{1}{2}\vec{\delta t}_{p}^\mathrm{T}C_{pp}^{-1}\vec{\delta t}_{p}\right)}}{\sqrt{2\pi C_{pp}}} =\prod_{p=1}^{N_p} p(\vec{\delta t}_p|\vec{\eta}),
    \end{align}
    where $\{\vec{\delta t}\}$ where is the set of timing residuals for all pulsars, $p(\vec{\delta t}_p|\vec{\eta})$ is the likelihood for a single pulsar $p$ with a set of timing residuals $\vec{\delta t}_p$, and $\vec{\eta}$ are model hyperparameters describing variables like spectral-shape parameters, etc. However, this factorization is not exploited to full effect within the production-level \texttt{enterprise} analysis pipeline \citep{enterprise}, which carries this out as a serialized calculation over pulsars. Parallelizing the computation over $N_p$ processors would theoretically remove the likelihood computation's dependence on the number of pulsars, while being numerically equivalent to an analysis that uses the production-level PTA likelihood.
    
    \subsection{\label{sec:refit}Factorized likelihood methods}
    The factorized likelihood (FL) approach makes possible a class of techniques where \autoref{eq:model2a} is computed in parallel across pulsars, with re-weighted posterior distributions from each pulsar combined in post-processing to calculate the likelihood for the array. Evaluation of the likelihood becomes approximately scale invariant with $N_p$. \citet{Taylor2022-nq} modeled a power-law SGWB strain spectrum with a fixed spectral index of $\alpha=-2/3$ to recover posteriors on the SGWB strain amplitude for each pulsar. The posteriors on the strain amplitude were represented by histograms, re-weighted by the single-pulsar parameter priors, then multiplied across pulsars with a suitable prior for the final posterior calculation.
    
    This fixed-index FL technique (along with variants) has already been adopted as a new tool in large analysis campaigns from NANOGrav \citep{Arzoumanian2020-br,2023ApJ...951L...8A}, the Parkes Pulsar Timing Array \citep{2023ApJ...951L...6R,Goncharov_2021}, and IPTA \citep{iptadr2}, as well as other studies \citep{Johnson_2022,2023ApJ...951..121S}, to accelerate parameter estimation and cross-validation. However as discussed earlier in \autoref{sec:intro}, there are many reasons why the SGWB strain spectrum could deviate from this simple fixed-index power-law model. We therefore require a more flexible and generalized FL approach that would allow for inference of physically-motivated SGWB spectral models.
    
    A General Factorized Likelihood (GFL) approach is possible by fitting spectral models to the \textit{free-spectrum}, a minimally-modeled Bayesian spectral characterization of pulsar timing data \citep{lentati, taylor13}. The free-spectrum recovers the joint posterior of the red-noise power spectrum at all sampling frequencies, parameterized by the coefficient $\rho$, such that
    \begin{equation} \label{eq:freespec}
        \rho_{k}^2 := \frac{\langle\vec{a_k}^\mathrm{T}\vec{a_k}\rangle}{T} = \frac{S(f_k)}{T},
    \end{equation}
    where $k$ is the Fourier frequency-bin index and $S$ is the power spectral density of the timing residuals induced by red processes. Typically, a free-spectrum analysis is conducted with a uniform prior on $\log_{10}\rho$, with posteriors jointly recovered at all sampling frequencies. In the following we assume that there is independence between sampling frequencies, thus no covariance between them. Pulsar timing analyses deal with unevenly sampled observations, so we will assess the strength of this assumption in our tests.

    Refitting spectral models to Bayesian free-spectra can be done at various levels; one can $(i)$ perform a PTA Bayesian free-spectrum analysis, followed by refitting on the frequency-factorized PTA free-spectrum, or $(ii)$ perform free-spectral analysis on individual pulsars, which are then combined into a frequency- and pulsar-factorized likelihood against which spectral models are fit. The general scheme for $(i)$ is as follows. Translating $h_\mathrm{c}$ into $\rho$-space gives
    \begin{equation} \label{eq:rho_pl}
        \rho_k^2 = \frac{h_\mathrm{c}(f_k)^2}{12\pi^2f_k^3 T} = \frac{A^2}{12\pi^2T}\left(\frac{f_k}{f_\mathrm{1yr^{-1}}}\right)^{-\gamma},
    \end{equation}
    where $\gamma=3-2\alpha=13/3$ for the idealized SMBHB population. We form a likelihood by computing the probability that a given model is supported by the free-spectrum at each frequency:
    \begin{align} \label{eq:FL}
        p(\{\vec{\delta t}\} | \vec\eta) &= \int \mathrm{d}\vec{\rho}\,\, p(\{\vec{\delta t}\} | \vec{\rho}) \, p(\vec{\rho} | \vec\eta)  \nonumber \\
        & \propto \int \mathrm{d}\vec{\rho}\,\, \frac{p(\vec{\rho} | \{\vec{\delta t}\})}{p(\vec{\rho})} \times p(\vec{\rho} | \vec\eta) \\
        &\approx \prod_{k=1}^{N_f} \int \mathrm{d}\rho_k\,\, \frac{p(\rho_k | \{\vec{\delta t}\})}{p(\rho_k)} \times p(\rho_k | \vec\eta) \nonumber  
    \end{align}
    where $p(\rho_k)$ is the prior probability of $\rho_k$ in the free-spectrum analysis, $p(\rho_k | \{\vec{\delta t}\})$ is the marginal posterior probability density of $\rho_k$ that is sampled using MCMC techniques, and $p(\rho_k | \vec\eta)$ is the probability of $\rho_k$ under a parametrized spectral model, such as \autoref{eq:rho_pl}. In all cases considered here, the spectral model maps precisely to a value of $\rho$ at each frequency, in which case the integral in \autoref{eq:FL} is trivial. However, the more general form shown allows for models that have intrinsic spread, e.g., where there is an expected form of the spectrum due to a population of SMBHBs, and population finiteness induces departures in a given realization \citep{Burke-Spolaor2018-io, 2023ApJ...952L..37A}. We note that in $(i)$, the PTA free-spectrum need not necessarily use only autocorrelation information in the PTA likelihood; $p(\rho_k | \{\vec{\delta t}\})$ is not yet factorized over pulsars, hence an inter-pulsar--correlated free-spectral analysis may be performed that accounts for HD correlations, and this would still allow a frequency-factorized refitting analysis to be subsequently performed.

    Finally, (ii), the extension to factorize the likelihood over pulsars simply requires that the right-hand side of \autoref{eq:FL} is modified to have an additional product over pulsars. However, by doing so, we must explicitly make the usual FL assumption of conditioning spectral characterization on the PTA autocorrelation information under a CURN model:
    \begin{equation}
    	 p(\{\vec{\delta t}\} | \vec\eta) \appropto \prod_{p=1}^{N_p}\prod_{k=1}^{N_f} \int \mathrm{d}\rho_k\,\, \frac{p(\rho_k | \vec{\delta t}_p)}{p(\rho_k)} \times p(\rho_k | \vec\eta) .
    \end{equation}
    To compute probabilities of a spectral model with a given set of hyper-parameters $\vec\eta$ under the free-spectral likelihoods $p(\{\vec{\delta t}\} | \rho_k)$, we use optimized density estimation with MCMC samples. There are already a number of examples in the literature of fitting SGWB spectra to a free-spectrum of a PTA \citep[e.g.,][]{2018PhRvL.120r1101C,Ratzinger_2021, deng}. It is favored over analyzing the full likelihood because it is fast, since the data structures that we are fitting to are no longer the timing residuals themselves, but a compressed data representation in terms of a red process at each GW frequency. Other timing-residual contributions from the timing model, uncorrelated and correlated white noise, and interstellar-medium effects, are marginalized over.

    The simplest density estimation technique is to bin our free-spectrum MCMC samples as histograms, just like in the FL method. This recreates the probability densities of the free-spectra posteriors within bins of $\log_{10} \rho$. To faithfully reconstruct the original distribution, an appropriate choice of bin width must be made. If the width of the bins is too large, the histogram will be \textit{oversmoothed}, perhaps removing important fluctuations in the actual distribution. In contrast, if the bin width is too narrow, the histogram will \textit{undersmooth} the data, creating a density reconstruction that captures all of the fluctuations in the data that are a result of statistical sampling randomness and not due to the underlying distribution. There are several standard rules-of-thumb for finding the optimal bin width for a histogram given some data, such as by using Scott's Normal Reference Rule \citep{scott} or the Freedman-Diaconis Rule \citep{freedman1981histogram}, which are tuned for an underlying normal distribution. Unfortunately, histograms do not result in a continuous distribution from which probability densities can be extracted, causing some loss of data in-between bins, particularly if the bin width is wide.
    
    An alternative method is to use Kernel Density Estimators (KDEs) \citep{kde1, kde2}. A KDE recreates a distribution by replacing each sample with a normalized, symmetric, strictly positive, real-valued function called a kernel (also known as a window function). If samples $(x_1, x_2, \ldots, x_N)$ are extracted from an unknown distribution $f$, the density estimate $\hat{f}$ of a KDE is given by
    \begin{equation} \label{eq:KDE}
        \hat{f}(x) = \frac{1}{Nh} \sum_{i=1}^{N} K\left(\frac{x-x_i}{h}\right),
    \end{equation}
    where $K$ is the kernel function, and $h$ is the bandwidth of the KDE. As with histograms, an appropriate bandwidth must be chosen to avoid creating an under- or over-smoothed estimator. The kernel function itself is also a choice to be decided. In this paper, we use an \textit{Epanechnikov kernel} \citep{epanechnikov}, and select bandwidths using the \textit{Sheather-Jones Plug-in Selector} \citep{sj}. Further details on these choices, and KDEs in general, are given in Appendix~\ref{appendix:kde}.

    Some free-spectrum posteriors may be poorly constrained and show non-negligible support for $\log_{10}\rho$ down to their lower prior boundary. The corresponding likelihood would effectively be constant if the boundary were lowered to $-\infty$. To ensure accurate KDE reconstruction at the boundary, we mirror the data about the boundary to create the KDE, and then cut off the KDE at the boundary. Any proposed spectral model in our refitting scheme that goes below the boundary is given the same probability as spectra at the boundary.
    \subsection{\label{sec:pipe}Refit pipelines}
    
        We refit parametrized spectral models against these optimized KDE representations of PTA and pulsar free-spectra using MCMC techniques. A typical algorithm is as follows: $(1)$ an iteration of the MCMC proposes a set of parameters for the spectral model, from which we calculate our $\log_{10}\rho$ coefficients at all GW sampling frequencies; $(2)$ we then find the probability of these model $\log_{10}\rho$ values under the free-spectrum likelihoods at each frequency---and, if applicable, for each pulsar---given our KDEs\footnote{KDE objects are memory intensive, and extracting the probability density function of a point from every KDE object at each MCMC iteration would slow down computation. However, KDEs are continuous, therefore before conducting the MCMC, we extract an array of probabilities along a grid of $\log_{10}\rho$ that is intentionally finer than the KDE bandwidth. This allows us to implement \texttt{numpy} vectorization techniques to accelerate the computation of the likelihood. When a set of $\log_{10}\rho$ is calculated, we look up its probability within the pre-calculated KDE density array across frequencies (and pulsars, where relevant).}, and take their product to compute the total likelihood. The employed Metropolis-Hastings algorithm will then reject or accept those parameters accordingly. We repeat this until the MCMC has sufficiently sampled the parameter space of the spectral model and converged to the target posterior.
        
        In this paper, we explore two possible types of refits:

        \begin{enumerate}[label=\textbf{$(\roman*)$},wide=0pt]
        \item PTA free-spectrum refit: This involves refitting spectral models against the PTA free-spectrum, which requires an initial analysis using the full PTA likelihood as a one-time cost. The PTA free-spectrum analysis describes each each pulsar with a timing model, white noise, and power-law intrinsic red noise, with a free-spectrum common process across the entire array. Note that the model of inter-pulsar correlations for this common process can be CURN (uncorrelated), HD (SGWB-correlated), or other, since this refit technique involves only a factorization over frequencies. While we can refit SGWB spectral models to different numbers of frequencies with the PTA free-spectrum, we cannot refit using different combinations of pulsars without recomputing the PTA free-spectrum.
        \item[$(ii)$] Generalized Factorized Likelihood (GFL) Lite: In preparation for our goal of a complete generalization of the factorized likelihood technique, we introduce and study an intermediate analysis approach here called GFL Lite. Each pulsar is analyzed independently in parallel, with a model composed of a timing model, white noise, power-law intrinsic red noise, and a free-spectrum that acts as a proxy for the common process in each pulsar. Given the implicit factorization of the PTA likelihood over pulsars, GFL Lite assumes CURN as an inter-pulsar correlation model. We then refit a common spectral model to the free-spectra of an ensemble of all (or a subset of) pulsars to recreate the PTA common process. This method allows us to fit a common signal to different combinations of pulsars and frequencies quickly. However, the per-pulsar intrinsic red noise model cannot be refit. This method is labeled as `Lite' because the full GFL technique will also be capable of refitting per-pulsar intrinsic red noise models. Plans and prospects for full GFL are discussed in \autoref{sec:discuss}.
        \end{enumerate}

        \noindent
        
    \subsection{\label{sec:timing}Pipeline profiling}
    
    In addition to these techniques being modular and flexible, we are also motivated by the prospects of significantly accelerating spectral characterization of the SGWB with PTA data, especially where many repeated studies and simulations are required. As discussed in \autoref{sec:current}, the full PTA likelihood with a CURN model should scale $\propto N_p$ because of the required inversion of a block-diagonal PTA data covariance matrix, and should scale as $\propto N_p^3$ for the full likelihood with an HD model because of the additional off-diagonal structure of the data covariance matrix.

\begin{figure}
        \centering
        \includegraphics[width=\columnwidth]{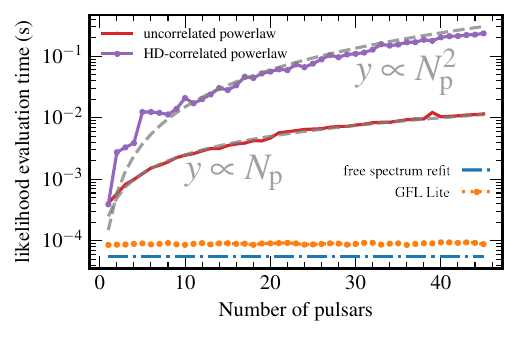}
        \caption{The likelihood evaluation time as a function of the number of pulsars on a simulated data set, ran on an AMD EPYC 7702 64-core processor. The CURN (red) and HD-correlated (purple) full PTA likelihoods scale $\propto N_p$ and $\propto N_p^2$ respectively. GFL Lite (orange) is scale-independent with the number of pulsars. The PTA free-spectrum refit (dashed blue) is the most rapid method, being $10^2-10^4 \times$ faster than the CURN and HD-correlated full PTA likelihoods.}
        \label{fig:profiler}
    \end{figure}

    Before carrying out a suite of simulations to compare the accuracy of our refit parameter estimation with the full PTA likelihood, we profiled our analyses on a simulated $N_p$-pulsar PTA data set that contains an injected SGWB signal (as detailed later in \autoref{sec:sims}). The timing profiles are shown in \autoref{fig:profiler}. For a simulated 45-pulsar data set, the mean likelihood evaluation time for the CURN full likelihood was $0.012$~seconds, and $0.24$~seconds for the HD full likelihood. The CURN full likelihood scales as expected with the number of pulsars. However the HD full likelihood scales $\propto N_p^2$, rather than the expected $\propto N_p^3$. As explained in \citet{johnson2023nanograv}, this is due to the use of sparse matrix algebra. The exact scaling depends on details such as memory transfer, sparse matrix representation transforms, parallel computation across CPU cores, and matrix layout, all of which differ depending on the exact PTA analysis that is performed. But empirically, this typically results in a $\propto N_p^2$ dependence. The 45-pulsar PTA free-spectrum refit likelihood takes $53$~microseconds while the GFL Lite likelihood takes $88$~microseconds. These are $226$ and $136$ times faster than the CURN full likelihood respectively. The GFL Lite likelihood evaluation is sub-linear as the number of pulsars increases. The PTA free-spectrum is the fastest; however, a new free-spectrum must be produced if we wish to change the number of pulsars in the array.
\section{\label{sec:results}Results}

    We present the results of our analyses of 100 simulated PTA data sets that contain injected SGWB signals, comparing the performance of the full PTA likelihood to our refit techniques. In each analysis, we model intrinsic red noise as a $10$ frequency power-law in each pulsar in addition to a $10$ frequency power-law common process, unless otherwise specified. These frequencies are linearly spaced from $1/T$ to $10/T$, where $T$ is the total observing time of the array. We assess the ability of the PTA free-spectrum refit and GFL Lite techniques to recover SGWB parameter posteriors that are comparable to the full likelihood, and investigate tolerance factors.

    To quantify the difference between the posteriors recovered by our techniques compared to the full likelihood, we use the \textit{Hellinger distance} \citep{hellinger}, a measure of the similarity between two probability distributions. 
    The Hellinger distance is bounded $0 \leq H \leq 1$, where $H=0$ implies that distributions are identical, while $H=1$ implies that they do not have any overlap and are completely different distributions. For our refit techniques to be robust and accurate, we seek Hellinger distances to be low with respect to results from using the full likelihood. See Appendix~\ref{appendix:hellinger} for more details and guiding values for interpretation. We compare Hellinger distances between the 2D posteriors, as well as the 1D marginalized posteriors for each parameter in a power-law spectral model for the SGWB signal, $\gamma$ and $\log_{10}A$, as defined in \autoref{eq:rho_pl}.
    
    \subsection{\label{sec:sims}Simulations}
    
    Our simulated data set creation follows \citet{astro4cast}. The pulsar data sets are based on the observational timestamps and TOA uncertainties from the 45 pulsars of the NANOGrav $12.5$~year data set \citep{Arzoumanian2020-br}. We extended the timespan of the data set by drawing new TOAs and uncertainties from the distributions of the final year of each pulsar's observations to form a $15$~year data set. However, we kept the number of pulsars fixed, rather than adding new ones over time. We injected intrinsic red noise in each pulsar at linearly-spaced frequencies of $1/T$ to $10/T$, where $T=15$~years. The injected spectral characteristics of a pulsar's intrinsic red noise were based on measured values taken from a CURN search in the NANOGrav 12.5 year data set.
    
\begin{figure}
            \centering
            \includegraphics[width=\columnwidth]{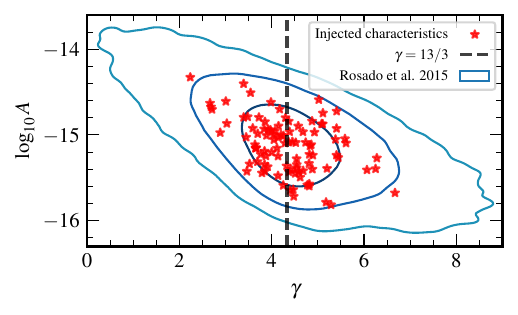}
            \caption{The blue regions are the 68, 95, and 99\% credible regions of the distribution of SGWB spectral characteristics from the $234,000$ SMBHB population realizations of \citet{rosado}. We randomly selected $100$ SGWB realizations from this distribution (red markers). The dashed black line designates $\gamma=13/3$, the realization-averaged expected SGWB spectral index from a population of SMBHBs.}
            \label{fig:middleton}
        \end{figure}

    Finally, 100 SGWB signal realizations were injected into 100 copies of our simulated PTA data set. We randomly drew SGWB spectra from a bank of $234,000$ that had been fit to SMBHB population realizations \citep{rosado} (see also \citep{middleton}). \autoref{fig:middleton} shows the total distribution of SGWB spectral characteristics in blue, and the spectral characteristics injected into our simulations in red. Typical PTA analyses use priors of $\gamma\in[0, 7]$ and $\log_{10}A\in[-18, -12]$. Following this convention, we also ensured that the randomly-drawn SGWB spectral characteristics satisfied these prior constraints. Unless otherwise stated, all models search for a CURN process to ensure the most fair comparison between the full PTA likelihood and the refitting techniques.
    
    \subsection{\label{sec:model2a}Parameter Estimation Fidelity}
    
\begin{figure*}
        \centering
        \includegraphics[width=\textwidth]{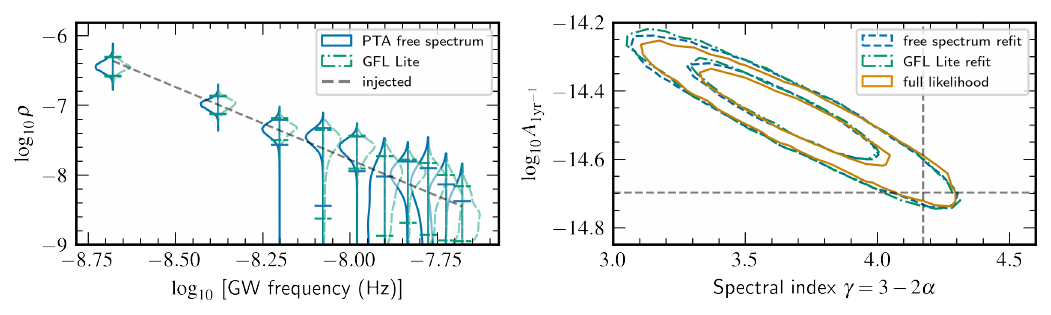}
        \caption{\textit{Left panel:} A comparison of the free-spectrum from a full PTA likelihood analysis (blue) with a product of the per-pulsar free-spectra from the GFL Lite pipeline (green) on a simulated data set. The two violins are nearly identical and follow the injected SGWB power-law (grey line). \textit{Right panel:} Posteriors of a 10 frequency power-law analysis with the full likelihood (orange), PTA free-spectrum refit (blue), and GFL Lite methods (green), for the simulated data set shown in the left panel. Credible regions enclose $68\%$ and $95\%$ of the posterior. The injected SGWB spectral characteristics are shown as the dashed grey lines, with $\log_{10}A=-14.7$ and $\gamma=4.17$. The PTA free-spectrum refit and GFL Lite posteriors match well to the full likelihood.}
        \label{fig:corner}
    \end{figure*}

    We choose one of our simulations as a case study of our refitting techniques. The chosen simulation has spectral characteristics comparable to the CURN detected in the NANOGrav $12.5$~year data set, and has one of the smallest Hellinger distances between the uncorrelated full-likelihood power-law analysis and the PTA free-spectrum refit. As a first exploration, given that each GFL-Lite per-pulsar free-spectrum has already been marginalized over intrinsic red noise parameters, the combined product of those likelihood distributions across pulsars should be consistent with the PTA free-spectrum. This is shown in the left panel of \autoref{fig:corner}, where there is broad consistency between the techniques. 
    
\begin{figure*}
        \centering
        \includegraphics[width=\textwidth]{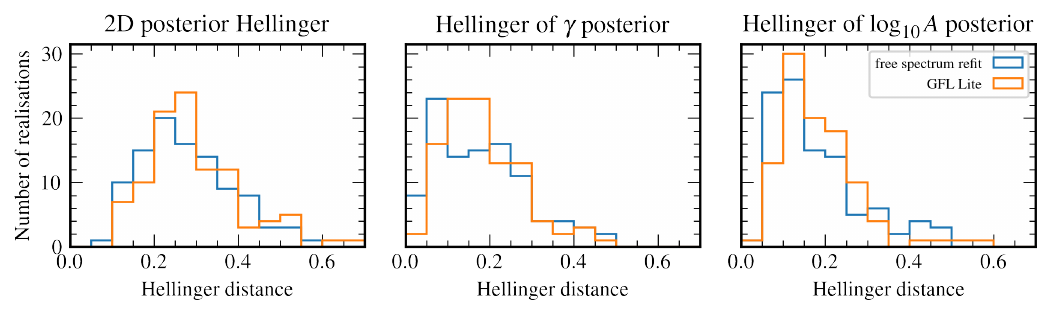}
        \caption{Hellinger distances between the posteriors of the full likelihood and for each refitting technique for all 100 SGWB data set realizations. Both methods have a similar distribution of Hellinger distances, thereby demonstrating similar performance when compared to the full PTA likelihood analysis.}
        \label{fig:1Dhellinger}
    \end{figure*}

    The comparison of power-law--model posterior distributions for our case-study simulation is shown in the right panel of \autoref{fig:corner}, where credible regions correspond to 68\% and 95\% levels for the spatially-uncorrelated full-likelihood, the PTA free-spectrum refit, and the GFL Lite analysis. 
    Both refit methods perform well, recovering posteriors consistent with the full production-level PTA likelihood, with both achieving 2D Hellinger distances of 0.10. \textcolor{red}{}{The 1D-marginalized posteriors on $\log_{10}A$ and $\gamma$ have distances with respect to the full PTA likelihood of $0.06$ and $0.07$ for the PTA free-spectrum, and $0.09$ and $0.05$ for GFL Lite. In this case, the PTA free-spectrum refit and GFL Lite performances are on par}. We see a similar consistency when comparing the Hellinger distances of all 100 data set realizations. The distributions of Hellinger distances for the 2D and 1D marginalized posteriors are shown in \autoref{fig:1Dhellinger}, from which we quote the median, $16^\mathrm{th}$ percentile, and $84^\mathrm{th}$ percentile values. The 2D Hellinger distances between the PTA free-spectrum refit and the full likelihood are $0.26\substack{0.40 \\ 0.17}$, while GFL Lite has 2D Hellinger distances of $0.27\substack{0.40 \\ 0.20}$. We conclude that the PTA free-spectrum refit and GFL Lite analysis are consistent with each other.
    
    To better understand the origin of discrepancies between our refit methods and the full likelihood, we investigate the magnitude of inter-frequency correlations in the Bayesian free-spectrum posteriors, using \textit{Pearson's correlation coefficient} \citep{pearson1895vii}. If inter-frequency correlations are weak, the correlation matrix of the posterior samples should be mostly diagonal in structure. Pearson's correlation coefficient quantifies how `diagonal' a correlation matrix is, with a coefficient of $1$ indicating a perfectly diagonal matrix (i.e., no inter-frequency correlations), and lower values indicating off-diagonal structure (i.e., inter-frequency correlations). In the limit that there are no inter-frequency correlations, Pearson's correlation coefficient becomes unity, and our approximation becomes an identity. The median, $16^\mathrm{th}$, and $84^\mathrm{th}$ percentile values of this coefficient across all 100 realizations of the PTA free-spectra is $0.92\substack{0.98 \\ 0.86}$, suggesting weak correlations between GW frequencies. We also compute the coefficient for all $45$ per-pulsar free-spectra from the GFL Lite pipeline across all $100$ simulation realizations, giving $0.99\substack{1.0 \\ 0.91}$; per-pulsar free-spectra appear to be uncorrelated across frequencies. Hence our assumption throughout of independence between frequencies is justified, and suggests that information being lost from our refit pipelines is through the compounding of small inaccuracies in our density estimators.
    
\begin{figure}
        \centering
        \includegraphics[width=\columnwidth]{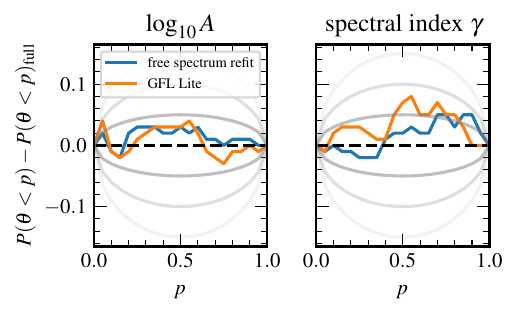}
        \caption{The difference in $p$--$p$ plots between the full likelihood and the PTA free-spectrum refit or GFL Lite. Equivalent recovery would show zero for all $p\%$ credible regions. The PTA free-spectrum refit is centered close to zero and mostly within the $1\sigma$ confidence region, where grey curves show $1\sigma$, $2\sigma$, $3\sigma$ regions. GFL Lite is also close to zero, and mostly within the $2\sigma$ confidence region for both parameters.}
        \label{fig:pp-plot}
    \end{figure}

    Finally, we test the efficacy of Bayesian recovery between our proposed methods and the full likelihood with $p$--$p$ plots, as shown in \autoref{fig:pp-plot}. If we were to draw our injected spectral characteristics from the same priors as employed in our Bayesian analysis, then we would expect to recover our injections within the $p\%$-credible region for $p\%$ of our simulations. However in our analyses---even with the full likelihood---we see bias, causing deviation from the diagonal $p$--$p$ plot, since we drew our injected characteristics from the SMBHB populations of \citet{rosado}, and other analysis approximations. Instead, we compare the relative efficacy of our refit methods to the full likelihood analysis by taking the difference in $p$--$p$ recovery between the full likelihood and our refit methods. A perfect comparison would give zero difference for all $p$. The PTA free-spectrum refit has the smallest differences from the full likelihood, showing deviations around zero mostly within a $1\sigma$ confidence interval, where $\sigma=\sqrt{p(1-p)/100}$ is the binomial standard error for a sample of 100 realizations \citep{bilby}. GFL Lite shows more deviation from the full likelihood than the PTA free-spectrum refit, but these remain typically within a $1\sigma$ confidence interval on $\log_{10}A$, and within $2\sigma$ for the spectral index.

    For more context, we compare \autoref{fig:pp-plot} to a gallery of toy univariate distribution comparisons in \autoref{fig:pp-explanation}. For the PTA free-spectrum refit, the $p$-$p$ plot for $\log_{10}A$ is similar to the center top panel of \autoref{fig:pp-explanation}, which suggests this method may underestimate $\log_{10}A$ relative to the full PTA likelihood. Meanwhile, the spectral index recovery appears similar to the left middle panel, suggesting that the width of the recovered $\gamma$ posterior is narrower than the full PTA likelihood. By contrast, the $p$-$p$ plot for $\log_{10}A$ and $\gamma$ for GFL Lite appear similar to the right middle panel and top center panel of \autoref{fig:pp-explanation} respectively, suggesting a typically wider recovered $\log_{10}A$ posterior, and a slightly underestimated $\gamma$ recovery. These are again likely due to compounding of small inaccuracies in our density estimators over many frequencies (and pulsars). However, overall these refitting methods achieve excellent parameter posterior recovery when compared to the full PTA likelihood.

    \begin{figure*}
        \centering
        \includegraphics[width=\textwidth]{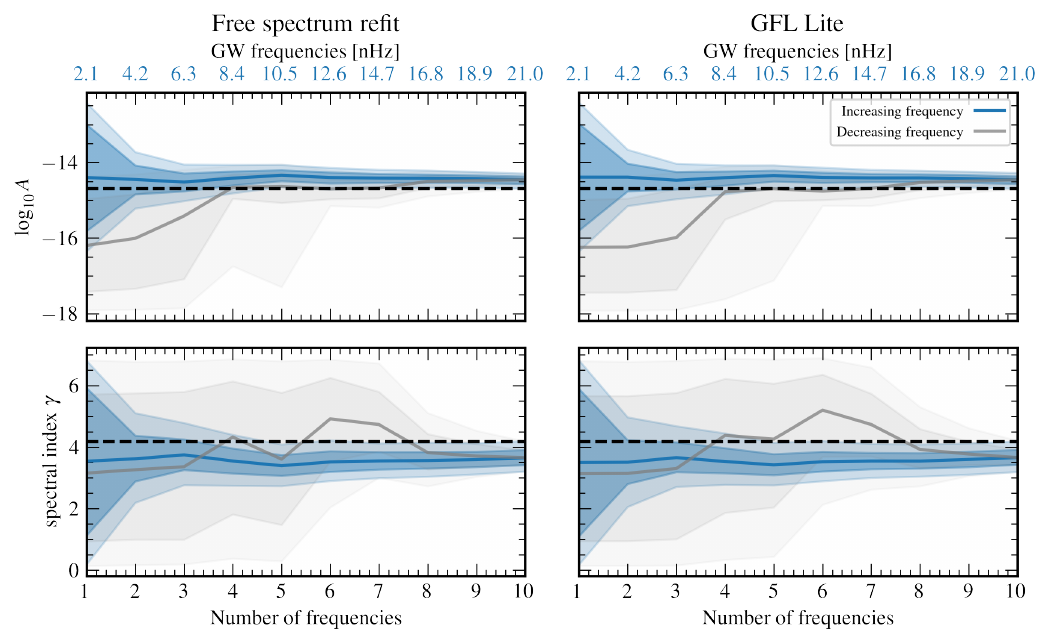}
        \caption{The median, 1-$\sigma$, and 2-$\sigma$ posterior credible constraints on $\log_{10}A$, $\gamma$ for a power-law process as a function of the number of modeled frequencies, $N_f$. The blue regions signify the constraints from fitting to the lowest frequency upwards (where these frequencies are explicitly shown in blue on the top $x$-axis)}, while the grey signifies fitting from the tenth frequency downwards.  As the number of frequencies increase, the posteriors become more constrained towards the injected parameter values (dashed black lines). For the PTA free-spectrum refit, we see the expected behavior of the blue contours constraining the parameters more quickly than the grey. Qualitatively, GFL Lite (right column) performs as well as the PTA free-spectrum refit (left column).
        \label{fig:perfreq}
    \end{figure*}

    \subsection{\label{sec:model_select}Model Selection}
    
    We now explore the efficacy of our refitting techniques for spectral model selection. The SGWB spectrum is typically modeled as a power-law, but other astrophysical and cosmological phenomena, and potentially even noise contamination, may influence its inferred shape. We would like to test whether these models better fit PTA data than a simple power-law, and make astrophysical and cosmological interpretations from their spectral characteristics.

    Model selection with the current production-level PTA analysis pipeline is challenging given the relatively slow computation time of the PTA likelihood compared to the size of the parameter space that must be searched over. We must compare the Bayesian evidence of our data given our hypothesis models, $p(\vec{\delta t}|\mathcal{H})$, to derive a Bayes factor $\mathcal{B}_{12}=p(\vec{\delta t}|\mathcal{H}_1)/p(\vec{\delta t}|\mathcal{H}_2)$, and interpret those values to reject or accept $\mathcal{H}_1$ over $\mathcal{H}_2$. The interpretation is problem-specific, but some rules-of-thumb are given in \citet{kass1995bayes}. In PTA analysis, model selection is typically conducted via calculating the Savage-Dickey density ratio \citep{dickey1971weighted} for low-contrast nested models, or with product-space sampling for mildly-disjoint nested models \citep[see, e.g.,][]{carlin1995bayesian,godsill2001relationship,aggarwal2019nanograv}.

    One model selection technique that is currently impractical for production-level PTA analyses on large arrays ($\gtrsim 40$~pulsars) is nested sampling, for which one analyzes each model separately to compute the Bayesian evidence \citep{skilling2004nested,2021arXiv210109675B}). Nested sampling is computationally expensive and cannot be realistically used with the full PTA likelihood given the combination of parameter dimensionality and slow evaluation time for larger arrays. In the PTA literature, nested sampling has been used before, but only for a small collection of pulsars \citep{chen2021common}. Our new techniques now make spectral model selection via nested sampling feasible for larger PTAs.
    \begin{table}
        \begin{tabular}{c|c|c}
            \hline
            Disfavored model & Favored model & $\mathcal{B}$ \\
            \hline
            \hline
             broken power-law & power-law & $21.1 \pm 6.0$ \\
             turnover & power-law & $1.71 \pm 0.44$ \\
             $t$-process & power-law & $50.7 \pm 11.8$ \\ \hline
        \end{tabular}
        \caption{Bayes factors for different 10-frequency CURN spectral models compared to a power-law when refitted to a PTA free-spectrum via the \texttt{Ultranest} nested sampler \citep{ultranest}. As expected, a power-law model is favored over every other tested model.}
        \label{table:table}
    \end{table}

        \autoref{table:table} compares Bayes factors between various spectral models and the injected power-law behavior from the same case-study simulation as \autoref{fig:corner}, using the PTA free-spectrum refitting technique. A \textit{broken power-law} has power-law behavior at low frequencies that then transitions into (in this case) a flat spectrum at higher frequencies in order to account for a white-noise floor in real data. This is used often in production-level analyses as a data-driven way of identifying the optimal number of frequencies with which to model a common red-noise process such that the inference is not biased by white noise \cite{Arzoumanian2020-br}. A \textit{turnover model} is similar in spirit to the broken power-law---in that it is effectively two power-laws connected by a bend---but motivated as a way to model low-frequency SGWB spectral attenuation from a binary population's interactions with their respective galaxy environments \cite{Sampson2015-qj,Arzoumanian_2016}. A \textit{$t$-process} model has an underlying power-law behavior, but with per-frequency deviations that are constrained by an inverse-Gamma prior. This is used to account for spectral fuzziness owing to noise conflation with the CURN, or potentially even binary-population finiteness influencing the spectral shape \cite{arzoumanian2020multimessenger}. Unsurprisingly, the power-law is the most favored model, since it is the injected spectrum. However, the power-law is only slightly favored over the turnover model with $\mathcal{B}=1.71$. This is because we allowed the range of turnover frequencies to be in any of the $10$ modeled GW frequency bins. The model favored the lowest frequency bin, which made it behave mostly like a power-law. The broken power-law's bend frequency was also allowed to vary across all frequencies, however it is much less favored than the power-law because its spectral index at frequencies greater than the bend frequency is fixed at zero, which the data do not support. Similarly, the injected power-law signal is so strong that any noise-induced deviations from it are small, thereby disfavoring the $t$-process model.
    
        Using our spectral refitting techniques, it is now possible to systematically explore the evidence for various realistic SGWB spectra in PTA data. We however emphasize that this is currently only for spectral model selection; necessary developments for performing model selection between inter-pulsar correlated models (e.g., monopole, dipole, Hellings--Downs), or to assess evidence for the presence of a CURN process over only intrinsic per-pulsar noise, are discussed in \autoref{sec:discuss}.

    \subsection{Evolution of Bayesian spectral constraints with number of GW frequencies and pulsars}

    Given that spectral characterization is now trivial with our refitting techniques, we use our simulations to study how the Bayesian inference of spectral characteristics evolves with the number of modeled GW frequencies and pulsars.
    
    \subsubsection{Dependence on number of GW frequencies} \label{subsec:specchar_numfreq}
    
    In \autoref{fig:perfreq} we recover the Bayesian posterior for a CURN power-law process on our case-study simulation as a function of the number of modeled GW frequencies. Typically, we fit a common-process model to the $N_f$ lowest GW frequencies; this is shown by the blue regions. However, we may also fit a power-law to our highest $N_f$ frequencies, given by the grey contours. This gives a comparison between the information content of the highest versus lowest frequencies. The SGWB spectrum from astrophysical or cosmological sources is expected to be red, with more power at lower frequencies. Hence, PTAs should be more sensitive to the SGWB at frequencies of $\sim 1/T$ than at higher frequencies, where intrinsic per-pulsar red noise and white noise can dominate \cite{moore2015estimating, hazboun}. Therefore we expect the blue contours to converge toward the lines of injected values faster than the grey contours; we see this for both the PTA free-spectrum refit and GFL Lite techniques, where the posterior spread in recovered parameters decreases significantly after only two frequencies. The grey contours (representing fitting from higher frequencies downwards) remain wide for a larger number of modeled frequencies, where both techniques require eight frequencies to converge on the spectral index $\gamma$, while the recovered amplitude converges on the injection after only $4$ frequencies. As expected, PTAs derive most information on SGWB spectral characteristics from the lowest analyzed GW frequencies, by virtue of the fact that red noise processes have more power there.

 	\subsubsection{Dependence on number of pulsars}
	
	\begin{figure}
        \centering
        \includegraphics[width=\columnwidth]{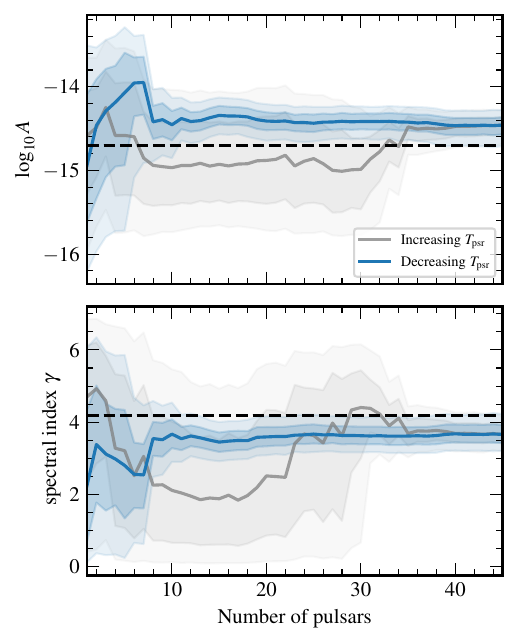}
        \caption{The median, 1-$\sigma$, and 2-$\sigma$ posterior credible constraints on $\log_{10}A$, $\gamma$ for a power-law process as a function of the number of modeled pulsars, $N_p$. The blue regions signify posterior constraints of a $10$ frequency power-law CURN fitted to the $N_p$-pulsars with the longest observing timespans, while the grey regions are the corresponding constraints from the $N_p$ shortest-timed pulsars. The black dashed line denotes the injected SGWB spectral characteristics.}
        \label{fig:perpsr}
    \end{figure}

    We may also analyze the SGWB parameter posterior recovery as a function of the number of pulsars $N_p$ in our PTA (\autoref{fig:perpsr}), this time using the GFL Lite technique. Given the large number of combinations with which $N_p$ pulsars can be chosen from the array of $45$, we only look at two sets of analyses, where we either add pulsars by decreasing or increasing timespan. Similar to \autoref{subsec:specchar_numfreq}, pulsars with longer observational timespans should be more informative of lower GW frequencies, where the signal is expected to be strongest. Therefore, we expect---and indeed, observe---that the blue contours converge on the injected parameter values  faster than the grey contours, requiring only the $\sim$eight longest-timespan pulsars before the median and posterior credible regions of the recovered spectral characteristics become approximately constant. By contrast, the $\sim 35$ shortest-timed pulsars are required to recover the same precision as those eight longest-timed pulsars. 
    
    \subsubsection{Characterization through the effective number of pulsars}\label{sec:neff}

    From these analyses, it is clear that not all pulsars and frequencies contribute equally toward spectral characterization. Frequencies with more noise than others will be down-weighted in spectral model fitting, as will pulsars whose overall noise level exceeds that of others. Using the GFL Lite free-spectrum of each pulsar, we can calculate the \textit{effective number of pulsars} $N_\mathrm{eff}$ in an $N_p$-pulsar PTA searching for an $N_f$ frequency power-law SGWB spectrum. We adapt and modify Eq.\ (8) in \citet{Cornish2015-ly} to the case of spectral characterization, also accounting for the uncertainty on the free-spectrum measurements:
    \begin{equation} \label{eq:neff}
        N_\mathrm{eff} = \frac{\sum_{p=1}^{N_p}\sum_{k=1}^{N_f} 1/\sigma(\log_{10}\rho_{p,k})^2}{\mathrm{max}_{1\leq p\leq N_p}\sum_{k=1}^{N_f} 1/\sigma(\log_{10}\rho_{p,k})^{2}},
    \end{equation}
    where $\sigma$ is measurement uncertainty. The free-spectrum posteriors $\log_{10}\rho_{p,k}$ come from the $p$-th pulsar and $k$-th frequency of the GFL Lite free-spectrum pipeline. We estimate the measurement uncertainty of the posterior of the $p$-th pulsar and $k$-th frequency with $\sigma_\mathrm{G}$, a rank-based estimate of the standard deviation to account for distribution non-Gaussianity, $\sigma_\mathrm{G}\approx0.7413\times\mathrm{IQR}$, where $\mathrm{IQR}$ is the interquartile range, and the prefactor originates from computing the IQR of a Gaussian \citep{ivezic2020statistics}. However, some posteriors are prior-dominated and uninformative, and estimating the standard deviation will return, at worst, that of the prior. We determine which pulsar and frequency posteriors are uninformative by computing the Savage-Dickey density ratio \citep{dickey1971weighted}, which, in this case, is used to estimate the Bayes factor between a model with and without a CURN process in a given pulsar, at a given frequency. $\mathcal{B}> 1$ suggests that a CURN process is supported, while if $\mathcal{B}< 1$, we determine that it is uninformative and set $\sigma=\infty$. The normalization of \autoref{eq:neff} ensures that $N_\mathrm{eff}\geq1$ for all $N_p$ and $N_f$. Therefore $N_\mathrm{eff}$ is the effective number of pulsars relative to the most constrained (i.e., least noisy, and therefore most informative) pulsar for spectral characterization in the array. For a PTA with heterogeneous pulsar spectral uncertainties, $N_\mathrm{eff}<N_p$, while a PTA with homogeneous uncertainties would have $N_\mathrm{eff}=N_p$. 
    
\begin{figure}
        \centering
        \includegraphics[width=\columnwidth]{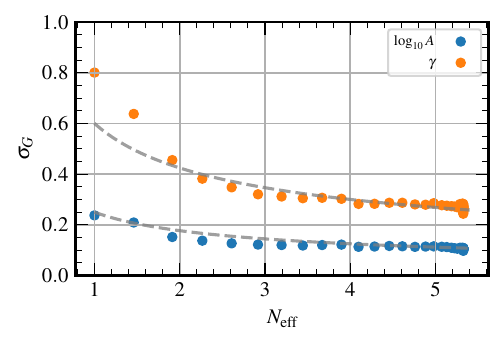}
        \caption{The relationship between the effective number of pulsars in a PTA, $N_\mathrm{eff}$, and the uncertainty on the spectral parameters (see text for a definition of $\sigma_\mathrm{G}$), derived using GFL Lite. We increase $N_p$ and keep the number of GW frequencies as $10$. The recovered parameter uncertainties scale approximately as the expected $1/\sqrt{N_\mathrm{eff}}$ for both $\log_{10}A$ and $\gamma$.}
        \label{fig:sigma}
    \end{figure}

    \autoref{fig:sigma} shows the relationship between the power-law parameter uncertainties as a function of $N_\mathrm{eff}$. We fitted a 10-frequency, $N_p$-pulsar powerlaw with the GFL Lite pipeline, adding pulsars in order of the greatest to smallest value of $\sum_k^{N_f} 1/ \sigma_\mathrm{G}(\log_{10}\rho_{p,k})^{2}$, i.e., in order of most-constrained to least-constrained pulsar spectrum. We computed the marginalized posterior uncertainty on both power-law parameters using the rank-based standard-deviation estimate $\sigma_\mathrm{G}$, defined es earlier. We see here that increasing the number of real pulsars increases the effective number of pulsars in the PTA, and decreases $\sigma_\mathrm{G}$ for both parameters. These studies allow us to posit a general relationship for spectral constraints in Bayesian PTA analyses, where $\sigma_\mathrm{G}\propto 1/\sqrt{N_\mathrm{eff}}$, as one may expect for a standard-deviation-type quantity computed from a data sample.

    \subsection{Recreating the results of the NANOGrav $12.5$-year data set} \label{sec:realdata}
    
    \begin{figure}
        \centering
        \includegraphics[width=\columnwidth]{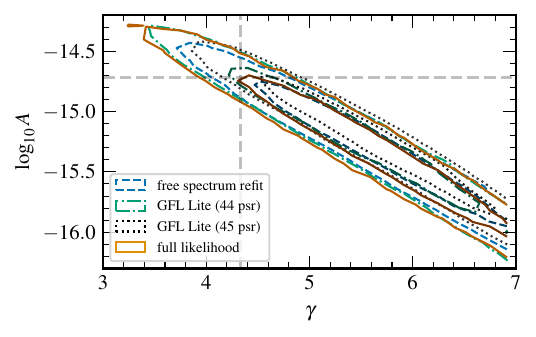}
        \caption{Fitting a $5$ frequency power-law to the NANOGrav $12.5$-year data set via the PTA free-spectrum refit technique (blue) and GFL Lite analysis (green and dotted black). We compare our analyses to the published posterior (orange). We find excellent agreement between the published result and the PTA free-spectrum refit, which attains a Hellinger distance of $H=0.13$. The 45-pulsar GFL Lite analysis does not recover the full likelihood posterior as well ($H=0.31$). However, removing one mismodeled pulsar results in better performance (green, $H=0.12$) -- see text for details.}
        \label{fig:12p5}
    \end{figure}

    We now apply our refitting techniques to the NANOGrav $12.5$-year data set to assess performance against published results. Analysis of the NANOGrav $12.5$-year data set did not find significant evidence for Hellings \& Downs inter-pulsar correlations, however, there was strong evidence for a CURN process. The posterior probability density for an analysis with a $5$ frequency power-law CURN process (including $30$ frequency power-law per-pulsar intrinsic red noise) is shown in \autoref{fig:12p5}, along with a PTA free-spectrum refit and GFL Lite analysis on this data set. The PTA free-spectrum refit is consistent with the published full likelihood with a Hellinger distance of $H=0.13$.
    
    For GFL Lite, we modeled a 5-frequency free-spectrum and 30-frequency power-law (to model the intrinsic red noise), and refit to the $5$ free-spectrum posteriors. We found that, because there is excess unmodelled noise in the real data set\footnote{The potential for model misspecification in pulsar timing datasets when only simple noise models are used has now been recognized. Ameliorating this requires custom noise modeling. This has been challenging to incorporate in large-array studies, but is recognized as the correct path forward.}, modeling a greater number of frequencies with the free-spectrum in each individual pulsar resulted in noise corruption, causing the free-spectrum to be conflated with intrinsic red noise in some pulsars. This is not a problem in the PTA free-spectrum refit, where the strength of the CURN from all of the pulsars inhibits the potential conflation with intrinsic red noise in pulsars that have misspecified noise models. Keeping the GFL Lite free-spectrum to just $5$ frequencies, and allowing the intrinsic red noise to be informed by $30$ frequencies, attempts to limit this confusion. Unfortunately for pulsar B1855+09, the power-law is a poor model for its intrinsic red noise, resulting in the free-spectrum posterior recovering the strong intrinsic red noise of this pulsar rather than the CURN. When a $5$ frequency GFL Lite refit is conducted, this pulsar is influential, causing the GFL Lite refit posterior to appear slightly offset from that of the full PTA likelihood in \autoref{fig:12p5}, with a Hellinger distance of $H=0.31$. Removing this pulsar results in a more consistent refit, with a Hellinger distance of just $H=0.12$. For the remainder, unless otherwise specified, we conduct the GFL Lite refit with just 44 pulsars. Improving the modelling of B1855+09 is beyond the scope of this paper and we discuss how we can improve analysis of pulsars like it in \autoref{sec:discuss}.

    \autoref{table:table12p5} shows the results of model selection for various 5-frequency spectral models with the PTA free-spectrum refit via nested sampling, a technique that estimates the Bayesian evidence of a model, and which we introduced in \autoref{sec:model_select}. We see that a varied-$\gamma$ power-law is barely favored over a $\gamma=13/3$ power-law, and $\gamma=13/3$ is not ruled out by these data. There is a little more evidence to favor a power-law over broken power-law, turnover, and $t$-process spectra, however none of these are substantial.
    \begin{table}
       \begin{tabular}{c|c|c}
            \hline
            Disfavored model & Favored model & $\mathcal{B}$ \\
            \hline
            \hline
             $\gamma\!=\!13/3$ power-law & power-law & $1.17\pm0.40$\\
             broken power-law & power-law & $1.82\pm0.34$\\
             turnover & power-law & $2.23\pm0.57$\\
             $t$-process & power-law & $1.83\pm0.57$\\ \hline
        \end{tabular}
        \caption{Bayes factors $\mathcal{B}$ for different 5-frequency common-process spectral models compared to a power-law, when refitted to a PTA free-spectrum for the NANOGrav $12.5$-year data set. The power-law has varied spectral index $\gamma$ unless stated.}
        \label{table:table12p5}
    \end{table}
    
    \begin{figure*}
        \centering
        \includegraphics[width=\textwidth]{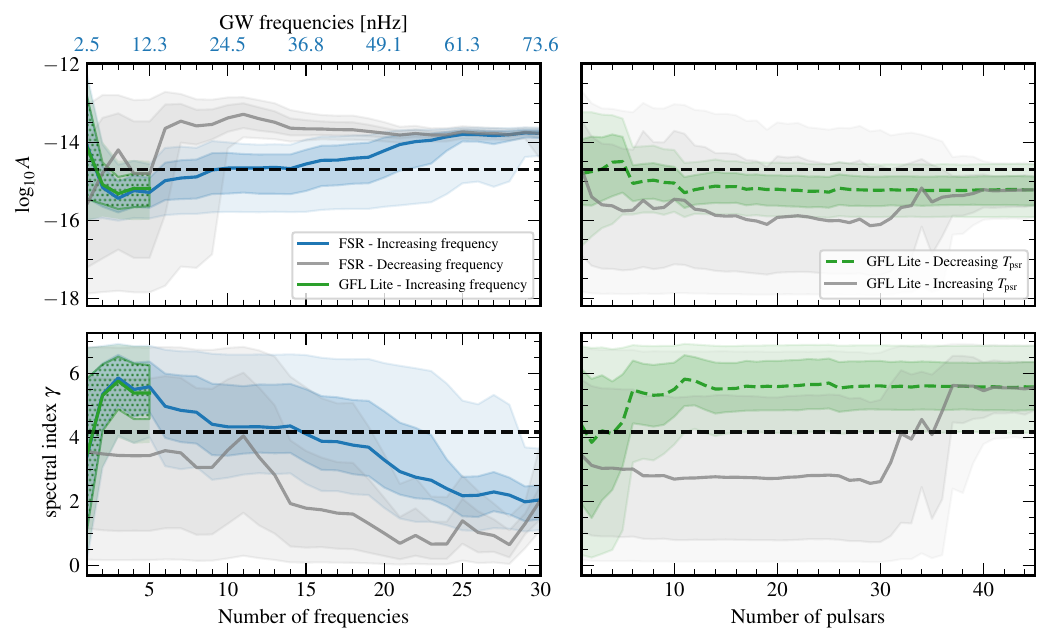}
        \caption{\textit{Left:} The median, 1-$\sigma$, and 2-$\sigma$ posterior credible constraints on $\log_{10}A$, $\gamma$ for an $N_f$ frequency power-law as a function of number of GW frequencies in the NANOGrav $12.5$-year data set. In blue, we show increasing number of frequencies from the lowest bin and increase upwards to $f=30/T$ for the free-spectrum refit, and in green, we show the same analysis for GFL Lite upwards to $5/T$ which is consistent with the blue contour. In grey, we show addition of frequencies from $f=30/T$ downwards for the free-spectrum refit. We observe a similar shallowing of the spectrum as \citet{Arzoumanian2020-br} when a larger number of frequencies are modeled because of the contribution of white noise. \textit{Right:} A 5-frequency power-law is fit to an increasing number of pulsars in the NANOGrav $12.5$-year data set, where green regions show constraints from adding pulsars in longest- to shortest-timed order. The blue posteriors are well constrained after $\sim14$ pulsars, while the grey posterior require $\sim36$ pulsars out of 45 to be constrained. Hence, the longest 14 pulsars are the most important for spectral characterization in this data set.}
        \label{fig:perfreq_12.5}
    \end{figure*}

    We also characterize the spectral recovery as a function of the number of modeled GW frequencies and pulsars. The PTA free-spectrum refit to increasing numbers of low GW frequencies (blue) in the left panel of \autoref{fig:perfreq_12.5} shows a similar ``shallowing'' of the spectrum as seen in \citet{Arzoumanian2020-br}, where $\gamma$ trends toward $\sim2-3$, potentially due to coupling with unmodeled excess higher frequency noise. Meanwhile, increasing the number of frequencies from $f=30/T$ downwards tends to have a broad, unconstrained posterior for all frequencies consistent with low $\gamma$ i.e. a flatter power spectrum typified by white noise. Fitting up to the first 5 frequencies, GFL Lite is consistent with the PTA free-spectrum refit. In the right panel, we use GFL Lite to fit a $5$ frequency power-law to an increasing number of pulsars (including B1855+09), added by decreasing pulsar timespan (green) and increasing pulsar timespan (grey). As with \autoref{fig:perpsr}, the green posterior is constrained after relatively few pulsars are added ($\sim 14$),\footnote{The 14 pulsars are J1744-1134, J1455-3330, J1012+5307, B1937+21, J2145-0750, J1909-3744, J1918-0642, J1643-1224, J2317+1439, B1855+09, J1713+0747, J0030+0451, J1640+2224 and J0613-0200. It is likely that removing B1855+09 would result in better constraining power.} while the grey posterior is unconstrained, particularly for $\gamma$, until the final 10 pulsars are added.

\begin{figure}
        \centering
        \includegraphics[width=\columnwidth]{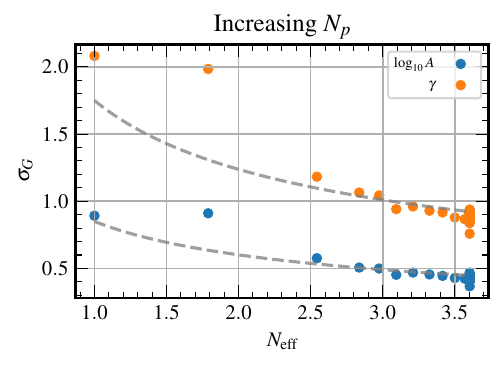}
        \caption{Similar to \autoref{fig:sigma}, using the NANOGrav 12.5 year data set, minus B1855+09}. We fit a 5-frequency power-law to an increasing number of pulsars in the array. About 12 pulsars ($N_\mathrm{eff}\sim3.6$) inform the spectral characteristics of the CURN, with a scaling of $1/\sqrt{N_\mathrm{eff}}$. We use the single pulsar free-spectra from GFL Lite for calculating $N_\mathrm{eff}$.
        \label{fig:perfreqpsr12.5}
    \end{figure}

    Finally, we also analyze the measurement uncertainty of $\{\log_{10}A, \gamma\}$ as a function of the effective number of pulsars, $N_\mathrm{eff}$, using the GFL Lite technique. \autoref{fig:perfreqpsr12.5} shows $\sigma_\mathrm{G}$ of parameters for a $5$ frequency power-law model as a function of $N_\mathrm{eff}$ as we increase the number of pulsars in the array in order of greatest to the smallest value of $\sum_k^{N_f} 1/ \sigma_\mathrm{G}(\log_{10}\rho_{p,k})^{2}$. We use the same methods as \autoref{sec:neff}. Again, we observe an approximate $1/\sqrt{N_\mathrm{eff}}$ scaling relationship. We need at most 4 pulsars that are equivalent to the best modeled pulsar in order to effectively recover the spectral characteristics of the CURN from this data set. We also notice that the real data set has fewer numbers of effective pulsars than our earlier simulated data sets, due to the data model of the simulations being entirely known and prescribed.

\section{\label{sec:discuss}Conclusions \& Future Prospects}

    We have developed a set of rapid and robust spectral refitting techniques that operate on posterior samples from pulsar and PTA Bayesian periodogram analyses, where the power spectral density is jointly modeled by free parameters at each GW frequency (sometimes referred to in the PTA literature as \textit{free-spectrum} analyses). This is a generalization of our previously-developed \textit{Factorized Likelihood} (FL) technique \citep{Taylor2022-nq}, where GW background amplitude posteriors for fixed power-law spectral index models are combined in post-processing, under the assumption that spectral characterization is mostly driven by autocorrelation information in the PTA covariance matrix. The main limitation of FL was its conditioning on a GW-background spectral model with a fixed power-law spectral index. Our new formulations loosen that assumption, allowing for refitting and inference of arbitrary spectral models.
    
    In order of generality, we assessed the performance of: a model that refits on a Bayesian PTA free-spectrum (\textit{PTA free-spectrum refit}); and one that refits on the combination of per-pulsar free-spectra, which act as proxies for the CURN signal in each pulsar, with intrinsic per-pulsar red noise modeled separately (\textit{GFL Lite}). These techniques are several orders of magnitude faster in evaluating their likelihood functions when compared to the production-level PTA pipeline, and also scale much more favorably when adding new pulsars. These gains in speed and scalability will be important in safeguarding PTA analyses from future bottlenecks, as significantly more data and pulsars are added to arrays through IPTA combinations and high-cadence observations in MeerTime \citep{bailes2018meertime}, CHIME \citep{chime}, and (farther in the future) the SKA \citep{ska}.
    
    We assessed the fidelity of parameter estimation using a set of $100$ realistic PTA data sets based on the NANOGrav $12.5$-year data set that is extended into the future, and into which realizations of a GW background are injected with power-law spectral characteristics based on supermassive black-hole binary population models. Through Hellinger-distance comparisons---which assess the distance between probability distributions---we found that the \textit{PTA free-spectrum refit} and \textit{GFL Lite} analyses are equivalent in performance, and consistent with the full production-level PTA likelihood analysis. While equivalent, we recommend using these refit methods in the following cases. For the PTA free-spectrum refit, it should be used when one is analyzing the evolution of spectral characterization with the number of GW frequencies, because combined influence of the PTA will make it less likely to confuse a CURN process with high-frequency white and/or mismodelled noise. If available, this technique can also be used to refit on PTA free-spectral posteriors that have an assumed inter-pulsar correlation signature. For example, in the case study presented in \autoref{fig:corner}, the Hellinger distance (closer to zero is better) of the PTA free-spectrum refit was $0.06$ when refitting on the HD-correlated free-spectrum, compared to $0.10$ from the CURN free-spectrum. Additionally, we showed that the PTA free-spectrum refit technique allows us to trivially perform spectral-model selection. The \textit{GFL Lite} technique should be used in studying how different subsets of frequencies and pulsars---e.g., long-baseline pulsars versus short-baseline pulsars---affect the spectral characterization of a CURN process (which is used as an approximate spectral model of the GW background). Care must be taken to ensure that, at the single pulsar level, the CURN process and intrinsic pulsar noise are not being conflated, as this will reduce the accuracy of the refitted posterior when compared to the full PTA likelihood. GFL Lite is more sensitive to noise misspecification than the PTA free-spectrum refit, since the latter has the benefit of many other pulsars to mitigate the impact of noise and CURN conflation.

    We plan for a further generalization of the GFL Lite method, called GFL,  which will have the advantage of allowing trivial changes to the spectral models and priors of the intrinsic red noise in each pulsar. This method will enable quick GW-background analyses in the presence of advanced per-pulsar red-noise models that are customized to each pulsar, which is currently not tractable with the production-level PTA pipeline for large arrays. As shown in \autoref{sec:realdata}, model misspecification of intrinsic red noise results in an inaccurate refitted posterior; advanced noise modeling, in concert with GFL, will improve SGWB spectral characterization for more pulsars and numbers of GW frequencies \citep{Agazie_2023}. We also suspect that the main loss of information and fidelity at the moment is through the sampling and representation of the distribution tails of the per-pulsar free-spectral posteriors. A potential solution to this is to use Gibbs techniques and to draw directly from the analytic conditional posteriors of our free-spectral parameters, which has been shown to have better tail sampling \citep{geman1984stochastic, PhysRevD.108.063008}. This work will appear in a future publication.

    Improvements to the representation of the posterior densities could be achieved through alternative KDE kernel functions that have more gradual drop-offs in support. If information is being lost in our density estimation, then performance gains may be made through multivariate KDEs across frequencies, or other higher-dimensional density estimation techniques based on neural network architectures, such as normalizing flows \citep{rezende2015variational}. Another avenue is based on likelihood reweighting techniques, where an approximate distribution that is easier to sample is used to generate many random draws, then a subsequent reweighting stage updates these samples based on their support under the correct (potentially computationally-expensive) distribution \citep[see, e.g.,][for a recent PTA application]{Hourihane_Meyers_Johnson_Chatziioannou_Vallisneri_2022}. Given the speed with which GFL refit analyses can be conducted, we could subsequently reweight these samples to match the full PTA likelihood. While this procedure will add extra computation time, it would still be quite a bit faster than a full pipeline analysis.

    We also envision that future development of GFL-style refitting techniques will include inter-pulsar correlations, which would be the zenith of stochastic GW-background modeling through compressed sufficient statistics. While our current techniques are based on power-spectrum modeling, we would need to recover the Fourier coefficients of the timing residuals in order to retain phase information among the pulsars. We would then need to accurately represent the likelihood distribution of these Fourier coefficients, using density estimation techniques, to act as sufficient statistics for inter-pulsar correlation studies. There is ongoing development along these lines to replace the current production-level PTA pipeline and ensure that future Bayesian PTA analyses with significantly larger data sets will continue to be tractable.

    The new techniques presented in this paper will have several immediate benefits for astrophysical- and cosmological-model testing with PTA data. The demographics and dynamics of supermassive black-hole binary populations is encoded in the amplitude and shape of the GW characteristic strain spectrum in the PTA band. Our techniques offer a path to use intermediate data products (i.e., Bayesian free-spectrum posteriors) for rapid spectral parameter estimation and model selection. Likewise, several potential sources of early-Universe GW-background signals give rise to strain spectra that deviate from the expected form of the supermassive black-hole binary population signal, e.g., a phase transition may produce a more peaked spectrum than the power-law expected from binaries.
    
    We plan to use our fast and flexible techniques to study milestones for PTA spectral estimation, such as what can be inferred in the near future about SMBHB populations, and the conditions under which cosmological background signals could be inferred beneath a dominant astrophysical signal. Answering these questions, and developing the spectral-estimation techniques with which they are addressed, are key to illuminating the path for PTA science in the next decade.

    \subsection{\label{sec:software}Software}
        The introduced refit methods are featured in a new analysis suite called \href{https://github.com/astrolamb/ceffyl}{\texttt{ceffyl}} for quick model selection and parameter estimation of spectra given PTA data. This is achieved by creating condensed data products representing the Bayesian spectra of a PTA's timing residuals. The data are represented by highly optimized KDEs from which we can extract probabilities to form Bayesian likelihoods to estimate our PTA likelihoods and to rapidly recover posteriors to our models. The suite employs code from \texttt{enterprise} \citep{enterprise}, which was also used to create our free-spectra and the full likelihood posteriors to which our analyses were compared. The PTA free-spectrum refit method is featured in the wrapper code, \href{https://github.com/andrea-mitridate/PTArcade}{\texttt{PTArcade}} \citep{Mitridate2023oar}. We conducted parameter estimation via MCMC with \texttt{PTMCMC} \citep{justin_ellis_2017_1037579}, which utilizes parallel tempering and empirical proposal distributions for more efficient sampling of the parameter space, while the nested sampler \texttt{UltraNest} \citep{ultranest} is used for model selection. To calculate the relevant KDE bandwidths, we translated the Sheather-Jones algorithm from an R implementation \citep{R} into Python; this code is now contained within the \texttt{ceffyl} suite. The KDEs are created using the \texttt{FFTKDE} method in \texttt{KDEpy} \citep{tommy}, and we use \texttt{ChainConsumer} \citep{Hinton2016} to create our corner plots to compare posteriors. The suite of PTA simulations were created with \texttt{libstempo} \citep{2020ascl.soft02017V}.

\begin{acknowledgements}
We thank our colleagues in NANOGrav and the International Pulsar Timing Array for fruitful discussions and feedback during the development of this technique. We particularly thank Alberto Sesana for providing the distributions of power-law fit parameters to characteristic strain spectra from SMBHB population realizations, based on \citet{rosado}; Kyle Gersbach for conversations on using the Pearson correlation coefficient; David Wright for help to make the \texttt{ceffyl} software suite easily installable; Xavier Siemens and Michele Vallisnerifor stimulating discussions; and Joe Romano for creating a gallery of $p$-$p$ plots that inspired our \autoref{fig:pp-explanation}. SRT acknowledges support from NSF AST-2007993, the NANOGrav NSF Physics Frontier Center \#2020265, and an NSF CAREER \#2146016.  WGL is supported by the NANOGrav NSF Physics Frontier Center \#2020265, and acknowledges travel support from the Divison of Gravitational Physics (DGRAV) to present this work at the APS April 2022 meeting, and from Vanderbilt University's Graduate Student Council. This work was conducted in part using the resources of the Advanced Computing Center for Research and Education (ACCRE) at Vanderbilt University, Nashville, TN. This work was performed in part at Aspen Center for Physics, which is supported by National Science Foundation grant PHY-2210452.
\end{acknowledgements}

\bibliography{apssamp}

\providecommand{\noopsort}[1]{}\providecommand{\singleletter}[1]{#1}%
\begin{thebibliography}{83}%
\makeatletter
\providecommand \@ifxundefined [1]{%
 \@ifx{#1\undefined}
}%
\providecommand \@ifnum [1]{%
 \ifnum #1\expandafter \@firstoftwo
 \else \expandafter \@secondoftwo
 \fi
}%
\providecommand \@ifx [1]{%
 \ifx #1\expandafter \@firstoftwo
 \else \expandafter \@secondoftwo
 \fi
}%
\providecommand \natexlab [1]{#1}%
\providecommand \enquote  [1]{``#1''}%
\providecommand \bibnamefont  [1]{#1}%
\providecommand \bibfnamefont [1]{#1}%
\providecommand \citenamefont [1]{#1}%
\providecommand \href@noop [0]{\@secondoftwo}%
\providecommand \href [0]{\begingroup \@sanitize@url \@href}%
\providecommand \@href[1]{\@@startlink{#1}\@@href}%
\providecommand \@@href[1]{\endgroup#1\@@endlink}%
\providecommand \@sanitize@url [0]{\catcode `\\12\catcode `\$12\catcode
  `\&12\catcode `\#12\catcode `\^12\catcode `\_12\catcode `\%12\relax}%
\providecommand \@@startlink[1]{}%
\providecommand \@@endlink[0]{}%
\providecommand \url  [0]{\begingroup\@sanitize@url \@url }%
\providecommand \@url [1]{\endgroup\@href {#1}{\urlprefix }}%
\providecommand \urlprefix  [0]{URL }%
\providecommand \Eprint [0]{\href }%
\providecommand \doibase [0]{https://doi.org/}%
\providecommand \selectlanguage [0]{\@gobble}%
\providecommand \bibinfo  [0]{\@secondoftwo}%
\providecommand \bibfield  [0]{\@secondoftwo}%
\providecommand \translation [1]{[#1]}%
\providecommand \BibitemOpen [0]{}%
\providecommand \bibitemStop [0]{}%
\providecommand \bibitemNoStop [0]{.\EOS\space}%
\providecommand \EOS [0]{\spacefactor3000\relax}%
\providecommand \BibitemShut  [1]{\csname bibitem#1\endcsname}%
\let\auto@bib@innerbib\@empty
\bibitem [{\citenamefont {{Foster}}\ and\ \citenamefont
  {{Backer}}(1990)}]{1990ApJ...361..300F}%
  \BibitemOpen
  \bibfield  {author} {\bibinfo {author} {\bibfnamefont {R.~S.}\ \bibnamefont
  {{Foster}}}\ and\ \bibinfo {author} {\bibfnamefont {D.~C.}\ \bibnamefont
  {{Backer}}},\ }\bibfield  {title} {\bibinfo {title} {{Constructing a Pulsar
  Timing Array}},\ }\href {https://doi.org/10.1086/169195} {\bibfield
  {journal} {\bibinfo  {journal} {\apj}\ }\textbf {\bibinfo {volume} {361}},\
  \bibinfo {pages} {300} (\bibinfo {year} {1990})}\BibitemShut {NoStop}%
\bibitem [{\citenamefont {Agazie}\ \emph
  {et~al.}(2023{\natexlab{a}})\citenamefont {Agazie}, \citenamefont
  {Anumarlapudi}, \citenamefont {Archibald}, \citenamefont {Arzoumanian},
  \citenamefont {Baker}, \citenamefont {B{\'e}csy}, \citenamefont {Blecha},
  \citenamefont {Brazier}, \citenamefont {Brook}, \citenamefont {Burke-Spolaor}
  \emph {et~al.}}]{2023ApJ...951L...8A}%
  \BibitemOpen
  \bibfield  {author} {\bibinfo {author} {\bibfnamefont {G.}~\bibnamefont
  {Agazie}}, \bibinfo {author} {\bibfnamefont {A.}~\bibnamefont
  {Anumarlapudi}}, \bibinfo {author} {\bibfnamefont {A.~M.}\ \bibnamefont
  {Archibald}}, \bibinfo {author} {\bibfnamefont {Z.}~\bibnamefont
  {Arzoumanian}}, \bibinfo {author} {\bibfnamefont {P.~T.}\ \bibnamefont
  {Baker}}, \bibinfo {author} {\bibfnamefont {B.}~\bibnamefont {B{\'e}csy}},
  \bibinfo {author} {\bibfnamefont {L.}~\bibnamefont {Blecha}}, \bibinfo
  {author} {\bibfnamefont {A.}~\bibnamefont {Brazier}}, \bibinfo {author}
  {\bibfnamefont {P.~R.}\ \bibnamefont {Brook}}, \bibinfo {author}
  {\bibfnamefont {S.}~\bibnamefont {Burke-Spolaor}}, \emph {et~al.},\
  }\bibfield  {title} {\bibinfo {title} {The nanograv 15 yr data set: Evidence
  for a gravitational-wave background},\ }\href@noop {} {\bibfield  {journal}
  {\bibinfo  {journal} {The Astrophysical Journal Letters}\ }\textbf {\bibinfo
  {volume} {951}},\ \bibinfo {pages} {L8} (\bibinfo {year}
  {2023}{\natexlab{a}})}\BibitemShut {NoStop}%
\bibitem [{\citenamefont {Antoniadis}\ \emph {et~al.}(2023)\citenamefont
  {Antoniadis}, \citenamefont {Arumugam}, \citenamefont {Arumugam},
  \citenamefont {Babak}, \citenamefont {Bagchi}, \citenamefont {Nielsen},
  \citenamefont {Bassa}, \citenamefont {Bathula}, \citenamefont {Berthereau},
  \citenamefont {Bonetti} \emph {et~al.}}]{2023arXiv230616214A}%
  \BibitemOpen
  \bibfield  {author} {\bibinfo {author} {\bibfnamefont {J.}~\bibnamefont
  {Antoniadis}}, \bibinfo {author} {\bibfnamefont {P.}~\bibnamefont
  {Arumugam}}, \bibinfo {author} {\bibfnamefont {S.}~\bibnamefont {Arumugam}},
  \bibinfo {author} {\bibfnamefont {S.}~\bibnamefont {Babak}}, \bibinfo
  {author} {\bibfnamefont {M.}~\bibnamefont {Bagchi}}, \bibinfo {author}
  {\bibfnamefont {A.-S.~B.}\ \bibnamefont {Nielsen}}, \bibinfo {author}
  {\bibfnamefont {C.}~\bibnamefont {Bassa}}, \bibinfo {author} {\bibfnamefont
  {A.}~\bibnamefont {Bathula}}, \bibinfo {author} {\bibfnamefont
  {A.}~\bibnamefont {Berthereau}}, \bibinfo {author} {\bibfnamefont
  {M.}~\bibnamefont {Bonetti}}, \emph {et~al.},\ }\bibfield  {title} {\bibinfo
  {title} {The second data release from the european pulsar timing array iii.
  search for gravitational wave signals},\ }\href@noop {} {\bibfield  {journal}
  {\bibinfo  {journal} {arXiv preprint arXiv:2306.16214}\ } (\bibinfo {year}
  {2023})}\BibitemShut {NoStop}%
\bibitem [{\citenamefont {Reardon}\ \emph {et~al.}(2023)\citenamefont
  {Reardon}, \citenamefont {Zic}, \citenamefont {Shannon}, \citenamefont
  {Hobbs}, \citenamefont {Bailes}, \citenamefont {Di~Marco}, \citenamefont
  {Kapur}, \citenamefont {Rogers}, \citenamefont {Thrane}, \citenamefont
  {Askew} \emph {et~al.}}]{2023ApJ...951L...6R}%
  \BibitemOpen
  \bibfield  {author} {\bibinfo {author} {\bibfnamefont {D.~J.}\ \bibnamefont
  {Reardon}}, \bibinfo {author} {\bibfnamefont {A.}~\bibnamefont {Zic}},
  \bibinfo {author} {\bibfnamefont {R.~M.}\ \bibnamefont {Shannon}}, \bibinfo
  {author} {\bibfnamefont {G.~B.}\ \bibnamefont {Hobbs}}, \bibinfo {author}
  {\bibfnamefont {M.}~\bibnamefont {Bailes}}, \bibinfo {author} {\bibfnamefont
  {V.}~\bibnamefont {Di~Marco}}, \bibinfo {author} {\bibfnamefont
  {A.}~\bibnamefont {Kapur}}, \bibinfo {author} {\bibfnamefont {A.~F.}\
  \bibnamefont {Rogers}}, \bibinfo {author} {\bibfnamefont {E.}~\bibnamefont
  {Thrane}}, \bibinfo {author} {\bibfnamefont {J.}~\bibnamefont {Askew}}, \emph
  {et~al.},\ }\bibfield  {title} {\bibinfo {title} {Search for an isotropic
  gravitational-wave background with the parkes pulsar timing array},\
  }\href@noop {} {\bibfield  {journal} {\bibinfo  {journal} {The Astrophysical
  Journal Letters}\ }\textbf {\bibinfo {volume} {951}},\ \bibinfo {pages} {L6}
  (\bibinfo {year} {2023})}\BibitemShut {NoStop}%
\bibitem [{\citenamefont {Xu}\ \emph {et~al.}(2023)\citenamefont {Xu},
  \citenamefont {Chen}, \citenamefont {Guo}, \citenamefont {Jiang},
  \citenamefont {Wang}, \citenamefont {Xu}, \citenamefont {Xue}, \citenamefont
  {Caballero}, \citenamefont {Yuan}, \citenamefont {Xu}, \citenamefont {Wang},
  \citenamefont {Hao}, \citenamefont {Luo}, \citenamefont {Lee}, \citenamefont
  {Han}, \citenamefont {Jiang}, \citenamefont {Shen}, \citenamefont {Wang},
  \citenamefont {Wang}, \citenamefont {Xu}, \citenamefont {Wu}, \citenamefont
  {Manchester}, \citenamefont {Qian}, \citenamefont {Guan}, \citenamefont
  {Huang}, \citenamefont {Sun},\ and\ \citenamefont
  {Zhu}}]{2023RAA....23g5024X}%
  \BibitemOpen
  \bibfield  {author} {\bibinfo {author} {\bibfnamefont {H.}~\bibnamefont
  {Xu}}, \bibinfo {author} {\bibfnamefont {S.}~\bibnamefont {Chen}}, \bibinfo
  {author} {\bibfnamefont {Y.}~\bibnamefont {Guo}}, \bibinfo {author}
  {\bibfnamefont {J.}~\bibnamefont {Jiang}}, \bibinfo {author} {\bibfnamefont
  {B.}~\bibnamefont {Wang}}, \bibinfo {author} {\bibfnamefont {J.}~\bibnamefont
  {Xu}}, \bibinfo {author} {\bibfnamefont {Z.}~\bibnamefont {Xue}}, \bibinfo
  {author} {\bibfnamefont {R.~N.}\ \bibnamefont {Caballero}}, \bibinfo {author}
  {\bibfnamefont {J.}~\bibnamefont {Yuan}}, \bibinfo {author} {\bibfnamefont
  {Y.}~\bibnamefont {Xu}}, \bibinfo {author} {\bibfnamefont {J.}~\bibnamefont
  {Wang}}, \bibinfo {author} {\bibfnamefont {L.}~\bibnamefont {Hao}}, \bibinfo
  {author} {\bibfnamefont {J.}~\bibnamefont {Luo}}, \bibinfo {author}
  {\bibfnamefont {K.}~\bibnamefont {Lee}}, \bibinfo {author} {\bibfnamefont
  {J.}~\bibnamefont {Han}}, \bibinfo {author} {\bibfnamefont {P.}~\bibnamefont
  {Jiang}}, \bibinfo {author} {\bibfnamefont {Z.}~\bibnamefont {Shen}},
  \bibinfo {author} {\bibfnamefont {M.}~\bibnamefont {Wang}}, \bibinfo {author}
  {\bibfnamefont {N.}~\bibnamefont {Wang}}, \bibinfo {author} {\bibfnamefont
  {R.}~\bibnamefont {Xu}}, \bibinfo {author} {\bibfnamefont {X.}~\bibnamefont
  {Wu}}, \bibinfo {author} {\bibfnamefont {R.}~\bibnamefont {Manchester}},
  \bibinfo {author} {\bibfnamefont {L.}~\bibnamefont {Qian}}, \bibinfo {author}
  {\bibfnamefont {X.}~\bibnamefont {Guan}}, \bibinfo {author} {\bibfnamefont
  {M.}~\bibnamefont {Huang}}, \bibinfo {author} {\bibfnamefont
  {C.}~\bibnamefont {Sun}},\ and\ \bibinfo {author} {\bibfnamefont
  {Y.}~\bibnamefont {Zhu}},\ }\bibfield  {title} {\bibinfo {title} {Searching
  for the nano-hertz stochastic gravitational wave background with the chinese
  pulsar timing array data release i},\ }\href
  {https://doi.org/10.1088/1674-4527/acdfa5} {\bibfield  {journal} {\bibinfo
  {journal} {Research in Astronomy and Astrophysics}\ }\textbf {\bibinfo
  {volume} {23}},\ \bibinfo {pages} {075024} (\bibinfo {year}
  {2023})}\BibitemShut {NoStop}%
\bibitem [{\citenamefont {Hellings}\ and\ \citenamefont
  {Downs}(1983)}]{hellings1983upper}%
  \BibitemOpen
  \bibfield  {author} {\bibinfo {author} {\bibfnamefont {R.}~\bibnamefont
  {Hellings}}\ and\ \bibinfo {author} {\bibfnamefont {G.}~\bibnamefont
  {Downs}},\ }\bibfield  {title} {\bibinfo {title} {Upper limits on the
  isotropic gravitational radiation background from pulsar timing analysis},\
  }\href@noop {} {\bibfield  {journal} {\bibinfo  {journal} {The Astrophysical
  Journal}\ }\textbf {\bibinfo {volume} {265}},\ \bibinfo {pages} {L39}
  (\bibinfo {year} {1983})}\BibitemShut {NoStop}%
\bibitem [{\citenamefont {Antoniadis}\ \emph {et~al.}(2022)\citenamefont
  {Antoniadis}, \citenamefont {Arzoumanian}, \citenamefont {Babak},
  \citenamefont {Bailes}, \citenamefont {Bak~Nielsen}, \citenamefont {Baker},
  \citenamefont {Bassa}, \citenamefont {Bécsy}, \citenamefont {Berthereau},
  \citenamefont {Bonetti}, \citenamefont {Brazier}, \citenamefont {Brook},
  \citenamefont {Burgay}, \citenamefont {Burke-Spolaor}, \citenamefont
  {Caballero}, \citenamefont {Casey-Clyde}, \citenamefont {Chalumeau},
  \citenamefont {Champion}, \citenamefont {Charisi}, \citenamefont
  {Chatterjee}, \citenamefont {Chen}, \citenamefont {Cognard}, \citenamefont
  {Cordes}, \citenamefont {Cornish}, \citenamefont {Crawford}, \citenamefont
  {Cromartie}, \citenamefont {Crowter}, \citenamefont {Dai}, \citenamefont
  {DeCesar}, \citenamefont {Demorest}, \citenamefont {Desvignes}, \citenamefont
  {Dolch}, \citenamefont {Drachler}, \citenamefont {Falxa}, \citenamefont
  {Ferrara}, \citenamefont {Fiore}, \citenamefont {Fonseca}, \citenamefont
  {Gair}, \citenamefont {Garver-Daniels}, \citenamefont {Goncharov},
  \citenamefont {Good}, \citenamefont {Graikou}, \citenamefont {Guillemot},
  \citenamefont {Guo}, \citenamefont {Hazboun}, \citenamefont {Hobbs},
  \citenamefont {Hu}, \citenamefont {Islo}, \citenamefont {Janssen},
  \citenamefont {Jennings}, \citenamefont {Johnson}, \citenamefont {Jones},
  \citenamefont {Kaiser}, \citenamefont {Kaplan}, \citenamefont {Karuppusamy},
  \citenamefont {Keith}, \citenamefont {Kelley}, \citenamefont {Kerr},
  \citenamefont {Key}, \citenamefont {Kramer}, \citenamefont {Lam},
  \citenamefont {Lamb}, \citenamefont {Lazio}, \citenamefont {Lee},
  \citenamefont {Lentati}, \citenamefont {Liu}, \citenamefont {Luo},
  \citenamefont {Lynch}, \citenamefont {Lyne}, \citenamefont {Madison},
  \citenamefont {Main}, \citenamefont {Manchester}, \citenamefont {McEwen},
  \citenamefont {McKee}, \citenamefont {McLaughlin}, \citenamefont
  {Mickaliger}, \citenamefont {Mingarelli}, \citenamefont {Ng}, \citenamefont
  {Nice}, \citenamefont {Osłowski}, \citenamefont {Parthasarathy},
  \citenamefont {Pennucci}, \citenamefont {Perera}, \citenamefont {Perrodin},
  \citenamefont {Petiteau}, \citenamefont {Pol}, \citenamefont {Porayko},
  \citenamefont {Possenti}, \citenamefont {Ransom}, \citenamefont {Ray},
  \citenamefont {Reardon}, \citenamefont {Russell}, \citenamefont {Samajdar},
  \citenamefont {Sampson}, \citenamefont {Sanidas}, \citenamefont {Sarkissian},
  \citenamefont {Schmitz}, \citenamefont {Schult}, \citenamefont {Sesana},
  \citenamefont {Shaifullah}, \citenamefont {Shannon}, \citenamefont
  {Shapiro-Albert}, \citenamefont {Siemens}, \citenamefont {Simon},
  \citenamefont {Smith}, \citenamefont {Speri}, \citenamefont {Spiewak},
  \citenamefont {Stairs}, \citenamefont {Stappers}, \citenamefont {Stinebring},
  \citenamefont {Swiggum}, \citenamefont {Taylor}, \citenamefont {Theureau},
  \citenamefont {Tiburzi}, \citenamefont {Vallisneri}, \citenamefont {van~der
  Wateren}, \citenamefont {Vecchio}, \citenamefont {Verbiest}, \citenamefont
  {Vigeland}, \citenamefont {Wahl}, \citenamefont {Wang}, \citenamefont {Wang},
  \citenamefont {Wang}, \citenamefont {Witt}, \citenamefont {Zhang},\ and\
  \citenamefont {Zhu}}]{iptadr2}%
  \BibitemOpen
  \bibfield  {author} {\bibinfo {author} {\bibfnamefont {J.}~\bibnamefont
  {Antoniadis}}, \bibinfo {author} {\bibfnamefont {Z.}~\bibnamefont
  {Arzoumanian}}, \bibinfo {author} {\bibfnamefont {S.}~\bibnamefont {Babak}},
  \bibinfo {author} {\bibfnamefont {M.}~\bibnamefont {Bailes}}, \bibinfo
  {author} {\bibfnamefont {A.-S.}\ \bibnamefont {Bak~Nielsen}}, \bibinfo
  {author} {\bibfnamefont {P.~T.}\ \bibnamefont {Baker}}, \bibinfo {author}
  {\bibfnamefont {C.~G.}\ \bibnamefont {Bassa}}, \bibinfo {author}
  {\bibfnamefont {B.}~\bibnamefont {Bécsy}}, \bibinfo {author} {\bibfnamefont
  {A.}~\bibnamefont {Berthereau}}, \bibinfo {author} {\bibfnamefont
  {M.}~\bibnamefont {Bonetti}}, \bibinfo {author} {\bibfnamefont
  {A.}~\bibnamefont {Brazier}}, \bibinfo {author} {\bibfnamefont {P.~R.}\
  \bibnamefont {Brook}}, \bibinfo {author} {\bibfnamefont {M.}~\bibnamefont
  {Burgay}}, \bibinfo {author} {\bibfnamefont {S.}~\bibnamefont
  {Burke-Spolaor}}, \bibinfo {author} {\bibfnamefont {R.~N.}\ \bibnamefont
  {Caballero}}, \bibinfo {author} {\bibfnamefont {J.~A.}\ \bibnamefont
  {Casey-Clyde}}, \bibinfo {author} {\bibfnamefont {A.}~\bibnamefont
  {Chalumeau}}, \bibinfo {author} {\bibfnamefont {D.~J.}\ \bibnamefont
  {Champion}}, \bibinfo {author} {\bibfnamefont {M.}~\bibnamefont {Charisi}},
  \bibinfo {author} {\bibfnamefont {S.}~\bibnamefont {Chatterjee}}, \bibinfo
  {author} {\bibfnamefont {S.}~\bibnamefont {Chen}}, \bibinfo {author}
  {\bibfnamefont {I.}~\bibnamefont {Cognard}}, \bibinfo {author} {\bibfnamefont
  {J.~M.}\ \bibnamefont {Cordes}}, \bibinfo {author} {\bibfnamefont {N.~J.}\
  \bibnamefont {Cornish}}, \bibinfo {author} {\bibfnamefont {F.}~\bibnamefont
  {Crawford}}, \bibinfo {author} {\bibfnamefont {H.~T.}\ \bibnamefont
  {Cromartie}}, \bibinfo {author} {\bibfnamefont {K.}~\bibnamefont {Crowter}},
  \bibinfo {author} {\bibfnamefont {S.}~\bibnamefont {Dai}}, \bibinfo {author}
  {\bibfnamefont {M.~E.}\ \bibnamefont {DeCesar}}, \bibinfo {author}
  {\bibfnamefont {P.~B.}\ \bibnamefont {Demorest}}, \bibinfo {author}
  {\bibfnamefont {G.}~\bibnamefont {Desvignes}}, \bibinfo {author}
  {\bibfnamefont {T.}~\bibnamefont {Dolch}}, \bibinfo {author} {\bibfnamefont
  {B.}~\bibnamefont {Drachler}}, \bibinfo {author} {\bibfnamefont
  {M.}~\bibnamefont {Falxa}}, \bibinfo {author} {\bibfnamefont {E.~C.}\
  \bibnamefont {Ferrara}}, \bibinfo {author} {\bibfnamefont {W.}~\bibnamefont
  {Fiore}}, \bibinfo {author} {\bibfnamefont {E.}~\bibnamefont {Fonseca}},
  \bibinfo {author} {\bibfnamefont {J.~R.}\ \bibnamefont {Gair}}, \bibinfo
  {author} {\bibfnamefont {N.}~\bibnamefont {Garver-Daniels}}, \bibinfo
  {author} {\bibfnamefont {B.}~\bibnamefont {Goncharov}}, \bibinfo {author}
  {\bibfnamefont {D.~C.}\ \bibnamefont {Good}}, \bibinfo {author}
  {\bibfnamefont {E.}~\bibnamefont {Graikou}}, \bibinfo {author} {\bibfnamefont
  {L.}~\bibnamefont {Guillemot}}, \bibinfo {author} {\bibfnamefont {Y.~J.}\
  \bibnamefont {Guo}}, \bibinfo {author} {\bibfnamefont {J.~S.}\ \bibnamefont
  {Hazboun}}, \bibinfo {author} {\bibfnamefont {G.}~\bibnamefont {Hobbs}},
  \bibinfo {author} {\bibfnamefont {H.}~\bibnamefont {Hu}}, \bibinfo {author}
  {\bibfnamefont {K.}~\bibnamefont {Islo}}, \bibinfo {author} {\bibfnamefont
  {G.~H.}\ \bibnamefont {Janssen}}, \bibinfo {author} {\bibfnamefont {R.~J.}\
  \bibnamefont {Jennings}}, \bibinfo {author} {\bibfnamefont {A.~D.}\
  \bibnamefont {Johnson}}, \bibinfo {author} {\bibfnamefont {M.~L.}\
  \bibnamefont {Jones}}, \bibinfo {author} {\bibfnamefont {A.~R.}\ \bibnamefont
  {Kaiser}}, \bibinfo {author} {\bibfnamefont {D.~L.}\ \bibnamefont {Kaplan}},
  \bibinfo {author} {\bibfnamefont {R.}~\bibnamefont {Karuppusamy}}, \bibinfo
  {author} {\bibfnamefont {M.~J.}\ \bibnamefont {Keith}}, \bibinfo {author}
  {\bibfnamefont {L.~Z.}\ \bibnamefont {Kelley}}, \bibinfo {author}
  {\bibfnamefont {M.}~\bibnamefont {Kerr}}, \bibinfo {author} {\bibfnamefont
  {J.~S.}\ \bibnamefont {Key}}, \bibinfo {author} {\bibfnamefont
  {M.}~\bibnamefont {Kramer}}, \bibinfo {author} {\bibfnamefont {M.~T.}\
  \bibnamefont {Lam}}, \bibinfo {author} {\bibfnamefont {W.~G.}\ \bibnamefont
  {Lamb}}, \bibinfo {author} {\bibfnamefont {T.~J.~W.}\ \bibnamefont {Lazio}},
  \bibinfo {author} {\bibfnamefont {K.~J.}\ \bibnamefont {Lee}}, \bibinfo
  {author} {\bibfnamefont {L.}~\bibnamefont {Lentati}}, \bibinfo {author}
  {\bibfnamefont {K.}~\bibnamefont {Liu}}, \bibinfo {author} {\bibfnamefont
  {J.}~\bibnamefont {Luo}}, \bibinfo {author} {\bibfnamefont {R.~S.}\
  \bibnamefont {Lynch}}, \bibinfo {author} {\bibfnamefont {A.~G.}\ \bibnamefont
  {Lyne}}, \bibinfo {author} {\bibfnamefont {D.~R.}\ \bibnamefont {Madison}},
  \bibinfo {author} {\bibfnamefont {R.~A.}\ \bibnamefont {Main}}, \bibinfo
  {author} {\bibfnamefont {R.~N.}\ \bibnamefont {Manchester}}, \bibinfo
  {author} {\bibfnamefont {A.}~\bibnamefont {McEwen}}, \bibinfo {author}
  {\bibfnamefont {J.~W.}\ \bibnamefont {McKee}}, \bibinfo {author}
  {\bibfnamefont {M.~A.}\ \bibnamefont {McLaughlin}}, \bibinfo {author}
  {\bibfnamefont {M.~B.}\ \bibnamefont {Mickaliger}}, \bibinfo {author}
  {\bibfnamefont {C.~M.~F.}\ \bibnamefont {Mingarelli}}, \bibinfo {author}
  {\bibfnamefont {C.}~\bibnamefont {Ng}}, \bibinfo {author} {\bibfnamefont
  {D.~J.}\ \bibnamefont {Nice}}, \bibinfo {author} {\bibfnamefont
  {S.}~\bibnamefont {Osłowski}}, \bibinfo {author} {\bibfnamefont
  {A.}~\bibnamefont {Parthasarathy}}, \bibinfo {author} {\bibfnamefont {T.~T.}\
  \bibnamefont {Pennucci}}, \bibinfo {author} {\bibfnamefont {B.~B.~P.}\
  \bibnamefont {Perera}}, \bibinfo {author} {\bibfnamefont {D.}~\bibnamefont
  {Perrodin}}, \bibinfo {author} {\bibfnamefont {A.}~\bibnamefont {Petiteau}},
  \bibinfo {author} {\bibfnamefont {N.~S.}\ \bibnamefont {Pol}}, \bibinfo
  {author} {\bibfnamefont {N.~K.}\ \bibnamefont {Porayko}}, \bibinfo {author}
  {\bibfnamefont {A.}~\bibnamefont {Possenti}}, \bibinfo {author}
  {\bibfnamefont {S.~M.}\ \bibnamefont {Ransom}}, \bibinfo {author}
  {\bibfnamefont {P.~S.}\ \bibnamefont {Ray}}, \bibinfo {author} {\bibfnamefont
  {D.~J.}\ \bibnamefont {Reardon}}, \bibinfo {author} {\bibfnamefont {C.~J.}\
  \bibnamefont {Russell}}, \bibinfo {author} {\bibfnamefont {A.}~\bibnamefont
  {Samajdar}}, \bibinfo {author} {\bibfnamefont {L.~M.}\ \bibnamefont
  {Sampson}}, \bibinfo {author} {\bibfnamefont {S.}~\bibnamefont {Sanidas}},
  \bibinfo {author} {\bibfnamefont {J.~M.}\ \bibnamefont {Sarkissian}},
  \bibinfo {author} {\bibfnamefont {K.}~\bibnamefont {Schmitz}}, \bibinfo
  {author} {\bibfnamefont {L.}~\bibnamefont {Schult}}, \bibinfo {author}
  {\bibfnamefont {A.}~\bibnamefont {Sesana}}, \bibinfo {author} {\bibfnamefont
  {G.}~\bibnamefont {Shaifullah}}, \bibinfo {author} {\bibfnamefont {R.~M.}\
  \bibnamefont {Shannon}}, \bibinfo {author} {\bibfnamefont {B.~J.}\
  \bibnamefont {Shapiro-Albert}}, \bibinfo {author} {\bibfnamefont
  {X.}~\bibnamefont {Siemens}}, \bibinfo {author} {\bibfnamefont
  {J.}~\bibnamefont {Simon}}, \bibinfo {author} {\bibfnamefont {T.~L.}\
  \bibnamefont {Smith}}, \bibinfo {author} {\bibfnamefont {L.}~\bibnamefont
  {Speri}}, \bibinfo {author} {\bibfnamefont {R.}~\bibnamefont {Spiewak}},
  \bibinfo {author} {\bibfnamefont {I.~H.}\ \bibnamefont {Stairs}}, \bibinfo
  {author} {\bibfnamefont {B.~W.}\ \bibnamefont {Stappers}}, \bibinfo {author}
  {\bibfnamefont {D.~R.}\ \bibnamefont {Stinebring}}, \bibinfo {author}
  {\bibfnamefont {J.~K.}\ \bibnamefont {Swiggum}}, \bibinfo {author}
  {\bibfnamefont {S.~R.}\ \bibnamefont {Taylor}}, \bibinfo {author}
  {\bibfnamefont {G.}~\bibnamefont {Theureau}}, \bibinfo {author}
  {\bibfnamefont {C.}~\bibnamefont {Tiburzi}}, \bibinfo {author} {\bibfnamefont
  {M.}~\bibnamefont {Vallisneri}}, \bibinfo {author} {\bibfnamefont
  {E.}~\bibnamefont {van~der Wateren}}, \bibinfo {author} {\bibfnamefont
  {A.}~\bibnamefont {Vecchio}}, \bibinfo {author} {\bibfnamefont {J.~P.~W.}\
  \bibnamefont {Verbiest}}, \bibinfo {author} {\bibfnamefont {S.~J.}\
  \bibnamefont {Vigeland}}, \bibinfo {author} {\bibfnamefont {H.}~\bibnamefont
  {Wahl}}, \bibinfo {author} {\bibfnamefont {J.~B.}\ \bibnamefont {Wang}},
  \bibinfo {author} {\bibfnamefont {J.}~\bibnamefont {Wang}}, \bibinfo {author}
  {\bibfnamefont {L.}~\bibnamefont {Wang}}, \bibinfo {author} {\bibfnamefont
  {C.~A.}\ \bibnamefont {Witt}}, \bibinfo {author} {\bibfnamefont
  {S.}~\bibnamefont {Zhang}},\ and\ \bibinfo {author} {\bibfnamefont {X.~J.}\
  \bibnamefont {Zhu}},\ }\bibfield  {title} {\bibinfo {title} {The
  international pulsar timing array second data release: Search for an
  isotropic gravitational wave background},\ }\href
  {https://doi.org/10.1093/mnras/stab3418} {\bibfield  {journal} {\bibinfo
  {journal} {Monthly Notices of the Royal Astronomical Society}\ }\textbf
  {\bibinfo {volume} {510}},\ \bibinfo {pages} {4873–4887} (\bibinfo {year}
  {2022})}\BibitemShut {NoStop}%
\bibitem [{\citenamefont {Romano}\ \emph {et~al.}(2021)\citenamefont {Romano},
  \citenamefont {Hazboun}, \citenamefont {Siemens},\ and\ \citenamefont
  {Archibald}}]{Romano2020-mq}%
  \BibitemOpen
  \bibfield  {author} {\bibinfo {author} {\bibfnamefont {J.~D.}\ \bibnamefont
  {Romano}}, \bibinfo {author} {\bibfnamefont {J.~S.}\ \bibnamefont {Hazboun}},
  \bibinfo {author} {\bibfnamefont {X.}~\bibnamefont {Siemens}},\ and\ \bibinfo
  {author} {\bibfnamefont {A.~M.}\ \bibnamefont {Archibald}},\ }\bibfield
  {title} {\bibinfo {title} {Common-spectrum process versus cross-correlation
  for gravitational-wave searches using pulsar timing arrays},\ }\href@noop {}
  {\bibfield  {journal} {\bibinfo  {journal} {Physical Review D}\ }\textbf
  {\bibinfo {volume} {103}},\ \bibinfo {pages} {063027} (\bibinfo {year}
  {2021})}\BibitemShut {NoStop}%
\bibitem [{\citenamefont {Pol}\ \emph {et~al.}(2021)\citenamefont {Pol},
  \citenamefont {Taylor}, \citenamefont {Kelley}, \citenamefont {Vigeland},
  \citenamefont {Simon}, \citenamefont {Chen}, \citenamefont {Arzoumanian},
  \citenamefont {Baker}, \citenamefont {B{\'e}csy}, \citenamefont {Brazier}
  \emph {et~al.}}]{astro4cast}%
  \BibitemOpen
  \bibfield  {author} {\bibinfo {author} {\bibfnamefont {N.~S.}\ \bibnamefont
  {Pol}}, \bibinfo {author} {\bibfnamefont {S.~R.}\ \bibnamefont {Taylor}},
  \bibinfo {author} {\bibfnamefont {L.~Z.}\ \bibnamefont {Kelley}}, \bibinfo
  {author} {\bibfnamefont {S.~J.}\ \bibnamefont {Vigeland}}, \bibinfo {author}
  {\bibfnamefont {J.}~\bibnamefont {Simon}}, \bibinfo {author} {\bibfnamefont
  {S.}~\bibnamefont {Chen}}, \bibinfo {author} {\bibfnamefont {Z.}~\bibnamefont
  {Arzoumanian}}, \bibinfo {author} {\bibfnamefont {P.~T.}\ \bibnamefont
  {Baker}}, \bibinfo {author} {\bibfnamefont {B.}~\bibnamefont {B{\'e}csy}},
  \bibinfo {author} {\bibfnamefont {A.}~\bibnamefont {Brazier}}, \emph
  {et~al.},\ }\bibfield  {title} {\bibinfo {title} {Astrophysics milestones for
  pulsar timing array gravitational-wave detection},\ }\href@noop {} {\bibfield
   {journal} {\bibinfo  {journal} {The Astrophysical Journal Letters}\ }\textbf
  {\bibinfo {volume} {911}},\ \bibinfo {pages} {L34} (\bibinfo {year}
  {2021})}\BibitemShut {NoStop}%
\bibitem [{\citenamefont {Arzoumanian}\ \emph
  {et~al.}(2020{\natexlab{a}})\citenamefont {Arzoumanian}, \citenamefont
  {Baker}, \citenamefont {Blumer}, \citenamefont {B{\'e}csy}, \citenamefont
  {Brazier}, \citenamefont {Brook}, \citenamefont {Burke-Spolaor},
  \citenamefont {Chatterjee}, \citenamefont {Chen}, \citenamefont {Cordes}
  \emph {et~al.}}]{Arzoumanian2020-br}%
  \BibitemOpen
  \bibfield  {author} {\bibinfo {author} {\bibfnamefont {Z.}~\bibnamefont
  {Arzoumanian}}, \bibinfo {author} {\bibfnamefont {P.~T.}\ \bibnamefont
  {Baker}}, \bibinfo {author} {\bibfnamefont {H.}~\bibnamefont {Blumer}},
  \bibinfo {author} {\bibfnamefont {B.}~\bibnamefont {B{\'e}csy}}, \bibinfo
  {author} {\bibfnamefont {A.}~\bibnamefont {Brazier}}, \bibinfo {author}
  {\bibfnamefont {P.~R.}\ \bibnamefont {Brook}}, \bibinfo {author}
  {\bibfnamefont {S.}~\bibnamefont {Burke-Spolaor}}, \bibinfo {author}
  {\bibfnamefont {S.}~\bibnamefont {Chatterjee}}, \bibinfo {author}
  {\bibfnamefont {S.}~\bibnamefont {Chen}}, \bibinfo {author} {\bibfnamefont
  {J.~M.}\ \bibnamefont {Cordes}}, \emph {et~al.},\ }\bibfield  {title}
  {\bibinfo {title} {The nanograv 12.5 yr data set: search for an isotropic
  stochastic gravitational-wave background},\ }\href@noop {} {\bibfield
  {journal} {\bibinfo  {journal} {The Astrophysical Journal Letters}\ }\textbf
  {\bibinfo {volume} {905}},\ \bibinfo {pages} {L34} (\bibinfo {year}
  {2020}{\natexlab{a}})}\BibitemShut {NoStop}%
\bibitem [{\citenamefont {Goncharov}\ \emph {et~al.}(2021)\citenamefont
  {Goncharov}, \citenamefont {Shannon}, \citenamefont {Reardon}, \citenamefont
  {Hobbs}, \citenamefont {Zic}, \citenamefont {Bailes}, \citenamefont
  {Cury{\l}o}, \citenamefont {Dai}, \citenamefont {Kerr}, \citenamefont
  {Lower}, \citenamefont {Manchester}, \citenamefont {Mandow}, \citenamefont
  {Middleton}, \citenamefont {Miles}, \citenamefont {Parthasarathy},
  \citenamefont {Thrane}, \citenamefont {Thyagarajan}, \citenamefont {Xue},
  \citenamefont {Zhu}, \citenamefont {Cameron}, \citenamefont {Feng},
  \citenamefont {Luo}, \citenamefont {Russell}, \citenamefont {Sarkissian},
  \citenamefont {Spiewak}, \citenamefont {Wang}, \citenamefont {Wang},
  \citenamefont {Zhang},\ and\ \citenamefont {Zhang}}]{Goncharov_2021}%
  \BibitemOpen
  \bibfield  {author} {\bibinfo {author} {\bibfnamefont {B.}~\bibnamefont
  {Goncharov}}, \bibinfo {author} {\bibfnamefont {R.~M.}\ \bibnamefont
  {Shannon}}, \bibinfo {author} {\bibfnamefont {D.~J.}\ \bibnamefont
  {Reardon}}, \bibinfo {author} {\bibfnamefont {G.}~\bibnamefont {Hobbs}},
  \bibinfo {author} {\bibfnamefont {A.}~\bibnamefont {Zic}}, \bibinfo {author}
  {\bibfnamefont {M.}~\bibnamefont {Bailes}}, \bibinfo {author} {\bibfnamefont
  {M.}~\bibnamefont {Cury{\l}o}}, \bibinfo {author} {\bibfnamefont
  {S.}~\bibnamefont {Dai}}, \bibinfo {author} {\bibfnamefont {M.}~\bibnamefont
  {Kerr}}, \bibinfo {author} {\bibfnamefont {M.~E.}\ \bibnamefont {Lower}},
  \bibinfo {author} {\bibfnamefont {R.~N.}\ \bibnamefont {Manchester}},
  \bibinfo {author} {\bibfnamefont {R.}~\bibnamefont {Mandow}}, \bibinfo
  {author} {\bibfnamefont {H.}~\bibnamefont {Middleton}}, \bibinfo {author}
  {\bibfnamefont {M.~T.}\ \bibnamefont {Miles}}, \bibinfo {author}
  {\bibfnamefont {A.}~\bibnamefont {Parthasarathy}}, \bibinfo {author}
  {\bibfnamefont {E.}~\bibnamefont {Thrane}}, \bibinfo {author} {\bibfnamefont
  {N.}~\bibnamefont {Thyagarajan}}, \bibinfo {author} {\bibfnamefont
  {X.}~\bibnamefont {Xue}}, \bibinfo {author} {\bibfnamefont {X.-J.}\
  \bibnamefont {Zhu}}, \bibinfo {author} {\bibfnamefont {A.~D.}\ \bibnamefont
  {Cameron}}, \bibinfo {author} {\bibfnamefont {Y.}~\bibnamefont {Feng}},
  \bibinfo {author} {\bibfnamefont {R.}~\bibnamefont {Luo}}, \bibinfo {author}
  {\bibfnamefont {C.~J.}\ \bibnamefont {Russell}}, \bibinfo {author}
  {\bibfnamefont {J.}~\bibnamefont {Sarkissian}}, \bibinfo {author}
  {\bibfnamefont {R.}~\bibnamefont {Spiewak}}, \bibinfo {author} {\bibfnamefont
  {S.}~\bibnamefont {Wang}}, \bibinfo {author} {\bibfnamefont {J.~B.}\
  \bibnamefont {Wang}}, \bibinfo {author} {\bibfnamefont {L.}~\bibnamefont
  {Zhang}},\ and\ \bibinfo {author} {\bibfnamefont {S.}~\bibnamefont {Zhang}},\
  }\bibfield  {title} {\bibinfo {title} {On the evidence for a common-spectrum
  process in the search for the nanohertz gravitational-wave background with
  the parkes pulsar timing array},\ }\href
  {https://doi.org/10.3847/2041-8213/ac17f4} {\bibfield  {journal} {\bibinfo
  {journal} {The Astrophysical Journal Letters}\ }\textbf {\bibinfo {volume}
  {917}},\ \bibinfo {pages} {L19} (\bibinfo {year} {2021})}\BibitemShut
  {NoStop}%
\bibitem [{\citenamefont {Chen}\ \emph
  {et~al.}(2021{\natexlab{a}})\citenamefont {Chen}, \citenamefont {Caballero},
  \citenamefont {Guo}, \citenamefont {Chalumeau}, \citenamefont {Liu},
  \citenamefont {Shaifullah}, \citenamefont {Lee}, \citenamefont {Babak},
  \citenamefont {Desvignes}, \citenamefont {Parthasarathy}, \citenamefont {Hu},
  \citenamefont {van der Wateren}, \citenamefont {Antoniadis}, \citenamefont
  {Bak Nielsen}, \citenamefont {Bassa}, \citenamefont {Berthereau},
  \citenamefont {Burgay}, \citenamefont {Champion}, \citenamefont {Cognard},
  \citenamefont {Falxa}, \citenamefont {Ferdman}, \citenamefont {Freire},
  \citenamefont {Gair}, \citenamefont {Graikou}, \citenamefont {Guillemot},
  \citenamefont {Jang}, \citenamefont {Janssen}, \citenamefont {Karuppusamy},
  \citenamefont {Keith}, \citenamefont {Kramer}, \citenamefont {Liu},
  \citenamefont {Lyne}, \citenamefont {Main}, \citenamefont {McKee},
  \citenamefont {Mickaliger}, \citenamefont {Perera}, \citenamefont {Perrodin},
  \citenamefont {Petiteau}, \citenamefont {Porayko}, \citenamefont {Possenti},
  \citenamefont {Samajdar}, \citenamefont {Sanidas}, \citenamefont {Sesana},
  \citenamefont {Speri}, \citenamefont {Stappers}, \citenamefont {Theureau},
  \citenamefont {Tiburzi}, \citenamefont {Vecchio}, \citenamefont {Verbiest},
  \citenamefont {Wang}, \citenamefont {Wang},\ and\ \citenamefont
  {Xu}}]{10.1093/mnras/stab2833}%
  \BibitemOpen
  \bibfield  {author} {\bibinfo {author} {\bibfnamefont {S.}~\bibnamefont
  {Chen}}, \bibinfo {author} {\bibfnamefont {R.~N.}\ \bibnamefont {Caballero}},
  \bibinfo {author} {\bibfnamefont {Y.~J.}\ \bibnamefont {Guo}}, \bibinfo
  {author} {\bibfnamefont {A.}~\bibnamefont {Chalumeau}}, \bibinfo {author}
  {\bibfnamefont {K.}~\bibnamefont {Liu}}, \bibinfo {author} {\bibfnamefont
  {G.}~\bibnamefont {Shaifullah}}, \bibinfo {author} {\bibfnamefont {K.~J.}\
  \bibnamefont {Lee}}, \bibinfo {author} {\bibfnamefont {S.}~\bibnamefont
  {Babak}}, \bibinfo {author} {\bibfnamefont {G.}~\bibnamefont {Desvignes}},
  \bibinfo {author} {\bibfnamefont {A.}~\bibnamefont {Parthasarathy}}, \bibinfo
  {author} {\bibfnamefont {H.}~\bibnamefont {Hu}}, \bibinfo {author}
  {\bibfnamefont {E.}~\bibnamefont {van der Wateren}}, \bibinfo {author}
  {\bibfnamefont {J.}~\bibnamefont {Antoniadis}}, \bibinfo {author}
  {\bibfnamefont {A.-S.}\ \bibnamefont {Bak Nielsen}}, \bibinfo {author}
  {\bibfnamefont {C.~G.}\ \bibnamefont {Bassa}}, \bibinfo {author}
  {\bibfnamefont {A.}~\bibnamefont {Berthereau}}, \bibinfo {author}
  {\bibfnamefont {M.}~\bibnamefont {Burgay}}, \bibinfo {author} {\bibfnamefont
  {D.~J.}\ \bibnamefont {Champion}}, \bibinfo {author} {\bibfnamefont
  {I.}~\bibnamefont {Cognard}}, \bibinfo {author} {\bibfnamefont
  {M.}~\bibnamefont {Falxa}}, \bibinfo {author} {\bibfnamefont {R.~D.}\
  \bibnamefont {Ferdman}}, \bibinfo {author} {\bibfnamefont {P.~C.~C.}\
  \bibnamefont {Freire}}, \bibinfo {author} {\bibfnamefont {J.~R.}\
  \bibnamefont {Gair}}, \bibinfo {author} {\bibfnamefont {E.}~\bibnamefont
  {Graikou}}, \bibinfo {author} {\bibfnamefont {L.}~\bibnamefont {Guillemot}},
  \bibinfo {author} {\bibfnamefont {J.}~\bibnamefont {Jang}}, \bibinfo {author}
  {\bibfnamefont {G.~H.}\ \bibnamefont {Janssen}}, \bibinfo {author}
  {\bibfnamefont {R.}~\bibnamefont {Karuppusamy}}, \bibinfo {author}
  {\bibfnamefont {M.~J.}\ \bibnamefont {Keith}}, \bibinfo {author}
  {\bibfnamefont {M.}~\bibnamefont {Kramer}}, \bibinfo {author} {\bibfnamefont
  {X.~J.}\ \bibnamefont {Liu}}, \bibinfo {author} {\bibfnamefont {A.~G.}\
  \bibnamefont {Lyne}}, \bibinfo {author} {\bibfnamefont {R.~A.}\ \bibnamefont
  {Main}}, \bibinfo {author} {\bibfnamefont {J.~W.}\ \bibnamefont {McKee}},
  \bibinfo {author} {\bibfnamefont {M.~B.}\ \bibnamefont {Mickaliger}},
  \bibinfo {author} {\bibfnamefont {B.~B.~P.}\ \bibnamefont {Perera}}, \bibinfo
  {author} {\bibfnamefont {D.}~\bibnamefont {Perrodin}}, \bibinfo {author}
  {\bibfnamefont {A.}~\bibnamefont {Petiteau}}, \bibinfo {author}
  {\bibfnamefont {N.~K.}\ \bibnamefont {Porayko}}, \bibinfo {author}
  {\bibfnamefont {A.}~\bibnamefont {Possenti}}, \bibinfo {author}
  {\bibfnamefont {A.}~\bibnamefont {Samajdar}}, \bibinfo {author}
  {\bibfnamefont {S.~A.}\ \bibnamefont {Sanidas}}, \bibinfo {author}
  {\bibfnamefont {A.}~\bibnamefont {Sesana}}, \bibinfo {author} {\bibfnamefont
  {L.}~\bibnamefont {Speri}}, \bibinfo {author} {\bibfnamefont {B.~W.}\
  \bibnamefont {Stappers}}, \bibinfo {author} {\bibfnamefont {G.}~\bibnamefont
  {Theureau}}, \bibinfo {author} {\bibfnamefont {C.}~\bibnamefont {Tiburzi}},
  \bibinfo {author} {\bibfnamefont {A.}~\bibnamefont {Vecchio}}, \bibinfo
  {author} {\bibfnamefont {J.~P.~W.}\ \bibnamefont {Verbiest}}, \bibinfo
  {author} {\bibfnamefont {J.}~\bibnamefont {Wang}}, \bibinfo {author}
  {\bibfnamefont {L.}~\bibnamefont {Wang}},\ and\ \bibinfo {author}
  {\bibfnamefont {H.}~\bibnamefont {Xu}},\ }\bibfield  {title} {\bibinfo
  {title} {{Common-red-signal analysis with 24-yr high-precision timing of the
  European Pulsar Timing Array: inferences in the stochastic gravitational-wave
  background search}},\ }\href {https://doi.org/10.1093/mnras/stab2833}
  {\bibfield  {journal} {\bibinfo  {journal} {Monthly Notices of the Royal
  Astronomical Society}\ }\textbf {\bibinfo {volume} {508}},\ \bibinfo {pages}
  {4970} (\bibinfo {year} {2021}{\natexlab{a}})},\ \Eprint
  {https://arxiv.org/abs/https://academic.oup.com/mnras/article-pdf/508/4/4970/40979667/stab2833.pdf}
  {https://academic.oup.com/mnras/article-pdf/508/4/4970/40979667/stab2833.pdf}
  \BibitemShut {NoStop}%
\bibitem [{\citenamefont {Phinney}(2001)}]{Phinney_2001}%
  \BibitemOpen
  \bibfield  {author} {\bibinfo {author} {\bibfnamefont {E.~S.}\ \bibnamefont
  {Phinney}},\ }\bibfield  {title} {\bibinfo {title} {A practical theorem on
  gravitational wave backgrounds}} (\bibinfo {year} {2001})\BibitemShut
  {NoStop}%
\bibitem [{\citenamefont {Burke-Spolaor}\ \emph {et~al.}(2019)\citenamefont
  {Burke-Spolaor}, \citenamefont {Taylor}, \citenamefont {Charisi},
  \citenamefont {Dolch}, \citenamefont {Hazboun}, \citenamefont {Holgado},
  \citenamefont {Kelley}, \citenamefont {Lazio}, \citenamefont {Madison},
  \citenamefont {McMann} \emph {et~al.}}]{Burke-Spolaor2018-io}%
  \BibitemOpen
  \bibfield  {author} {\bibinfo {author} {\bibfnamefont {S.}~\bibnamefont
  {Burke-Spolaor}}, \bibinfo {author} {\bibfnamefont {S.~R.}\ \bibnamefont
  {Taylor}}, \bibinfo {author} {\bibfnamefont {M.}~\bibnamefont {Charisi}},
  \bibinfo {author} {\bibfnamefont {T.}~\bibnamefont {Dolch}}, \bibinfo
  {author} {\bibfnamefont {J.~S.}\ \bibnamefont {Hazboun}}, \bibinfo {author}
  {\bibfnamefont {A.~M.}\ \bibnamefont {Holgado}}, \bibinfo {author}
  {\bibfnamefont {L.~Z.}\ \bibnamefont {Kelley}}, \bibinfo {author}
  {\bibfnamefont {T.~J.~W.}\ \bibnamefont {Lazio}}, \bibinfo {author}
  {\bibfnamefont {D.~R.}\ \bibnamefont {Madison}}, \bibinfo {author}
  {\bibfnamefont {N.}~\bibnamefont {McMann}}, \emph {et~al.},\ }\bibfield
  {title} {\bibinfo {title} {The astrophysics of nanohertz gravitational
  waves},\ }\href@noop {} {\bibfield  {journal} {\bibinfo  {journal} {The
  Astronomy and astrophysics review}\ }\textbf {\bibinfo {volume} {27}},\
  \bibinfo {pages} {1} (\bibinfo {year} {2019})}\BibitemShut {NoStop}%
\bibitem [{\citenamefont {Agazie}\ \emph
  {et~al.}(2023{\natexlab{b}})\citenamefont {Agazie}, \citenamefont
  {Anumarlapudi}, \citenamefont {Archibald}, \citenamefont {Baker},
  \citenamefont {B{\'e}csy}, \citenamefont {Blecha}, \citenamefont {Bonilla},
  \citenamefont {Brazier}, \citenamefont {Brook}, \citenamefont {Burke-Spolaor}
  \emph {et~al.}}]{2023ApJ...952L..37A}%
  \BibitemOpen
  \bibfield  {author} {\bibinfo {author} {\bibfnamefont {G.}~\bibnamefont
  {Agazie}}, \bibinfo {author} {\bibfnamefont {A.}~\bibnamefont
  {Anumarlapudi}}, \bibinfo {author} {\bibfnamefont {A.~M.}\ \bibnamefont
  {Archibald}}, \bibinfo {author} {\bibfnamefont {P.~T.}\ \bibnamefont
  {Baker}}, \bibinfo {author} {\bibfnamefont {B.}~\bibnamefont {B{\'e}csy}},
  \bibinfo {author} {\bibfnamefont {L.}~\bibnamefont {Blecha}}, \bibinfo
  {author} {\bibfnamefont {A.}~\bibnamefont {Bonilla}}, \bibinfo {author}
  {\bibfnamefont {A.}~\bibnamefont {Brazier}}, \bibinfo {author} {\bibfnamefont
  {P.~R.}\ \bibnamefont {Brook}}, \bibinfo {author} {\bibfnamefont
  {S.}~\bibnamefont {Burke-Spolaor}}, \emph {et~al.},\ }\bibfield  {title}
  {\bibinfo {title} {The nanograv 15 yr data set: Constraints on supermassive
  black hole binaries from the gravitational-wave background},\ }\href@noop {}
  {\bibfield  {journal} {\bibinfo  {journal} {The Astrophysical Journal
  Letters}\ }\textbf {\bibinfo {volume} {952}},\ \bibinfo {pages} {L37}
  (\bibinfo {year} {2023}{\natexlab{b}})}\BibitemShut {NoStop}%
\bibitem [{\citenamefont {Sesana}(2013)}]{Sesana_2013}%
  \BibitemOpen
  \bibfield  {author} {\bibinfo {author} {\bibfnamefont {A.}~\bibnamefont
  {Sesana}},\ }\bibfield  {title} {\bibinfo {title} {Insights into the
  astrophysics of supermassive black hole binaries from pulsar timing
  observations},\ }\href {https://doi.org/10.1088/0264-9381/30/22/224014}
  {\bibfield  {journal} {\bibinfo  {journal} {Classical and Quantum Gravity}\
  }\textbf {\bibinfo {volume} {30}},\ \bibinfo {pages} {224014} (\bibinfo
  {year} {2013})}\BibitemShut {NoStop}%
\bibitem [{\citenamefont {Sampson}\ \emph {et~al.}(2015)\citenamefont
  {Sampson}, \citenamefont {Cornish},\ and\ \citenamefont
  {McWilliams}}]{Sampson2015-qj}%
  \BibitemOpen
  \bibfield  {author} {\bibinfo {author} {\bibfnamefont {L.}~\bibnamefont
  {Sampson}}, \bibinfo {author} {\bibfnamefont {N.~J.}\ \bibnamefont
  {Cornish}},\ and\ \bibinfo {author} {\bibfnamefont {S.~T.}\ \bibnamefont
  {McWilliams}},\ }\bibfield  {title} {\bibinfo {title} {Constraining the
  solution to the last parsec problem with pulsar timing},\ }\href@noop {}
  {\bibfield  {journal} {\bibinfo  {journal} {Physical Review D}\ }\textbf
  {\bibinfo {volume} {91}},\ \bibinfo {pages} {084055} (\bibinfo {year}
  {2015})}\BibitemShut {NoStop}%
\bibitem [{\citenamefont {Taylor}\ \emph {et~al.}(2017)\citenamefont {Taylor},
  \citenamefont {Simon},\ and\ \citenamefont {Sampson}}]{2017PhRvL.118r1102T}%
  \BibitemOpen
  \bibfield  {author} {\bibinfo {author} {\bibfnamefont {S.~R.}\ \bibnamefont
  {Taylor}}, \bibinfo {author} {\bibfnamefont {J.}~\bibnamefont {Simon}},\ and\
  \bibinfo {author} {\bibfnamefont {L.}~\bibnamefont {Sampson}},\ }\bibfield
  {title} {\bibinfo {title} {Constraints on the dynamical environments of
  supermassive black-hole binaries using pulsar-timing arrays},\ }\href@noop {}
  {\bibfield  {journal} {\bibinfo  {journal} {Physical Review Letters}\
  }\textbf {\bibinfo {volume} {118}},\ \bibinfo {pages} {181102} (\bibinfo
  {year} {2017})}\BibitemShut {NoStop}%
\bibitem [{\citenamefont {Sesana}\ \emph {et~al.}(2008)\citenamefont {Sesana},
  \citenamefont {Vecchio},\ and\ \citenamefont
  {Colacino}}]{sesana_vecchio_colacino}%
  \BibitemOpen
  \bibfield  {author} {\bibinfo {author} {\bibfnamefont {A.}~\bibnamefont
  {Sesana}}, \bibinfo {author} {\bibfnamefont {A.}~\bibnamefont {Vecchio}},\
  and\ \bibinfo {author} {\bibfnamefont {C.~N.}\ \bibnamefont {Colacino}},\
  }\bibfield  {title} {\bibinfo {title} {{The stochastic gravitational-wave
  background from massive black hole binary systems: implications for
  observations with Pulsar Timing Arrays}},\ }\href
  {https://doi.org/10.1111/j.1365-2966.2008.13682.x} {\bibfield  {journal}
  {\bibinfo  {journal} {Monthly Notices of the Royal Astronomical Society}\
  }\textbf {\bibinfo {volume} {390}},\ \bibinfo {pages} {192} (\bibinfo {year}
  {2008})},\ \Eprint
  {https://arxiv.org/abs/https://academic.oup.com/mnras/article-pdf/390/1/192/2959688/mnras0390-0192.pdf}
  {https://academic.oup.com/mnras/article-pdf/390/1/192/2959688/mnras0390-0192.pdf}
  \BibitemShut {NoStop}%
\bibitem [{\citenamefont {Roebber}\ \emph {et~al.}(2016)\citenamefont
  {Roebber}, \citenamefont {Holder}, \citenamefont {Holz},\ and\ \citenamefont
  {Warren}}]{Roebber_2016}%
  \BibitemOpen
  \bibfield  {author} {\bibinfo {author} {\bibfnamefont {E.}~\bibnamefont
  {Roebber}}, \bibinfo {author} {\bibfnamefont {G.}~\bibnamefont {Holder}},
  \bibinfo {author} {\bibfnamefont {D.~E.}\ \bibnamefont {Holz}},\ and\
  \bibinfo {author} {\bibfnamefont {M.}~\bibnamefont {Warren}},\ }\bibfield
  {title} {\bibinfo {title} {Cosmic variance in the nanohertz gravitational
  wave background},\ }\href {https://doi.org/10.3847/0004-637X/819/2/163}
  {\bibfield  {journal} {\bibinfo  {journal} {The Astrophysical Journal}\
  }\textbf {\bibinfo {volume} {819}},\ \bibinfo {pages} {163} (\bibinfo {year}
  {2016})}\BibitemShut {NoStop}%
\bibitem [{\citenamefont {Kelley}\ \emph {et~al.}(2017)\citenamefont {Kelley},
  \citenamefont {Blecha}, \citenamefont {Hernquist}, \citenamefont {Sesana},\
  and\ \citenamefont {Taylor}}]{kelley_17}%
  \BibitemOpen
  \bibfield  {author} {\bibinfo {author} {\bibfnamefont {L.~Z.}\ \bibnamefont
  {Kelley}}, \bibinfo {author} {\bibfnamefont {L.}~\bibnamefont {Blecha}},
  \bibinfo {author} {\bibfnamefont {L.}~\bibnamefont {Hernquist}}, \bibinfo
  {author} {\bibfnamefont {A.}~\bibnamefont {Sesana}},\ and\ \bibinfo {author}
  {\bibfnamefont {S.~R.}\ \bibnamefont {Taylor}},\ }\bibfield  {title}
  {\bibinfo {title} {{The gravitational wave background from massive black hole
  binaries in Illustris: spectral features and time to detection with pulsar
  timing arrays}},\ }\href {https://doi.org/10.1093/mnras/stx1638} {\bibfield
  {journal} {\bibinfo  {journal} {Monthly Notices of the Royal Astronomical
  Society}\ }\textbf {\bibinfo {volume} {471}},\ \bibinfo {pages} {4508}
  (\bibinfo {year} {2017})},\ \Eprint
  {https://arxiv.org/abs/https://academic.oup.com/mnras/article-pdf/471/4/4508/19609079/stx1638.pdf}
  {https://academic.oup.com/mnras/article-pdf/471/4/4508/19609079/stx1638.pdf}
  \BibitemShut {NoStop}%
\bibitem [{\citenamefont {Afzal}\ \emph {et~al.}(2023)\citenamefont {Afzal},
  \citenamefont {Agazie}, \citenamefont {Anumarlapudi}, \citenamefont
  {Archibald}, \citenamefont {Arzoumanian}, \citenamefont {Baker},
  \citenamefont {B{\'e}csy}, \citenamefont {Blanco-Pillado}, \citenamefont
  {Blecha}, \citenamefont {Boddy} \emph {et~al.}}]{2023ApJ...951L..11A}%
  \BibitemOpen
  \bibfield  {author} {\bibinfo {author} {\bibfnamefont {A.}~\bibnamefont
  {Afzal}}, \bibinfo {author} {\bibfnamefont {G.}~\bibnamefont {Agazie}},
  \bibinfo {author} {\bibfnamefont {A.}~\bibnamefont {Anumarlapudi}}, \bibinfo
  {author} {\bibfnamefont {A.~M.}\ \bibnamefont {Archibald}}, \bibinfo {author}
  {\bibfnamefont {Z.}~\bibnamefont {Arzoumanian}}, \bibinfo {author}
  {\bibfnamefont {P.~T.}\ \bibnamefont {Baker}}, \bibinfo {author}
  {\bibfnamefont {B.}~\bibnamefont {B{\'e}csy}}, \bibinfo {author}
  {\bibfnamefont {J.~J.}\ \bibnamefont {Blanco-Pillado}}, \bibinfo {author}
  {\bibfnamefont {L.}~\bibnamefont {Blecha}}, \bibinfo {author} {\bibfnamefont
  {K.~K.}\ \bibnamefont {Boddy}}, \emph {et~al.},\ }\bibfield  {title}
  {\bibinfo {title} {The nanograv 15 yr data set: Search for signals from new
  physics},\ }\href@noop {} {\bibfield  {journal} {\bibinfo  {journal} {The
  Astrophysical Journal Letters}\ }\textbf {\bibinfo {volume} {951}},\ \bibinfo
  {pages} {L11} (\bibinfo {year} {2023})}\BibitemShut {NoStop}%
\bibitem [{\citenamefont {{Kibble}}(1976)}]{1986coco}%
  \BibitemOpen
  \bibfield  {author} {\bibinfo {author} {\bibfnamefont {T.~W.~B.}\
  \bibnamefont {{Kibble}}},\ }\bibfield  {title} {\bibinfo {title} {{Topology
  of Cosmic Domains and Strings}},\ }\href@noop {} {\bibfield  {journal}
  {\bibinfo  {journal} {Journal of Physics A}\ } (\bibinfo {year}
  {1976})}\BibitemShut {NoStop}%
\bibitem [{\citenamefont {Lasky}\ \emph {et~al.}(2016)\citenamefont {Lasky},
  \citenamefont {Mingarelli}, \citenamefont {Smith}, \citenamefont {Giblin},
  \citenamefont {Thrane}, \citenamefont {Reardon}, \citenamefont {Caldwell},
  \citenamefont {Bailes}, \citenamefont {Bhat}, \citenamefont {Burke-Spolaor},
  \citenamefont {Dai}, \citenamefont {Dempsey}, \citenamefont {Hobbs},
  \citenamefont {Kerr}, \citenamefont {Levin}, \citenamefont {Manchester},
  \citenamefont {Os\l{}owski}, \citenamefont {Ravi}, \citenamefont {Rosado},
  \citenamefont {Shannon}, \citenamefont {Spiewak}, \citenamefont {van
  Straten}, \citenamefont {Toomey}, \citenamefont {Wang}, \citenamefont {Wen},
  \citenamefont {You},\ and\ \citenamefont {Zhu}}]{Lasky_2016}%
  \BibitemOpen
  \bibfield  {author} {\bibinfo {author} {\bibfnamefont {P.~D.}\ \bibnamefont
  {Lasky}}, \bibinfo {author} {\bibfnamefont {C.~M.~F.}\ \bibnamefont
  {Mingarelli}}, \bibinfo {author} {\bibfnamefont {T.~L.}\ \bibnamefont
  {Smith}}, \bibinfo {author} {\bibfnamefont {J.~T.}\ \bibnamefont {Giblin}},
  \bibinfo {author} {\bibfnamefont {E.}~\bibnamefont {Thrane}}, \bibinfo
  {author} {\bibfnamefont {D.~J.}\ \bibnamefont {Reardon}}, \bibinfo {author}
  {\bibfnamefont {R.}~\bibnamefont {Caldwell}}, \bibinfo {author}
  {\bibfnamefont {M.}~\bibnamefont {Bailes}}, \bibinfo {author} {\bibfnamefont
  {N.~D.~R.}\ \bibnamefont {Bhat}}, \bibinfo {author} {\bibfnamefont
  {S.}~\bibnamefont {Burke-Spolaor}}, \bibinfo {author} {\bibfnamefont
  {S.}~\bibnamefont {Dai}}, \bibinfo {author} {\bibfnamefont {J.}~\bibnamefont
  {Dempsey}}, \bibinfo {author} {\bibfnamefont {G.}~\bibnamefont {Hobbs}},
  \bibinfo {author} {\bibfnamefont {M.}~\bibnamefont {Kerr}}, \bibinfo {author}
  {\bibfnamefont {Y.}~\bibnamefont {Levin}}, \bibinfo {author} {\bibfnamefont
  {R.~N.}\ \bibnamefont {Manchester}}, \bibinfo {author} {\bibfnamefont
  {S.}~\bibnamefont {Os\l{}owski}}, \bibinfo {author} {\bibfnamefont
  {V.}~\bibnamefont {Ravi}}, \bibinfo {author} {\bibfnamefont {P.~A.}\
  \bibnamefont {Rosado}}, \bibinfo {author} {\bibfnamefont {R.~M.}\
  \bibnamefont {Shannon}}, \bibinfo {author} {\bibfnamefont {R.}~\bibnamefont
  {Spiewak}}, \bibinfo {author} {\bibfnamefont {W.}~\bibnamefont {van
  Straten}}, \bibinfo {author} {\bibfnamefont {L.}~\bibnamefont {Toomey}},
  \bibinfo {author} {\bibfnamefont {J.}~\bibnamefont {Wang}}, \bibinfo {author}
  {\bibfnamefont {L.}~\bibnamefont {Wen}}, \bibinfo {author} {\bibfnamefont
  {X.}~\bibnamefont {You}},\ and\ \bibinfo {author} {\bibfnamefont
  {X.}~\bibnamefont {Zhu}},\ }\bibfield  {title} {\bibinfo {title}
  {Gravitational-wave cosmology across 29 decades in frequency},\ }\href
  {https://doi.org/10.1103/PhysRevX.6.011035} {\bibfield  {journal} {\bibinfo
  {journal} {Phys. Rev. X}\ }\textbf {\bibinfo {volume} {6}},\ \bibinfo {pages}
  {011035} (\bibinfo {year} {2016})}\BibitemShut {NoStop}%
\bibitem [{\citenamefont {Schwaller}(2015)}]{cpt2015}%
  \BibitemOpen
  \bibfield  {author} {\bibinfo {author} {\bibfnamefont {P.}~\bibnamefont
  {Schwaller}},\ }\bibfield  {title} {\bibinfo {title} {Gravitational waves
  from a dark phase transition},\ }\href@noop {} {\bibfield  {journal}
  {\bibinfo  {journal} {Physical Review Letters}\ }\textbf {\bibinfo {volume}
  {115}},\ \bibinfo {pages} {181101} (\bibinfo {year} {2015})}\BibitemShut
  {NoStop}%
\bibitem [{\citenamefont {Kaiser}\ \emph {et~al.}(2022)\citenamefont {Kaiser},
  \citenamefont {Pol}, \citenamefont {McLaughlin}, \citenamefont {Chen},
  \citenamefont {Hazboun}, \citenamefont {Kelley}, \citenamefont {Simon},
  \citenamefont {Taylor}, \citenamefont {Vigeland},\ and\ \citenamefont
  {Witt}}]{Kaiser2022-su}%
  \BibitemOpen
  \bibfield  {author} {\bibinfo {author} {\bibfnamefont {A.~R.}\ \bibnamefont
  {Kaiser}}, \bibinfo {author} {\bibfnamefont {N.~S.}\ \bibnamefont {Pol}},
  \bibinfo {author} {\bibfnamefont {M.~A.}\ \bibnamefont {McLaughlin}},
  \bibinfo {author} {\bibfnamefont {S.}~\bibnamefont {Chen}}, \bibinfo {author}
  {\bibfnamefont {J.~S.}\ \bibnamefont {Hazboun}}, \bibinfo {author}
  {\bibfnamefont {L.~Z.}\ \bibnamefont {Kelley}}, \bibinfo {author}
  {\bibfnamefont {J.}~\bibnamefont {Simon}}, \bibinfo {author} {\bibfnamefont
  {S.~R.}\ \bibnamefont {Taylor}}, \bibinfo {author} {\bibfnamefont {S.~J.}\
  \bibnamefont {Vigeland}},\ and\ \bibinfo {author} {\bibfnamefont {C.~A.}\
  \bibnamefont {Witt}},\ }\bibfield  {title} {\bibinfo {title} {Disentangling
  multiple stochastic gravitational wave background sources in pta data sets},\
  }\href@noop {} {\bibfield  {journal} {\bibinfo  {journal} {The Astrophysical
  Journal}\ }\textbf {\bibinfo {volume} {938}},\ \bibinfo {pages} {115}
  (\bibinfo {year} {2022})}\BibitemShut {NoStop}%
\bibitem [{\citenamefont {Arzoumanian}\ \emph {et~al.}(2018)\citenamefont
  {Arzoumanian}, \citenamefont {Baker}, \citenamefont {Brazier}, \citenamefont
  {Burke-Spolaor}, \citenamefont {Chamberlin}, \citenamefont {Chatterjee},
  \citenamefont {Christy}, \citenamefont {Cordes}, \citenamefont {Cornish},
  \citenamefont {Crawford} \emph {et~al.}}]{arzoumanian2018nanograv}%
  \BibitemOpen
  \bibfield  {author} {\bibinfo {author} {\bibfnamefont {Z.}~\bibnamefont
  {Arzoumanian}}, \bibinfo {author} {\bibfnamefont {P.}~\bibnamefont {Baker}},
  \bibinfo {author} {\bibfnamefont {A.}~\bibnamefont {Brazier}}, \bibinfo
  {author} {\bibfnamefont {S.}~\bibnamefont {Burke-Spolaor}}, \bibinfo {author}
  {\bibfnamefont {S.}~\bibnamefont {Chamberlin}}, \bibinfo {author}
  {\bibfnamefont {S.}~\bibnamefont {Chatterjee}}, \bibinfo {author}
  {\bibfnamefont {B.}~\bibnamefont {Christy}}, \bibinfo {author} {\bibfnamefont
  {J.~M.}\ \bibnamefont {Cordes}}, \bibinfo {author} {\bibfnamefont {N.~J.}\
  \bibnamefont {Cornish}}, \bibinfo {author} {\bibfnamefont {F.}~\bibnamefont
  {Crawford}}, \emph {et~al.},\ }\bibfield  {title} {\bibinfo {title} {The
  nanograv 11 year data set: pulsar-timing constraints on the stochastic
  gravitational-wave background},\ }\href@noop {} {\bibfield  {journal}
  {\bibinfo  {journal} {The Astrophysical Journal}\ }\textbf {\bibinfo {volume}
  {859}},\ \bibinfo {pages} {47} (\bibinfo {year} {2018})}\BibitemShut
  {NoStop}%
\bibitem [{\citenamefont {van Haasteren}\ and\ \citenamefont
  {Vallisneri}(2014)}]{2014PhRvD..90j4012V}%
  \BibitemOpen
  \bibfield  {author} {\bibinfo {author} {\bibfnamefont {R.}~\bibnamefont {van
  Haasteren}}\ and\ \bibinfo {author} {\bibfnamefont {M.}~\bibnamefont
  {Vallisneri}},\ }\bibfield  {title} {\bibinfo {title} {New advances in the
  gaussian-process approach to pulsar-timing data analysis},\ }\href@noop {}
  {\bibfield  {journal} {\bibinfo  {journal} {Physical Review D}\ }\textbf
  {\bibinfo {volume} {90}},\ \bibinfo {pages} {104012} (\bibinfo {year}
  {2014})}\BibitemShut {NoStop}%
\bibitem [{\citenamefont {van Haasteren}\ and\ \citenamefont
  {Vallisneri}(2015)}]{2015MNRAS.446.1170V}%
  \BibitemOpen
  \bibfield  {author} {\bibinfo {author} {\bibfnamefont {R.}~\bibnamefont {van
  Haasteren}}\ and\ \bibinfo {author} {\bibfnamefont {M.}~\bibnamefont
  {Vallisneri}},\ }\bibfield  {title} {\bibinfo {title} {Low-rank
  approximations for large stationary covariance matrices, as used in the
  bayesian and generalized-least-squares analysis of pulsar-timing data},\
  }\href@noop {} {\bibfield  {journal} {\bibinfo  {journal} {Monthly Notices of
  the Royal Astronomical Society}\ }\textbf {\bibinfo {volume} {446}},\
  \bibinfo {pages} {1170} (\bibinfo {year} {2015})}\BibitemShut {NoStop}%
\bibitem [{\citenamefont {Amiri}\ \emph {et~al.}(2021)\citenamefont {Amiri},
  \citenamefont {Bandura}, \citenamefont {Boyle}, \citenamefont {Brar},
  \citenamefont {Cliche}, \citenamefont {Crowter}, \citenamefont {Cubranic},
  \citenamefont {Demorest}, \citenamefont {Denman}, \citenamefont {Dobbs} \emph
  {et~al.}}]{2021ApJS..255....5C}%
  \BibitemOpen
  \bibfield  {author} {\bibinfo {author} {\bibfnamefont {M.}~\bibnamefont
  {Amiri}}, \bibinfo {author} {\bibfnamefont {K.}~\bibnamefont {Bandura}},
  \bibinfo {author} {\bibfnamefont {P.}~\bibnamefont {Boyle}}, \bibinfo
  {author} {\bibfnamefont {C.}~\bibnamefont {Brar}}, \bibinfo {author}
  {\bibfnamefont {J.-F.}\ \bibnamefont {Cliche}}, \bibinfo {author}
  {\bibfnamefont {K.}~\bibnamefont {Crowter}}, \bibinfo {author} {\bibfnamefont
  {D.}~\bibnamefont {Cubranic}}, \bibinfo {author} {\bibfnamefont
  {P.}~\bibnamefont {Demorest}}, \bibinfo {author} {\bibfnamefont
  {N.}~\bibnamefont {Denman}}, \bibinfo {author} {\bibfnamefont
  {M.}~\bibnamefont {Dobbs}}, \emph {et~al.},\ }\bibfield  {title} {\bibinfo
  {title} {The chime pulsar project: system overview},\ }\href@noop {}
  {\bibfield  {journal} {\bibinfo  {journal} {The Astrophysical Journal
  Supplement Series}\ }\textbf {\bibinfo {volume} {255}},\ \bibinfo {pages} {5}
  (\bibinfo {year} {2021})}\BibitemShut {NoStop}%
\bibitem [{\citenamefont {Taylor}\ \emph {et~al.}(2022)\citenamefont {Taylor},
  \citenamefont {Simon}, \citenamefont {Schult}, \citenamefont {Pol},\ and\
  \citenamefont {Lamb}}]{Taylor2022-nq}%
  \BibitemOpen
  \bibfield  {author} {\bibinfo {author} {\bibfnamefont {S.~R.}\ \bibnamefont
  {Taylor}}, \bibinfo {author} {\bibfnamefont {J.}~\bibnamefont {Simon}},
  \bibinfo {author} {\bibfnamefont {L.}~\bibnamefont {Schult}}, \bibinfo
  {author} {\bibfnamefont {N.}~\bibnamefont {Pol}},\ and\ \bibinfo {author}
  {\bibfnamefont {W.~G.}\ \bibnamefont {Lamb}},\ }\bibfield  {title} {\bibinfo
  {title} {A parallelized bayesian approach to accelerated gravitational-wave
  background characterization},\ }\href@noop {} {\bibfield  {journal} {\bibinfo
   {journal} {Physical Review D}\ }\textbf {\bibinfo {volume} {105}},\ \bibinfo
  {pages} {084049} (\bibinfo {year} {2022})}\BibitemShut {NoStop}%
\bibitem [{\citenamefont {Johnson}\ \emph {et~al.}(2022)\citenamefont
  {Johnson}, \citenamefont {Vigeland}, \citenamefont {Siemens},\ and\
  \citenamefont {Taylor}}]{Johnson_2022}%
  \BibitemOpen
  \bibfield  {author} {\bibinfo {author} {\bibfnamefont {A.~D.}\ \bibnamefont
  {Johnson}}, \bibinfo {author} {\bibfnamefont {S.~J.}\ \bibnamefont
  {Vigeland}}, \bibinfo {author} {\bibfnamefont {X.}~\bibnamefont {Siemens}},\
  and\ \bibinfo {author} {\bibfnamefont {S.~R.}\ \bibnamefont {Taylor}},\
  }\bibfield  {title} {\bibinfo {title} {Gravitational-wave statistics for
  pulsar timing arrays: Examining bias from using a finite number of pulsars},\
  }\href {https://doi.org/10.3847/1538-4357/ac6f5e} {\bibfield  {journal}
  {\bibinfo  {journal} {The Astrophysical Journal}\ }\textbf {\bibinfo {volume}
  {932}},\ \bibinfo {pages} {105} (\bibinfo {year} {2022})}\BibitemShut
  {NoStop}%
\bibitem [{\citenamefont {Cordes}\ and\ \citenamefont
  {Shannon}(2010)}]{cordes2010measurement}%
  \BibitemOpen
  \bibfield  {author} {\bibinfo {author} {\bibfnamefont {J.}~\bibnamefont
  {Cordes}}\ and\ \bibinfo {author} {\bibfnamefont {R.}~\bibnamefont
  {Shannon}},\ }\bibfield  {title} {\bibinfo {title} {A measurement model for
  precision pulsar timing},\ }\href@noop {} {\bibfield  {journal} {\bibinfo
  {journal} {arXiv preprint arXiv:1010.3785}\ } (\bibinfo {year}
  {2010})}\BibitemShut {NoStop}%
\bibitem [{\citenamefont {Keith}\ \emph {et~al.}(2013)\citenamefont {Keith},
  \citenamefont {Coles}, \citenamefont {Shannon}, \citenamefont {Hobbs},
  \citenamefont {Manchester}, \citenamefont {Bailes}, \citenamefont {Bhat},
  \citenamefont {Burke-Spolaor}, \citenamefont {Champion}, \citenamefont
  {Chaudhary} \emph {et~al.}}]{keith2013measurement}%
  \BibitemOpen
  \bibfield  {author} {\bibinfo {author} {\bibfnamefont {M.}~\bibnamefont
  {Keith}}, \bibinfo {author} {\bibfnamefont {W.}~\bibnamefont {Coles}},
  \bibinfo {author} {\bibfnamefont {R.}~\bibnamefont {Shannon}}, \bibinfo
  {author} {\bibfnamefont {G.}~\bibnamefont {Hobbs}}, \bibinfo {author}
  {\bibfnamefont {R.}~\bibnamefont {Manchester}}, \bibinfo {author}
  {\bibfnamefont {M.}~\bibnamefont {Bailes}}, \bibinfo {author} {\bibfnamefont
  {N.}~\bibnamefont {Bhat}}, \bibinfo {author} {\bibfnamefont {S.}~\bibnamefont
  {Burke-Spolaor}}, \bibinfo {author} {\bibfnamefont {D.}~\bibnamefont
  {Champion}}, \bibinfo {author} {\bibfnamefont {A.}~\bibnamefont {Chaudhary}},
  \emph {et~al.},\ }\bibfield  {title} {\bibinfo {title} {Measurement and
  correction of variations in interstellar dispersion in high-precision pulsar
  timing},\ }\href@noop {} {\bibfield  {journal} {\bibinfo  {journal} {Monthly
  Notices of the Royal Astronomical Society}\ }\textbf {\bibinfo {volume}
  {429}},\ \bibinfo {pages} {2161} (\bibinfo {year} {2013})}\BibitemShut
  {NoStop}%
\bibitem [{\citenamefont {Coles}\ \emph {et~al.}(2011)\citenamefont {Coles},
  \citenamefont {Hobbs}, \citenamefont {Champion}, \citenamefont {Manchester},\
  and\ \citenamefont {Verbiest}}]{coles2011pulsar}%
  \BibitemOpen
  \bibfield  {author} {\bibinfo {author} {\bibfnamefont {W.}~\bibnamefont
  {Coles}}, \bibinfo {author} {\bibfnamefont {G.}~\bibnamefont {Hobbs}},
  \bibinfo {author} {\bibfnamefont {D.}~\bibnamefont {Champion}}, \bibinfo
  {author} {\bibfnamefont {R.}~\bibnamefont {Manchester}},\ and\ \bibinfo
  {author} {\bibfnamefont {J.}~\bibnamefont {Verbiest}},\ }\bibfield  {title}
  {\bibinfo {title} {Pulsar timing analysis in the presence of correlated
  noise},\ }\href@noop {} {\bibfield  {journal} {\bibinfo  {journal} {Monthly
  Notices of the Royal Astronomical Society}\ }\textbf {\bibinfo {volume}
  {418}},\ \bibinfo {pages} {561} (\bibinfo {year} {2011})}\BibitemShut
  {NoStop}%
\bibitem [{\citenamefont {Ellis}\ \emph {et~al.}(2020)\citenamefont {Ellis},
  \citenamefont {Vallisneri}, \citenamefont {Taylor},\ and\ \citenamefont
  {Baker}}]{enterprise}%
  \BibitemOpen
  \bibfield  {author} {\bibinfo {author} {\bibfnamefont {J.~A.}\ \bibnamefont
  {Ellis}}, \bibinfo {author} {\bibfnamefont {M.}~\bibnamefont {Vallisneri}},
  \bibinfo {author} {\bibfnamefont {S.~R.}\ \bibnamefont {Taylor}},\ and\
  \bibinfo {author} {\bibfnamefont {P.~T.}\ \bibnamefont {Baker}},\ }\href
  {https://doi.org/10.5281/zenodo.4059815} {\bibinfo {title} {Enterprise:
  Enhanced numerical toolbox enabling a robust pulsar inference suite}},\
  \bibinfo {howpublished} {Zenodo} (\bibinfo {year} {2020})\BibitemShut
  {NoStop}%
\bibitem [{\citenamefont {Sun}\ \emph {et~al.}(2023)\citenamefont {Sun},
  \citenamefont {Baker}, \citenamefont {Johnson}, \citenamefont {Madison},\
  and\ \citenamefont {Siemens}}]{2023ApJ...951..121S}%
  \BibitemOpen
  \bibfield  {author} {\bibinfo {author} {\bibfnamefont {J.}~\bibnamefont
  {Sun}}, \bibinfo {author} {\bibfnamefont {P.~T.}\ \bibnamefont {Baker}},
  \bibinfo {author} {\bibfnamefont {A.~D.}\ \bibnamefont {Johnson}}, \bibinfo
  {author} {\bibfnamefont {D.~R.}\ \bibnamefont {Madison}},\ and\ \bibinfo
  {author} {\bibfnamefont {X.}~\bibnamefont {Siemens}},\ }\bibfield  {title}
  {\bibinfo {title} {Implementation of an efficient bayesian search for
  gravitational-wave bursts with memory in pulsar timing array data},\
  }\href@noop {} {\bibfield  {journal} {\bibinfo  {journal} {The Astrophysical
  Journal}\ }\textbf {\bibinfo {volume} {951}},\ \bibinfo {pages} {121}
  (\bibinfo {year} {2023})}\BibitemShut {NoStop}%
\bibitem [{\citenamefont {Lentati}\ \emph {et~al.}(2013)\citenamefont
  {Lentati}, \citenamefont {Alexander}, \citenamefont {Hobson}, \citenamefont
  {Taylor}, \citenamefont {Gair}, \citenamefont {Balan},\ and\ \citenamefont
  {van Haasteren}}]{lentati}%
  \BibitemOpen
  \bibfield  {author} {\bibinfo {author} {\bibfnamefont {L.}~\bibnamefont
  {Lentati}}, \bibinfo {author} {\bibfnamefont {P.}~\bibnamefont {Alexander}},
  \bibinfo {author} {\bibfnamefont {M.~P.}\ \bibnamefont {Hobson}}, \bibinfo
  {author} {\bibfnamefont {S.}~\bibnamefont {Taylor}}, \bibinfo {author}
  {\bibfnamefont {J.}~\bibnamefont {Gair}}, \bibinfo {author} {\bibfnamefont
  {S.~T.}\ \bibnamefont {Balan}},\ and\ \bibinfo {author} {\bibfnamefont
  {R.}~\bibnamefont {van Haasteren}},\ }\bibfield  {title} {\bibinfo {title}
  {Hyper-efficient model-independent bayesian method for the analysis of pulsar
  timing data},\ }\href@noop {} {\bibfield  {journal} {\bibinfo  {journal}
  {Phys. rev.}\ }\textbf {\bibinfo {volume} {87}} (\bibinfo {year}
  {2013})}\BibitemShut {NoStop}%
\bibitem [{\citenamefont {Taylor}\ \emph {et~al.}(2013)\citenamefont {Taylor},
  \citenamefont {Gair},\ and\ \citenamefont {Lentati}}]{taylor13}%
  \BibitemOpen
  \bibfield  {author} {\bibinfo {author} {\bibfnamefont {S.~R.}\ \bibnamefont
  {Taylor}}, \bibinfo {author} {\bibfnamefont {J.~R.}\ \bibnamefont {Gair}},\
  and\ \bibinfo {author} {\bibfnamefont {L.}~\bibnamefont {Lentati}},\
  }\bibfield  {title} {\bibinfo {title} {Weighing the evidence for a
  gravitational-wave background in the first international pulsar timing array
  data challenge},\ }\href {https://doi.org/10.1103/PhysRevD.87.044035}
  {\bibfield  {journal} {\bibinfo  {journal} {Phys. Rev. D}\ }\textbf {\bibinfo
  {volume} {87}},\ \bibinfo {pages} {044035} (\bibinfo {year}
  {2013})}\BibitemShut {NoStop}%
\bibitem [{\citenamefont {Cornish}\ \emph {et~al.}(2018)\citenamefont
  {Cornish}, \citenamefont {O’Beirne}, \citenamefont {Taylor},\ and\
  \citenamefont {Yunes}}]{2018PhRvL.120r1101C}%
  \BibitemOpen
  \bibfield  {author} {\bibinfo {author} {\bibfnamefont {N.~J.}\ \bibnamefont
  {Cornish}}, \bibinfo {author} {\bibfnamefont {L.}~\bibnamefont {O’Beirne}},
  \bibinfo {author} {\bibfnamefont {S.~R.}\ \bibnamefont {Taylor}},\ and\
  \bibinfo {author} {\bibfnamefont {N.}~\bibnamefont {Yunes}},\ }\bibfield
  {title} {\bibinfo {title} {Constraining alternative theories of gravity using
  pulsar timing arrays},\ }\href@noop {} {\bibfield  {journal} {\bibinfo
  {journal} {Physical review letters}\ }\textbf {\bibinfo {volume} {120}},\
  \bibinfo {pages} {181101} (\bibinfo {year} {2018})}\BibitemShut {NoStop}%
\bibitem [{\citenamefont {Ratzinger}\ and\ \citenamefont
  {Schwaller}(2021)}]{Ratzinger_2021}%
  \BibitemOpen
  \bibfield  {author} {\bibinfo {author} {\bibfnamefont {W.}~\bibnamefont
  {Ratzinger}}\ and\ \bibinfo {author} {\bibfnamefont {P.}~\bibnamefont
  {Schwaller}},\ }\bibfield  {title} {\bibinfo {title} {Whispers from the dark
  side: Confronting light new physics with {NANOGrav} data},\ }\bibfield
  {journal} {\bibinfo  {journal} {{SciPost} Physics}\ }\textbf {\bibinfo
  {volume} {10}},\ \href {https://doi.org/10.21468/scipostphys.10.2.047}
  {10.21468/scipostphys.10.2.047} (\bibinfo {year} {2021})\BibitemShut
  {NoStop}%
\bibitem [{\citenamefont {Wang}(2022)}]{deng}%
  \BibitemOpen
  \bibfield  {author} {\bibinfo {author} {\bibfnamefont {D.}~\bibnamefont
  {Wang}},\ }\href@noop {} {\bibinfo {title} {Novel physics with international
  pulsar timing array: Axionlike particles, domain walls and cosmic strings}}
  (\bibinfo {year} {2022})\BibitemShut {NoStop}%
\bibitem [{\citenamefont {Scott}(1979)}]{scott}%
  \BibitemOpen
  \bibfield  {author} {\bibinfo {author} {\bibfnamefont {D.~W.}\ \bibnamefont
  {Scott}},\ }\bibfield  {title} {\bibinfo {title} {{On optimal and data-based
  histograms}},\ }\href {https://doi.org/10.1093/biomet/66.3.605} {\bibfield
  {journal} {\bibinfo  {journal} {Biometrika}\ }\textbf {\bibinfo {volume}
  {66}},\ \bibinfo {pages} {605} (\bibinfo {year} {1979})},\ \Eprint
  {https://arxiv.org/abs/https://academic.oup.com/biomet/article-pdf/66/3/605/632347/66-3-605.pdf}
  {https://academic.oup.com/biomet/article-pdf/66/3/605/632347/66-3-605.pdf}
  \BibitemShut {NoStop}%
\bibitem [{\citenamefont {Freedman}\ and\ \citenamefont
  {Diaconis}(1981)}]{freedman1981histogram}%
  \BibitemOpen
  \bibfield  {author} {\bibinfo {author} {\bibfnamefont {D.}~\bibnamefont
  {Freedman}}\ and\ \bibinfo {author} {\bibfnamefont {P.}~\bibnamefont
  {Diaconis}},\ }\bibfield  {title} {\bibinfo {title} {On the histogram as a
  density estimator: L 2 theory},\ }\href@noop {} {\bibfield  {journal}
  {\bibinfo  {journal} {Zeitschrift f{\"u}r Wahrscheinlichkeitstheorie und
  verwandte Gebiete}\ }\textbf {\bibinfo {volume} {57}},\ \bibinfo {pages}
  {453} (\bibinfo {year} {1981})}\BibitemShut {NoStop}%
\bibitem [{\citenamefont {Parzen}(1962)}]{kde1}%
  \BibitemOpen
  \bibfield  {author} {\bibinfo {author} {\bibfnamefont {E.}~\bibnamefont
  {Parzen}},\ }\bibfield  {title} {\bibinfo {title} {{On Estimation of a
  Probability Density Function and Mode}},\ }\href
  {https://doi.org/10.1214/aoms/1177704472} {\bibfield  {journal} {\bibinfo
  {journal} {The Annals of Mathematical Statistics}\ }\textbf {\bibinfo
  {volume} {33}},\ \bibinfo {pages} {1065 } (\bibinfo {year}
  {1962})}\BibitemShut {NoStop}%
\bibitem [{\citenamefont {Rosenblatt}(1956)}]{kde2}%
  \BibitemOpen
  \bibfield  {author} {\bibinfo {author} {\bibfnamefont {M.}~\bibnamefont
  {Rosenblatt}},\ }\bibfield  {title} {\bibinfo {title} {{Remarks on Some
  Nonparametric Estimates of a Density Function}},\ }\href
  {https://doi.org/10.1214/aoms/1177728190} {\bibfield  {journal} {\bibinfo
  {journal} {The Annals of Mathematical Statistics}\ }\textbf {\bibinfo
  {volume} {27}},\ \bibinfo {pages} {832 } (\bibinfo {year}
  {1956})}\BibitemShut {NoStop}%
\bibitem [{\citenamefont {Epanechnikov}(1969)}]{epanechnikov}%
  \BibitemOpen
  \bibfield  {author} {\bibinfo {author} {\bibfnamefont {V.~A.}\ \bibnamefont
  {Epanechnikov}},\ }\bibfield  {title} {\bibinfo {title} {Non-parametric
  estimation of a multivariate probability density},\ }\href
  {https://doi.org/10.1137/1114019} {\bibfield  {journal} {\bibinfo  {journal}
  {Theory of Probability \& Its Applications}\ }\textbf {\bibinfo {volume}
  {14}},\ \bibinfo {pages} {153} (\bibinfo {year} {1969})},\ \Eprint
  {https://arxiv.org/abs/https://doi.org/10.1137/1114019}
  {https://doi.org/10.1137/1114019} \BibitemShut {NoStop}%
\bibitem [{\citenamefont {Sheather}\ and\ \citenamefont {Jones}(1991)}]{sj}%
  \BibitemOpen
  \bibfield  {author} {\bibinfo {author} {\bibfnamefont {S.~J.}\ \bibnamefont
  {Sheather}}\ and\ \bibinfo {author} {\bibfnamefont {M.~C.}\ \bibnamefont
  {Jones}},\ }\bibfield  {title} {\bibinfo {title} {A reliable data-based
  bandwidth selection method for kernel density estimation},\ }\href
  {http://www.jstor.org/stable/2345597} {\bibfield  {journal} {\bibinfo
  {journal} {Journal of the Royal Statistical Society. Series B
  (Methodological)}\ }\textbf {\bibinfo {volume} {53}},\ \bibinfo {pages} {683}
  (\bibinfo {year} {1991})}\BibitemShut {NoStop}%
\bibitem [{\citenamefont {Johnson}\ \emph {et~al.}(2023)\citenamefont
  {Johnson}, \citenamefont {Meyers}, \citenamefont {Baker}, \citenamefont
  {Cornish}, \citenamefont {Hazboun}, \citenamefont {Littenberg}, \citenamefont
  {Romano}, \citenamefont {Taylor}, \citenamefont {Vallisneri}, \citenamefont
  {Vigeland} \emph {et~al.}}]{johnson2023nanograv}%
  \BibitemOpen
  \bibfield  {author} {\bibinfo {author} {\bibfnamefont {A.~D.}\ \bibnamefont
  {Johnson}}, \bibinfo {author} {\bibfnamefont {P.~M.}\ \bibnamefont {Meyers}},
  \bibinfo {author} {\bibfnamefont {P.~T.}\ \bibnamefont {Baker}}, \bibinfo
  {author} {\bibfnamefont {N.~J.}\ \bibnamefont {Cornish}}, \bibinfo {author}
  {\bibfnamefont {J.~S.}\ \bibnamefont {Hazboun}}, \bibinfo {author}
  {\bibfnamefont {T.~B.}\ \bibnamefont {Littenberg}}, \bibinfo {author}
  {\bibfnamefont {J.~D.}\ \bibnamefont {Romano}}, \bibinfo {author}
  {\bibfnamefont {S.~R.}\ \bibnamefont {Taylor}}, \bibinfo {author}
  {\bibfnamefont {M.}~\bibnamefont {Vallisneri}}, \bibinfo {author}
  {\bibfnamefont {S.~J.}\ \bibnamefont {Vigeland}}, \emph {et~al.},\ }\bibfield
   {title} {\bibinfo {title} {The nanograv 15-year gravitational-wave
  background analysis pipeline},\ }\href@noop {} {\bibfield  {journal}
  {\bibinfo  {journal} {arXiv preprint arXiv:2306.16223}\ } (\bibinfo {year}
  {2023})}\BibitemShut {NoStop}%
\bibitem [{\citenamefont {Hellinger}(1909)}]{hellinger}%
  \BibitemOpen
  \bibfield  {author} {\bibinfo {author} {\bibfnamefont {E.~D.}\ \bibnamefont
  {Hellinger}},\ }\bibfield  {title} {\bibinfo {title} {Journal f\"{u}r die
  reine und angewandte mathematik},\ }\href@noop {} {\  (\bibinfo {year}
  {1909})}\BibitemShut {NoStop}%
\bibitem [{\citenamefont {Rosado}\ \emph {et~al.}(2015)\citenamefont {Rosado},
  \citenamefont {Sesana},\ and\ \citenamefont {Gair}}]{rosado}%
  \BibitemOpen
  \bibfield  {author} {\bibinfo {author} {\bibfnamefont {P.~A.}\ \bibnamefont
  {Rosado}}, \bibinfo {author} {\bibfnamefont {A.}~\bibnamefont {Sesana}},\
  and\ \bibinfo {author} {\bibfnamefont {J.}~\bibnamefont {Gair}},\ }\bibfield
  {title} {\bibinfo {title} {Expected properties of the first gravitational
  wave signal detected with pulsar timing arrays},\ }\href@noop {} {\bibfield
  {journal} {\bibinfo  {journal} {Monthly Notices of the Royal Astronomical
  Society}\ }\textbf {\bibinfo {volume} {451}},\ \bibinfo {pages} {2417}
  (\bibinfo {year} {2015})}\BibitemShut {NoStop}%
\bibitem [{\citenamefont {Middleton}\ \emph {et~al.}(2021)\citenamefont
  {Middleton}, \citenamefont {Sesana}, \citenamefont {Chen}, \citenamefont
  {Vecchio}, \citenamefont {Del~Pozzo},\ and\ \citenamefont
  {Rosado}}]{middleton}%
  \BibitemOpen
  \bibfield  {author} {\bibinfo {author} {\bibfnamefont {H.}~\bibnamefont
  {Middleton}}, \bibinfo {author} {\bibfnamefont {A.}~\bibnamefont {Sesana}},
  \bibinfo {author} {\bibfnamefont {S.}~\bibnamefont {Chen}}, \bibinfo {author}
  {\bibfnamefont {A.}~\bibnamefont {Vecchio}}, \bibinfo {author} {\bibfnamefont
  {W.}~\bibnamefont {Del~Pozzo}},\ and\ \bibinfo {author} {\bibfnamefont
  {P.~A.}\ \bibnamefont {Rosado}},\ }\bibfield  {title} {\bibinfo {title}
  {Massive black hole binary systems and the nanograv 12.5 yr results},\
  }\href@noop {} {\bibfield  {journal} {\bibinfo  {journal} {Monthly Notices of
  the Royal Astronomical Society: Letters}\ }\textbf {\bibinfo {volume}
  {502}},\ \bibinfo {pages} {L99} (\bibinfo {year} {2021})}\BibitemShut
  {NoStop}%
\bibitem [{\citenamefont {Pearson}(1895)}]{pearson1895vii}%
  \BibitemOpen
  \bibfield  {author} {\bibinfo {author} {\bibfnamefont {K.}~\bibnamefont
  {Pearson}},\ }\bibfield  {title} {\bibinfo {title} {Vii. note on regression
  and inheritance in the case of two parents},\ }\href@noop {} {\bibfield
  {journal} {\bibinfo  {journal} {proceedings of the royal society of London}\
  }\textbf {\bibinfo {volume} {58}},\ \bibinfo {pages} {240} (\bibinfo {year}
  {1895})}\BibitemShut {NoStop}%
\bibitem [{\citenamefont {Ashton}\ and\ \citenamefont {Talbot}(2021)}]{bilby}%
  \BibitemOpen
  \bibfield  {author} {\bibinfo {author} {\bibfnamefont {G.}~\bibnamefont
  {Ashton}}\ and\ \bibinfo {author} {\bibfnamefont {C.}~\bibnamefont
  {Talbot}},\ }\bibfield  {title} {\bibinfo {title} {Bilby-mcmc: an mcmc
  sampler for gravitational-wave inference},\ }\href@noop {} {\bibfield
  {journal} {\bibinfo  {journal} {Monthly Notices of the Royal Astronomical
  Society}\ }\textbf {\bibinfo {volume} {507}},\ \bibinfo {pages} {2037}
  (\bibinfo {year} {2021})}\BibitemShut {NoStop}%
\bibitem [{\citenamefont {Kass}\ and\ \citenamefont
  {Raftery}(1995)}]{kass1995bayes}%
  \BibitemOpen
  \bibfield  {author} {\bibinfo {author} {\bibfnamefont {R.~E.}\ \bibnamefont
  {Kass}}\ and\ \bibinfo {author} {\bibfnamefont {A.~E.}\ \bibnamefont
  {Raftery}},\ }\bibfield  {title} {\bibinfo {title} {Bayes factors},\
  }\href@noop {} {\bibfield  {journal} {\bibinfo  {journal} {Journal of the
  american statistical association}\ }\textbf {\bibinfo {volume} {90}},\
  \bibinfo {pages} {773} (\bibinfo {year} {1995})}\BibitemShut {NoStop}%
\bibitem [{\citenamefont {Dickey}(1971)}]{dickey1971weighted}%
  \BibitemOpen
  \bibfield  {author} {\bibinfo {author} {\bibfnamefont {J.~M.}\ \bibnamefont
  {Dickey}},\ }\bibfield  {title} {\bibinfo {title} {The weighted likelihood
  ratio, linear hypotheses on normal location parameters},\ }\href@noop {}
  {\bibfield  {journal} {\bibinfo  {journal} {The Annals of Mathematical
  Statistics}\ ,\ \bibinfo {pages} {204}} (\bibinfo {year} {1971})}\BibitemShut
  {NoStop}%
\bibitem [{\citenamefont {Carlin}\ and\ \citenamefont
  {Chib}(1995)}]{carlin1995bayesian}%
  \BibitemOpen
  \bibfield  {author} {\bibinfo {author} {\bibfnamefont {B.~P.}\ \bibnamefont
  {Carlin}}\ and\ \bibinfo {author} {\bibfnamefont {S.}~\bibnamefont {Chib}},\
  }\bibfield  {title} {\bibinfo {title} {Bayesian model choice via markov chain
  monte carlo methods},\ }\href@noop {} {\bibfield  {journal} {\bibinfo
  {journal} {Journal of the Royal Statistical Society: Series B
  (Methodological)}\ }\textbf {\bibinfo {volume} {57}},\ \bibinfo {pages} {473}
  (\bibinfo {year} {1995})}\BibitemShut {NoStop}%
\bibitem [{\citenamefont {Godsill}(2001)}]{godsill2001relationship}%
  \BibitemOpen
  \bibfield  {author} {\bibinfo {author} {\bibfnamefont {S.~J.}\ \bibnamefont
  {Godsill}},\ }\bibfield  {title} {\bibinfo {title} {On the relationship
  between markov chain monte carlo methods for model uncertainty},\ }\href@noop
  {} {\bibfield  {journal} {\bibinfo  {journal} {Journal of computational and
  graphical statistics}\ }\textbf {\bibinfo {volume} {10}},\ \bibinfo {pages}
  {230} (\bibinfo {year} {2001})}\BibitemShut {NoStop}%
\bibitem [{\citenamefont {Aggarwal}\ \emph {et~al.}(2019)\citenamefont
  {Aggarwal}, \citenamefont {Arzoumanian}, \citenamefont {Baker}, \citenamefont
  {Brazier}, \citenamefont {Brinson}, \citenamefont {Brook}, \citenamefont
  {Burke-Spolaor}, \citenamefont {Chatterjee}, \citenamefont {Cordes},
  \citenamefont {Cornish} \emph {et~al.}}]{aggarwal2019nanograv}%
  \BibitemOpen
  \bibfield  {author} {\bibinfo {author} {\bibfnamefont {K.}~\bibnamefont
  {Aggarwal}}, \bibinfo {author} {\bibfnamefont {Z.}~\bibnamefont
  {Arzoumanian}}, \bibinfo {author} {\bibfnamefont {P.}~\bibnamefont {Baker}},
  \bibinfo {author} {\bibfnamefont {A.}~\bibnamefont {Brazier}}, \bibinfo
  {author} {\bibfnamefont {M.}~\bibnamefont {Brinson}}, \bibinfo {author}
  {\bibfnamefont {P.}~\bibnamefont {Brook}}, \bibinfo {author} {\bibfnamefont
  {S.}~\bibnamefont {Burke-Spolaor}}, \bibinfo {author} {\bibfnamefont
  {S.}~\bibnamefont {Chatterjee}}, \bibinfo {author} {\bibfnamefont
  {J.}~\bibnamefont {Cordes}}, \bibinfo {author} {\bibfnamefont
  {N.}~\bibnamefont {Cornish}}, \emph {et~al.},\ }\bibfield  {title} {\bibinfo
  {title} {The nanograv 11 yr data set: limits on gravitational waves from
  individual supermassive black hole binaries},\ }\href@noop {} {\bibfield
  {journal} {\bibinfo  {journal} {The Astrophysical Journal}\ }\textbf
  {\bibinfo {volume} {880}},\ \bibinfo {pages} {116} (\bibinfo {year}
  {2019})}\BibitemShut {NoStop}%
\bibitem [{\citenamefont {Skilling}(2004)}]{skilling2004nested}%
  \BibitemOpen
  \bibfield  {author} {\bibinfo {author} {\bibfnamefont {J.}~\bibnamefont
  {Skilling}},\ }\bibfield  {title} {\bibinfo {title} {Nested sampling},\ }in\
  \href@noop {} {\emph {\bibinfo {booktitle} {Aip conference proceedings}}},\
  Vol.\ \bibinfo {volume} {735}\ (\bibinfo {organization} {American Institute
  of Physics},\ \bibinfo {year} {2004})\ pp.\ \bibinfo {pages}
  {395--405}\BibitemShut {NoStop}%
\bibitem [{\citenamefont {Buchner}(2023)}]{2021arXiv210109675B}%
  \BibitemOpen
  \bibfield  {author} {\bibinfo {author} {\bibfnamefont {J.}~\bibnamefont
  {Buchner}},\ }\bibfield  {title} {\bibinfo {title} {Nested sampling
  methods},\ }\href@noop {} {\bibfield  {journal} {\bibinfo  {journal}
  {Statistic Surveys}\ }\textbf {\bibinfo {volume} {17}},\ \bibinfo {pages}
  {169} (\bibinfo {year} {2023})}\BibitemShut {NoStop}%
\bibitem [{\citenamefont {Chen}\ \emph
  {et~al.}(2021{\natexlab{b}})\citenamefont {Chen}, \citenamefont {Caballero},
  \citenamefont {Guo}, \citenamefont {Chalumeau}, \citenamefont {Liu},
  \citenamefont {Shaifullah}, \citenamefont {Lee}, \citenamefont {Babak},
  \citenamefont {Desvignes}, \citenamefont {Parthasarathy} \emph
  {et~al.}}]{chen2021common}%
  \BibitemOpen
  \bibfield  {author} {\bibinfo {author} {\bibfnamefont {S.}~\bibnamefont
  {Chen}}, \bibinfo {author} {\bibfnamefont {R.}~\bibnamefont {Caballero}},
  \bibinfo {author} {\bibfnamefont {Y.}~\bibnamefont {Guo}}, \bibinfo {author}
  {\bibfnamefont {A.}~\bibnamefont {Chalumeau}}, \bibinfo {author}
  {\bibfnamefont {K.}~\bibnamefont {Liu}}, \bibinfo {author} {\bibfnamefont
  {G.}~\bibnamefont {Shaifullah}}, \bibinfo {author} {\bibfnamefont
  {K.}~\bibnamefont {Lee}}, \bibinfo {author} {\bibfnamefont {S.}~\bibnamefont
  {Babak}}, \bibinfo {author} {\bibfnamefont {G.}~\bibnamefont {Desvignes}},
  \bibinfo {author} {\bibfnamefont {A.}~\bibnamefont {Parthasarathy}}, \emph
  {et~al.},\ }\bibfield  {title} {\bibinfo {title} {Common-red-signal analysis
  with 24-yr high-precision timing of the european pulsar timing array:
  inferences in the stochastic gravitational-wave background search},\
  }\href@noop {} {\bibfield  {journal} {\bibinfo  {journal} {Monthly Notices of
  the Royal Astronomical Society}\ }\textbf {\bibinfo {volume} {508}},\
  \bibinfo {pages} {4970} (\bibinfo {year} {2021}{\natexlab{b}})}\BibitemShut
  {NoStop}%
\bibitem [{\citenamefont {{Buchner}}(2021)}]{ultranest}%
  \BibitemOpen
  \bibfield  {author} {\bibinfo {author} {\bibfnamefont {J.}~\bibnamefont
  {{Buchner}}},\ }\bibfield  {title} {\bibinfo {title} {{UltraNest - a robust,
  general purpose Bayesian inference engine}},\ }\href
  {https://doi.org/10.21105/joss.03001} {\bibfield  {journal} {\bibinfo
  {journal} {The Journal of Open Source Software}\ }\textbf {\bibinfo {volume}
  {6}},\ \bibinfo {eid} {3001} (\bibinfo {year} {2021})},\ \Eprint
  {https://arxiv.org/abs/2101.09604} {arXiv:2101.09604 [stat.CO]} \BibitemShut
  {NoStop}%
\bibitem [{\citenamefont {Arzoumanian}\ \emph {et~al.}(2016)\citenamefont
  {Arzoumanian}, \citenamefont {Brazier}, \citenamefont {Burke-Spolaor},
  \citenamefont {Chamberlin}, \citenamefont {Chatterjee}, \citenamefont
  {Christy}, \citenamefont {Cordes}, \citenamefont {Cornish}, \citenamefont
  {Crowter}, \citenamefont {Demorest}, \citenamefont {Deng}, \citenamefont
  {Dolch}, \citenamefont {Ellis}, \citenamefont {Ferdman}, \citenamefont
  {Fonseca}, \citenamefont {Garver-Daniels}, \citenamefont {Gonzalez},
  \citenamefont {Jenet}, \citenamefont {Jones}, \citenamefont {Jones},
  \citenamefont {Kaspi}, \citenamefont {Koop}, \citenamefont {Lam},
  \citenamefont {Lazio}, \citenamefont {Levin}, \citenamefont {Lommen},
  \citenamefont {Lorimer}, \citenamefont {Luo}, \citenamefont {Lynch},
  \citenamefont {Madison}, \citenamefont {McLaughlin}, \citenamefont
  {McWilliams}, \citenamefont {Mingarelli}, \citenamefont {Nice}, \citenamefont
  {Palliyaguru}, \citenamefont {Pennucci}, \citenamefont {Ransom},
  \citenamefont {Sampson}, \citenamefont {Sanidas}, \citenamefont {Sesana},
  \citenamefont {Siemens}, \citenamefont {Simon}, \citenamefont {Stairs},
  \citenamefont {Stinebring}, \citenamefont {Stovall}, \citenamefont {Swiggum},
  \citenamefont {Taylor}, \citenamefont {Vallisneri}, \citenamefont {van
  Haasteren}, \citenamefont {Wang}, \citenamefont {Zhu},\ and\ \citenamefont
  {Collaboration)}}]{Arzoumanian_2016}%
  \BibitemOpen
  \bibfield  {author} {\bibinfo {author} {\bibfnamefont {Z.}~\bibnamefont
  {Arzoumanian}}, \bibinfo {author} {\bibfnamefont {A.}~\bibnamefont
  {Brazier}}, \bibinfo {author} {\bibfnamefont {S.}~\bibnamefont
  {Burke-Spolaor}}, \bibinfo {author} {\bibfnamefont {S.~J.}\ \bibnamefont
  {Chamberlin}}, \bibinfo {author} {\bibfnamefont {S.}~\bibnamefont
  {Chatterjee}}, \bibinfo {author} {\bibfnamefont {B.}~\bibnamefont {Christy}},
  \bibinfo {author} {\bibfnamefont {J.~M.}\ \bibnamefont {Cordes}}, \bibinfo
  {author} {\bibfnamefont {N.~J.}\ \bibnamefont {Cornish}}, \bibinfo {author}
  {\bibfnamefont {K.}~\bibnamefont {Crowter}}, \bibinfo {author} {\bibfnamefont
  {P.~B.}\ \bibnamefont {Demorest}}, \bibinfo {author} {\bibfnamefont
  {X.}~\bibnamefont {Deng}}, \bibinfo {author} {\bibfnamefont {T.}~\bibnamefont
  {Dolch}}, \bibinfo {author} {\bibfnamefont {J.~A.}\ \bibnamefont {Ellis}},
  \bibinfo {author} {\bibfnamefont {R.~D.}\ \bibnamefont {Ferdman}}, \bibinfo
  {author} {\bibfnamefont {E.}~\bibnamefont {Fonseca}}, \bibinfo {author}
  {\bibfnamefont {N.}~\bibnamefont {Garver-Daniels}}, \bibinfo {author}
  {\bibfnamefont {M.~E.}\ \bibnamefont {Gonzalez}}, \bibinfo {author}
  {\bibfnamefont {F.}~\bibnamefont {Jenet}}, \bibinfo {author} {\bibfnamefont
  {G.}~\bibnamefont {Jones}}, \bibinfo {author} {\bibfnamefont {M.~L.}\
  \bibnamefont {Jones}}, \bibinfo {author} {\bibfnamefont {V.~M.}\ \bibnamefont
  {Kaspi}}, \bibinfo {author} {\bibfnamefont {M.}~\bibnamefont {Koop}},
  \bibinfo {author} {\bibfnamefont {M.~T.}\ \bibnamefont {Lam}}, \bibinfo
  {author} {\bibfnamefont {T.~J.~W.}\ \bibnamefont {Lazio}}, \bibinfo {author}
  {\bibfnamefont {L.}~\bibnamefont {Levin}}, \bibinfo {author} {\bibfnamefont
  {A.~N.}\ \bibnamefont {Lommen}}, \bibinfo {author} {\bibfnamefont {D.~R.}\
  \bibnamefont {Lorimer}}, \bibinfo {author} {\bibfnamefont {J.}~\bibnamefont
  {Luo}}, \bibinfo {author} {\bibfnamefont {R.~S.}\ \bibnamefont {Lynch}},
  \bibinfo {author} {\bibfnamefont {D.~R.}\ \bibnamefont {Madison}}, \bibinfo
  {author} {\bibfnamefont {M.~A.}\ \bibnamefont {McLaughlin}}, \bibinfo
  {author} {\bibfnamefont {S.~T.}\ \bibnamefont {McWilliams}}, \bibinfo
  {author} {\bibfnamefont {C.~M.~F.}\ \bibnamefont {Mingarelli}}, \bibinfo
  {author} {\bibfnamefont {D.~J.}\ \bibnamefont {Nice}}, \bibinfo {author}
  {\bibfnamefont {N.}~\bibnamefont {Palliyaguru}}, \bibinfo {author}
  {\bibfnamefont {T.~T.}\ \bibnamefont {Pennucci}}, \bibinfo {author}
  {\bibfnamefont {S.~M.}\ \bibnamefont {Ransom}}, \bibinfo {author}
  {\bibfnamefont {L.}~\bibnamefont {Sampson}}, \bibinfo {author} {\bibfnamefont
  {S.~A.}\ \bibnamefont {Sanidas}}, \bibinfo {author} {\bibfnamefont
  {A.}~\bibnamefont {Sesana}}, \bibinfo {author} {\bibfnamefont
  {X.}~\bibnamefont {Siemens}}, \bibinfo {author} {\bibfnamefont
  {J.}~\bibnamefont {Simon}}, \bibinfo {author} {\bibfnamefont {I.~H.}\
  \bibnamefont {Stairs}}, \bibinfo {author} {\bibfnamefont {D.~R.}\
  \bibnamefont {Stinebring}}, \bibinfo {author} {\bibfnamefont
  {K.}~\bibnamefont {Stovall}}, \bibinfo {author} {\bibfnamefont
  {J.}~\bibnamefont {Swiggum}}, \bibinfo {author} {\bibfnamefont {S.~R.}\
  \bibnamefont {Taylor}}, \bibinfo {author} {\bibfnamefont {M.}~\bibnamefont
  {Vallisneri}}, \bibinfo {author} {\bibfnamefont {R.}~\bibnamefont {van
  Haasteren}}, \bibinfo {author} {\bibfnamefont {Y.}~\bibnamefont {Wang}},
  \bibinfo {author} {\bibfnamefont {W.~W.}\ \bibnamefont {Zhu}},\ and\ \bibinfo
  {author} {\bibfnamefont {T.~N.}\ \bibnamefont {Collaboration)}},\ }\bibfield
  {title} {\bibinfo {title} {The nanograv nine-year data set: Limits on the
  isotropic stochastic gravitational wave background},\ }\href
  {https://doi.org/10.3847/0004-637X/821/1/13} {\bibfield  {journal} {\bibinfo
  {journal} {The Astrophysical Journal}\ }\textbf {\bibinfo {volume} {821}},\
  \bibinfo {pages} {13} (\bibinfo {year} {2016})}\BibitemShut {NoStop}%
\bibitem [{\citenamefont {Arzoumanian}\ \emph
  {et~al.}(2020{\natexlab{b}})\citenamefont {Arzoumanian}, \citenamefont
  {Baker}, \citenamefont {Brazier}, \citenamefont {Brook}, \citenamefont
  {Burke-Spolaor}, \citenamefont {B{\'e}csy}, \citenamefont {Charisi},
  \citenamefont {Chatterjee}, \citenamefont {Cordes}, \citenamefont {Cornish}
  \emph {et~al.}}]{arzoumanian2020multimessenger}%
  \BibitemOpen
  \bibfield  {author} {\bibinfo {author} {\bibfnamefont {Z.}~\bibnamefont
  {Arzoumanian}}, \bibinfo {author} {\bibfnamefont {P.~T.}\ \bibnamefont
  {Baker}}, \bibinfo {author} {\bibfnamefont {A.}~\bibnamefont {Brazier}},
  \bibinfo {author} {\bibfnamefont {P.~R.}\ \bibnamefont {Brook}}, \bibinfo
  {author} {\bibfnamefont {S.}~\bibnamefont {Burke-Spolaor}}, \bibinfo {author}
  {\bibfnamefont {B.}~\bibnamefont {B{\'e}csy}}, \bibinfo {author}
  {\bibfnamefont {M.}~\bibnamefont {Charisi}}, \bibinfo {author} {\bibfnamefont
  {S.}~\bibnamefont {Chatterjee}}, \bibinfo {author} {\bibfnamefont {J.~M.}\
  \bibnamefont {Cordes}}, \bibinfo {author} {\bibfnamefont {N.~J.}\
  \bibnamefont {Cornish}}, \emph {et~al.},\ }\bibfield  {title} {\bibinfo
  {title} {Multimessenger gravitational-wave searches with pulsar timing
  arrays: application to 3c 66b using the nanograv 11-year data set},\
  }\href@noop {} {\bibfield  {journal} {\bibinfo  {journal} {The Astrophysical
  Journal}\ }\textbf {\bibinfo {volume} {900}},\ \bibinfo {pages} {102}
  (\bibinfo {year} {2020}{\natexlab{b}})}\BibitemShut {NoStop}%
\bibitem [{\citenamefont {Moore}\ \emph {et~al.}(2015)\citenamefont {Moore},
  \citenamefont {Taylor},\ and\ \citenamefont {Gair}}]{moore2015estimating}%
  \BibitemOpen
  \bibfield  {author} {\bibinfo {author} {\bibfnamefont {C.~J.}\ \bibnamefont
  {Moore}}, \bibinfo {author} {\bibfnamefont {S.~R.}\ \bibnamefont {Taylor}},\
  and\ \bibinfo {author} {\bibfnamefont {J.~R.}\ \bibnamefont {Gair}},\
  }\bibfield  {title} {\bibinfo {title} {Estimating the sensitivity of pulsar
  timing arrays},\ }\href@noop {} {\bibfield  {journal} {\bibinfo  {journal}
  {Classical and Quantum Gravity}\ }\textbf {\bibinfo {volume} {32}},\ \bibinfo
  {pages} {055004} (\bibinfo {year} {2015})}\BibitemShut {NoStop}%
\bibitem [{\citenamefont {Hazboun}\ \emph {et~al.}(2019)\citenamefont
  {Hazboun}, \citenamefont {Romano},\ and\ \citenamefont {Smith}}]{hazboun}%
  \BibitemOpen
  \bibfield  {author} {\bibinfo {author} {\bibfnamefont {J.~S.}\ \bibnamefont
  {Hazboun}}, \bibinfo {author} {\bibfnamefont {J.~D.}\ \bibnamefont
  {Romano}},\ and\ \bibinfo {author} {\bibfnamefont {T.~L.}\ \bibnamefont
  {Smith}},\ }\bibfield  {title} {\bibinfo {title} {Realistic sensitivity
  curves for pulsar timing arrays},\ }\href
  {https://doi.org/10.1103/PhysRevD.100.104028} {\bibfield  {journal} {\bibinfo
   {journal} {Phys. Rev. D}\ }\textbf {\bibinfo {volume} {100}},\ \bibinfo
  {pages} {104028} (\bibinfo {year} {2019})}\BibitemShut {NoStop}%
\bibitem [{\citenamefont {Cornish}\ and\ \citenamefont
  {Sampson}(2016)}]{Cornish2015-ly}%
  \BibitemOpen
  \bibfield  {author} {\bibinfo {author} {\bibfnamefont {N.~J.}\ \bibnamefont
  {Cornish}}\ and\ \bibinfo {author} {\bibfnamefont {L.}~\bibnamefont
  {Sampson}},\ }\bibfield  {title} {\bibinfo {title} {Towards robust
  gravitational wave detection with pulsar timing arrays},\ }\href@noop {}
  {\bibfield  {journal} {\bibinfo  {journal} {Physical Review D}\ }\textbf
  {\bibinfo {volume} {93}},\ \bibinfo {pages} {104047} (\bibinfo {year}
  {2016})}\BibitemShut {NoStop}%
\bibitem [{\citenamefont {Ivezi{\'c}}\ \emph {et~al.}(2020)\citenamefont
  {Ivezi{\'c}}, \citenamefont {Connolly}, \citenamefont {VanderPlas},\ and\
  \citenamefont {Gray}}]{ivezic2020statistics}%
  \BibitemOpen
  \bibfield  {author} {\bibinfo {author} {\bibfnamefont {{\v{Z}}.}~\bibnamefont
  {Ivezi{\'c}}}, \bibinfo {author} {\bibfnamefont {A.~J.}\ \bibnamefont
  {Connolly}}, \bibinfo {author} {\bibfnamefont {J.~T.}\ \bibnamefont
  {VanderPlas}},\ and\ \bibinfo {author} {\bibfnamefont {A.}~\bibnamefont
  {Gray}},\ }\href@noop {} {\emph {\bibinfo {title} {Statistics, data mining,
  and machine learning in astronomy: a practical Python guide for the analysis
  of survey data}}}\ (\bibinfo  {publisher} {Princeton University Press},\
  \bibinfo {year} {2020})\BibitemShut {NoStop}%
\bibitem [{\citenamefont {Bailes}\ \emph {et~al.}(2018)\citenamefont {Bailes},
  \citenamefont {Barr}, \citenamefont {Bhat}, \citenamefont {Brink},
  \citenamefont {Buchner}, \citenamefont {Burgay}, \citenamefont {Camilo},
  \citenamefont {Champion}, \citenamefont {Hessels}, \citenamefont {Janssen}
  \emph {et~al.}}]{bailes2018meertime}%
  \BibitemOpen
  \bibfield  {author} {\bibinfo {author} {\bibfnamefont {M.}~\bibnamefont
  {Bailes}}, \bibinfo {author} {\bibfnamefont {E.}~\bibnamefont {Barr}},
  \bibinfo {author} {\bibfnamefont {N.}~\bibnamefont {Bhat}}, \bibinfo {author}
  {\bibfnamefont {J.}~\bibnamefont {Brink}}, \bibinfo {author} {\bibfnamefont
  {S.}~\bibnamefont {Buchner}}, \bibinfo {author} {\bibfnamefont
  {M.}~\bibnamefont {Burgay}}, \bibinfo {author} {\bibfnamefont
  {F.}~\bibnamefont {Camilo}}, \bibinfo {author} {\bibfnamefont
  {D.}~\bibnamefont {Champion}}, \bibinfo {author} {\bibfnamefont
  {J.}~\bibnamefont {Hessels}}, \bibinfo {author} {\bibfnamefont
  {G.}~\bibnamefont {Janssen}}, \emph {et~al.},\ }\bibfield  {title} {\bibinfo
  {title} {Meertime-the meerkat key science program on pulsar timing},\
  }\href@noop {} {\bibfield  {journal} {\bibinfo  {journal} {arXiv preprint
  arXiv:1803.07424}\ } (\bibinfo {year} {2018})}\BibitemShut {NoStop}%
\bibitem [{\citenamefont {Amiri}\ \emph {et~al.}(2018)\citenamefont {Amiri},
  \citenamefont {Bandura}, \citenamefont {Berger}, \citenamefont {Bhardwaj},
  \citenamefont {Boyce}, \citenamefont {Boyle}, \citenamefont {Brar},
  \citenamefont {Burhanpurkar}, \citenamefont {Chawla}, \citenamefont
  {Chowdhury} \emph {et~al.}}]{chime}%
  \BibitemOpen
  \bibfield  {author} {\bibinfo {author} {\bibfnamefont {M.}~\bibnamefont
  {Amiri}}, \bibinfo {author} {\bibfnamefont {K.}~\bibnamefont {Bandura}},
  \bibinfo {author} {\bibfnamefont {P.}~\bibnamefont {Berger}}, \bibinfo
  {author} {\bibfnamefont {M.}~\bibnamefont {Bhardwaj}}, \bibinfo {author}
  {\bibfnamefont {M.}~\bibnamefont {Boyce}}, \bibinfo {author} {\bibfnamefont
  {P.}~\bibnamefont {Boyle}}, \bibinfo {author} {\bibfnamefont
  {C.}~\bibnamefont {Brar}}, \bibinfo {author} {\bibfnamefont {M.}~\bibnamefont
  {Burhanpurkar}}, \bibinfo {author} {\bibfnamefont {P.}~\bibnamefont
  {Chawla}}, \bibinfo {author} {\bibfnamefont {J.}~\bibnamefont {Chowdhury}},
  \emph {et~al.},\ }\bibfield  {title} {\bibinfo {title} {The chime fast radio
  burst project: system overview},\ }\href@noop {} {\bibfield  {journal}
  {\bibinfo  {journal} {The Astrophysical Journal}\ }\textbf {\bibinfo {volume}
  {863}},\ \bibinfo {pages} {48} (\bibinfo {year} {2018})}\BibitemShut
  {NoStop}%
\bibitem [{\citenamefont {Carilli}\ and\ \citenamefont {Rawlings}(2004)}]{ska}%
  \BibitemOpen
  \bibfield  {author} {\bibinfo {author} {\bibfnamefont {C.}~\bibnamefont
  {Carilli}}\ and\ \bibinfo {author} {\bibfnamefont {S.}~\bibnamefont
  {Rawlings}},\ }\bibfield  {title} {\bibinfo {title} {Motivation, key science
  projects, standards and assumptions},\ }\href
  {https://doi.org/https://doi.org/10.1016/j.newar.2004.09.001} {\bibfield
  {journal} {\bibinfo  {journal} {New Astronomy Reviews}\ }\textbf {\bibinfo
  {volume} {48}},\ \bibinfo {pages} {979} (\bibinfo {year} {2004})},\ \bibinfo
  {note} {science with the Square Kilometre Array}\BibitemShut {NoStop}%
\bibitem [{\citenamefont {Agazie}\ \emph
  {et~al.}(2023{\natexlab{c}})\citenamefont {Agazie}, \citenamefont
  {Anumarlapudi}, \citenamefont {Archibald}, \citenamefont {Arzoumanian},
  \citenamefont {Baker}, \citenamefont {Bécsy}, \citenamefont {Blecha},
  \citenamefont {Brazier}, \citenamefont {Brook}, \citenamefont
  {Burke-Spolaor}, \citenamefont {Charisi}, \citenamefont {Chatterjee},
  \citenamefont {Cohen}, \citenamefont {Cordes}, \citenamefont {Cornish},
  \citenamefont {Crawford}, \citenamefont {Cromartie}, \citenamefont {Crowter},
  \citenamefont {DeCesar}, \citenamefont {Demorest}, \citenamefont {Dolch},
  \citenamefont {Drachler}, \citenamefont {Ferrara}, \citenamefont {Fiore},
  \citenamefont {Fonseca}, \citenamefont {Freedman}, \citenamefont
  {Garver-Daniels}, \citenamefont {Gentile}, \citenamefont {Glaser},
  \citenamefont {Good}, \citenamefont {Guertin}, \citenamefont {Gültekin},
  \citenamefont {Hazboun}, \citenamefont {Jennings}, \citenamefont {Johnson},
  \citenamefont {Jones}, \citenamefont {Kaiser}, \citenamefont {Kaplan},
  \citenamefont {Kelley}, \citenamefont {Kerr}, \citenamefont {Key},
  \citenamefont {Laal}, \citenamefont {Lam}, \citenamefont {Lamb},
  \citenamefont {Lazio}, \citenamefont {Lewandowska}, \citenamefont {Liu},
  \citenamefont {Lorimer}, \citenamefont {Luo}, \citenamefont {Lynch},
  \citenamefont {Ma}, \citenamefont {Madison}, \citenamefont {McEwen},
  \citenamefont {McKee}, \citenamefont {McLaughlin}, \citenamefont {McMann},
  \citenamefont {Meyers}, \citenamefont {Mingarelli}, \citenamefont
  {Mitridate}, \citenamefont {Ng}, \citenamefont {Nice}, \citenamefont {Ocker},
  \citenamefont {Olum}, \citenamefont {Pennucci}, \citenamefont {Perera},
  \citenamefont {Pol}, \citenamefont {Radovan}, \citenamefont {Ransom},
  \citenamefont {Ray}, \citenamefont {Romano}, \citenamefont {Sardesai},
  \citenamefont {Schmiedekamp}, \citenamefont {Schmiedekamp}, \citenamefont
  {Schmitz}, \citenamefont {Shapiro-Albert}, \citenamefont {Siemens},
  \citenamefont {Simon}, \citenamefont {Siwek}, \citenamefont {Stairs},
  \citenamefont {Stinebring}, \citenamefont {Stovall}, \citenamefont
  {Susobhanan}, \citenamefont {Swiggum}, \citenamefont {Taylor}, \citenamefont
  {Turner}, \citenamefont {Unal}, \citenamefont {Vallisneri}, \citenamefont
  {Vigeland}, \citenamefont {Wahl}, \citenamefont {Witt}, \citenamefont
  {Young},\ and\ \citenamefont {Collaboration}}]{Agazie_2023}%
  \BibitemOpen
  \bibfield  {author} {\bibinfo {author} {\bibfnamefont {G.}~\bibnamefont
  {Agazie}}, \bibinfo {author} {\bibfnamefont {A.}~\bibnamefont
  {Anumarlapudi}}, \bibinfo {author} {\bibfnamefont {A.~M.}\ \bibnamefont
  {Archibald}}, \bibinfo {author} {\bibfnamefont {Z.}~\bibnamefont
  {Arzoumanian}}, \bibinfo {author} {\bibfnamefont {P.~T.}\ \bibnamefont
  {Baker}}, \bibinfo {author} {\bibfnamefont {B.}~\bibnamefont {Bécsy}},
  \bibinfo {author} {\bibfnamefont {L.}~\bibnamefont {Blecha}}, \bibinfo
  {author} {\bibfnamefont {A.}~\bibnamefont {Brazier}}, \bibinfo {author}
  {\bibfnamefont {P.~R.}\ \bibnamefont {Brook}}, \bibinfo {author}
  {\bibfnamefont {S.}~\bibnamefont {Burke-Spolaor}}, \bibinfo {author}
  {\bibfnamefont {M.}~\bibnamefont {Charisi}}, \bibinfo {author} {\bibfnamefont
  {S.}~\bibnamefont {Chatterjee}}, \bibinfo {author} {\bibfnamefont
  {T.}~\bibnamefont {Cohen}}, \bibinfo {author} {\bibfnamefont {J.~M.}\
  \bibnamefont {Cordes}}, \bibinfo {author} {\bibfnamefont {N.~J.}\
  \bibnamefont {Cornish}}, \bibinfo {author} {\bibfnamefont {F.}~\bibnamefont
  {Crawford}}, \bibinfo {author} {\bibfnamefont {H.~T.}\ \bibnamefont
  {Cromartie}}, \bibinfo {author} {\bibfnamefont {K.}~\bibnamefont {Crowter}},
  \bibinfo {author} {\bibfnamefont {M.~E.}\ \bibnamefont {DeCesar}}, \bibinfo
  {author} {\bibfnamefont {P.~B.}\ \bibnamefont {Demorest}}, \bibinfo {author}
  {\bibfnamefont {T.}~\bibnamefont {Dolch}}, \bibinfo {author} {\bibfnamefont
  {B.}~\bibnamefont {Drachler}}, \bibinfo {author} {\bibfnamefont {E.~C.}\
  \bibnamefont {Ferrara}}, \bibinfo {author} {\bibfnamefont {W.}~\bibnamefont
  {Fiore}}, \bibinfo {author} {\bibfnamefont {E.}~\bibnamefont {Fonseca}},
  \bibinfo {author} {\bibfnamefont {G.~E.}\ \bibnamefont {Freedman}}, \bibinfo
  {author} {\bibfnamefont {N.}~\bibnamefont {Garver-Daniels}}, \bibinfo
  {author} {\bibfnamefont {P.~A.}\ \bibnamefont {Gentile}}, \bibinfo {author}
  {\bibfnamefont {J.}~\bibnamefont {Glaser}}, \bibinfo {author} {\bibfnamefont
  {D.~C.}\ \bibnamefont {Good}}, \bibinfo {author} {\bibfnamefont
  {L.}~\bibnamefont {Guertin}}, \bibinfo {author} {\bibfnamefont
  {K.}~\bibnamefont {Gültekin}}, \bibinfo {author} {\bibfnamefont {J.~S.}\
  \bibnamefont {Hazboun}}, \bibinfo {author} {\bibfnamefont {R.~J.}\
  \bibnamefont {Jennings}}, \bibinfo {author} {\bibfnamefont {A.~D.}\
  \bibnamefont {Johnson}}, \bibinfo {author} {\bibfnamefont {M.~L.}\
  \bibnamefont {Jones}}, \bibinfo {author} {\bibfnamefont {A.~R.}\ \bibnamefont
  {Kaiser}}, \bibinfo {author} {\bibfnamefont {D.~L.}\ \bibnamefont {Kaplan}},
  \bibinfo {author} {\bibfnamefont {L.~Z.}\ \bibnamefont {Kelley}}, \bibinfo
  {author} {\bibfnamefont {M.}~\bibnamefont {Kerr}}, \bibinfo {author}
  {\bibfnamefont {J.~S.}\ \bibnamefont {Key}}, \bibinfo {author} {\bibfnamefont
  {N.}~\bibnamefont {Laal}}, \bibinfo {author} {\bibfnamefont {M.~T.}\
  \bibnamefont {Lam}}, \bibinfo {author} {\bibfnamefont {W.~G.}\ \bibnamefont
  {Lamb}}, \bibinfo {author} {\bibfnamefont {T.~J.~W.}\ \bibnamefont {Lazio}},
  \bibinfo {author} {\bibfnamefont {N.}~\bibnamefont {Lewandowska}}, \bibinfo
  {author} {\bibfnamefont {T.}~\bibnamefont {Liu}}, \bibinfo {author}
  {\bibfnamefont {D.~R.}\ \bibnamefont {Lorimer}}, \bibinfo {author}
  {\bibfnamefont {J.}~\bibnamefont {Luo}}, \bibinfo {author} {\bibfnamefont
  {R.~S.}\ \bibnamefont {Lynch}}, \bibinfo {author} {\bibfnamefont {C.-P.}\
  \bibnamefont {Ma}}, \bibinfo {author} {\bibfnamefont {D.~R.}\ \bibnamefont
  {Madison}}, \bibinfo {author} {\bibfnamefont {A.}~\bibnamefont {McEwen}},
  \bibinfo {author} {\bibfnamefont {J.~W.}\ \bibnamefont {McKee}}, \bibinfo
  {author} {\bibfnamefont {M.~A.}\ \bibnamefont {McLaughlin}}, \bibinfo
  {author} {\bibfnamefont {N.}~\bibnamefont {McMann}}, \bibinfo {author}
  {\bibfnamefont {B.~W.}\ \bibnamefont {Meyers}}, \bibinfo {author}
  {\bibfnamefont {C.~M.~F.}\ \bibnamefont {Mingarelli}}, \bibinfo {author}
  {\bibfnamefont {A.}~\bibnamefont {Mitridate}}, \bibinfo {author}
  {\bibfnamefont {C.}~\bibnamefont {Ng}}, \bibinfo {author} {\bibfnamefont
  {D.~J.}\ \bibnamefont {Nice}}, \bibinfo {author} {\bibfnamefont {S.~K.}\
  \bibnamefont {Ocker}}, \bibinfo {author} {\bibfnamefont {K.~D.}\ \bibnamefont
  {Olum}}, \bibinfo {author} {\bibfnamefont {T.~T.}\ \bibnamefont {Pennucci}},
  \bibinfo {author} {\bibfnamefont {B.~B.~P.}\ \bibnamefont {Perera}}, \bibinfo
  {author} {\bibfnamefont {N.~S.}\ \bibnamefont {Pol}}, \bibinfo {author}
  {\bibfnamefont {H.~A.}\ \bibnamefont {Radovan}}, \bibinfo {author}
  {\bibfnamefont {S.~M.}\ \bibnamefont {Ransom}}, \bibinfo {author}
  {\bibfnamefont {P.~S.}\ \bibnamefont {Ray}}, \bibinfo {author} {\bibfnamefont
  {J.~D.}\ \bibnamefont {Romano}}, \bibinfo {author} {\bibfnamefont {S.~C.}\
  \bibnamefont {Sardesai}}, \bibinfo {author} {\bibfnamefont {A.}~\bibnamefont
  {Schmiedekamp}}, \bibinfo {author} {\bibfnamefont {C.}~\bibnamefont
  {Schmiedekamp}}, \bibinfo {author} {\bibfnamefont {K.}~\bibnamefont
  {Schmitz}}, \bibinfo {author} {\bibfnamefont {B.~J.}\ \bibnamefont
  {Shapiro-Albert}}, \bibinfo {author} {\bibfnamefont {X.}~\bibnamefont
  {Siemens}}, \bibinfo {author} {\bibfnamefont {J.}~\bibnamefont {Simon}},
  \bibinfo {author} {\bibfnamefont {M.~S.}\ \bibnamefont {Siwek}}, \bibinfo
  {author} {\bibfnamefont {I.~H.}\ \bibnamefont {Stairs}}, \bibinfo {author}
  {\bibfnamefont {D.~R.}\ \bibnamefont {Stinebring}}, \bibinfo {author}
  {\bibfnamefont {K.}~\bibnamefont {Stovall}}, \bibinfo {author} {\bibfnamefont
  {A.}~\bibnamefont {Susobhanan}}, \bibinfo {author} {\bibfnamefont {J.~K.}\
  \bibnamefont {Swiggum}}, \bibinfo {author} {\bibfnamefont {S.~R.}\
  \bibnamefont {Taylor}}, \bibinfo {author} {\bibfnamefont {J.~E.}\
  \bibnamefont {Turner}}, \bibinfo {author} {\bibfnamefont {C.}~\bibnamefont
  {Unal}}, \bibinfo {author} {\bibfnamefont {M.}~\bibnamefont {Vallisneri}},
  \bibinfo {author} {\bibfnamefont {S.~J.}\ \bibnamefont {Vigeland}}, \bibinfo
  {author} {\bibfnamefont {H.~M.}\ \bibnamefont {Wahl}}, \bibinfo {author}
  {\bibfnamefont {C.~A.}\ \bibnamefont {Witt}}, \bibinfo {author}
  {\bibfnamefont {O.}~\bibnamefont {Young}},\ and\ \bibinfo {author}
  {\bibfnamefont {T.~N.}\ \bibnamefont {Collaboration}},\ }\bibfield  {title}
  {\bibinfo {title} {The nanograv 15 yr data set: Detector characterization and
  noise budget},\ }\href {https://doi.org/10.3847/2041-8213/acda88} {\bibfield
  {journal} {\bibinfo  {journal} {The Astrophysical Journal Letters}\ }\textbf
  {\bibinfo {volume} {951}},\ \bibinfo {pages} {L10} (\bibinfo {year}
  {2023}{\natexlab{c}})}\BibitemShut {NoStop}%
\bibitem [{\citenamefont {Geman}\ and\ \citenamefont
  {Geman}(1984)}]{geman1984stochastic}%
  \BibitemOpen
  \bibfield  {author} {\bibinfo {author} {\bibfnamefont {S.}~\bibnamefont
  {Geman}}\ and\ \bibinfo {author} {\bibfnamefont {D.}~\bibnamefont {Geman}},\
  }\bibfield  {title} {\bibinfo {title} {Stochastic relaxation, gibbs
  distributions, and the bayesian restoration of images},\ }\href@noop {}
  {\bibfield  {journal} {\bibinfo  {journal} {IEEE Transactions on pattern
  analysis and machine intelligence}\ ,\ \bibinfo {pages} {721}} (\bibinfo
  {year} {1984})}\BibitemShut {NoStop}%
\bibitem [{\citenamefont {Laal}\ \emph {et~al.}(2023)\citenamefont {Laal},
  \citenamefont {Lamb}, \citenamefont {Romano}, \citenamefont {Siemens},
  \citenamefont {Taylor},\ and\ \citenamefont {van
  Haasteren}}]{PhysRevD.108.063008}%
  \BibitemOpen
  \bibfield  {author} {\bibinfo {author} {\bibfnamefont {N.}~\bibnamefont
  {Laal}}, \bibinfo {author} {\bibfnamefont {W.~G.}\ \bibnamefont {Lamb}},
  \bibinfo {author} {\bibfnamefont {J.~D.}\ \bibnamefont {Romano}}, \bibinfo
  {author} {\bibfnamefont {X.}~\bibnamefont {Siemens}}, \bibinfo {author}
  {\bibfnamefont {S.~R.}\ \bibnamefont {Taylor}},\ and\ \bibinfo {author}
  {\bibfnamefont {R.}~\bibnamefont {van Haasteren}},\ }\bibfield  {title}
  {\bibinfo {title} {Exploring the capabilities of gibbs sampling in pulsar
  timing arrays},\ }\href {https://doi.org/10.1103/PhysRevD.108.063008}
  {\bibfield  {journal} {\bibinfo  {journal} {Phys. Rev. D}\ }\textbf {\bibinfo
  {volume} {108}},\ \bibinfo {pages} {063008} (\bibinfo {year}
  {2023})}\BibitemShut {NoStop}%
\bibitem [{\citenamefont {Rezende}\ and\ \citenamefont
  {Mohamed}(2015)}]{rezende2015variational}%
  \BibitemOpen
  \bibfield  {author} {\bibinfo {author} {\bibfnamefont {D.}~\bibnamefont
  {Rezende}}\ and\ \bibinfo {author} {\bibfnamefont {S.}~\bibnamefont
  {Mohamed}},\ }\bibfield  {title} {\bibinfo {title} {Variational inference
  with normalizing flows},\ }in\ \href@noop {} {\emph {\bibinfo {booktitle}
  {International conference on machine learning}}}\ (\bibinfo {organization}
  {PMLR},\ \bibinfo {year} {2015})\ pp.\ \bibinfo {pages}
  {1530--1538}\BibitemShut {NoStop}%
\bibitem [{\citenamefont {Hourihane}\ \emph {et~al.}(2023)\citenamefont
  {Hourihane}, \citenamefont {Meyers}, \citenamefont {Johnson}, \citenamefont
  {Chatziioannou},\ and\ \citenamefont
  {Vallisneri}}]{Hourihane_Meyers_Johnson_Chatziioannou_Vallisneri_2022}%
  \BibitemOpen
  \bibfield  {author} {\bibinfo {author} {\bibfnamefont {S.}~\bibnamefont
  {Hourihane}}, \bibinfo {author} {\bibfnamefont {P.}~\bibnamefont {Meyers}},
  \bibinfo {author} {\bibfnamefont {A.}~\bibnamefont {Johnson}}, \bibinfo
  {author} {\bibfnamefont {K.}~\bibnamefont {Chatziioannou}},\ and\ \bibinfo
  {author} {\bibfnamefont {M.}~\bibnamefont {Vallisneri}},\ }\bibfield  {title}
  {\bibinfo {title} {Accurate characterization of the stochastic
  gravitational-wave background with pulsar timing arrays by likelihood
  reweighting},\ }\href@noop {} {\bibfield  {journal} {\bibinfo  {journal}
  {Physical Review D}\ }\textbf {\bibinfo {volume} {107}},\ \bibinfo {pages}
  {084045} (\bibinfo {year} {2023})}\BibitemShut {NoStop}%
\bibitem [{\citenamefont {Mitridate}\ \emph {et~al.}(2023)\citenamefont
  {Mitridate}, \citenamefont {Wright}, \citenamefont {von Eckardstein},
  \citenamefont {Schr\"oder}, \citenamefont {Nay}, \citenamefont {Olum},
  \citenamefont {Schmitz},\ and\ \citenamefont {Trickle}}]{Mitridate2023oar}%
  \BibitemOpen
  \bibfield  {author} {\bibinfo {author} {\bibfnamefont {A.}~\bibnamefont
  {Mitridate}}, \bibinfo {author} {\bibfnamefont {D.}~\bibnamefont {Wright}},
  \bibinfo {author} {\bibfnamefont {R.}~\bibnamefont {von Eckardstein}},
  \bibinfo {author} {\bibfnamefont {T.}~\bibnamefont {Schr\"oder}}, \bibinfo
  {author} {\bibfnamefont {J.}~\bibnamefont {Nay}}, \bibinfo {author}
  {\bibfnamefont {K.}~\bibnamefont {Olum}}, \bibinfo {author} {\bibfnamefont
  {K.}~\bibnamefont {Schmitz}},\ and\ \bibinfo {author} {\bibfnamefont
  {T.}~\bibnamefont {Trickle}},\ }\bibfield  {title} {\bibinfo {title}
  {{PTArcade}},\ }\href@noop {} {\  (\bibinfo {year} {2023})},\ \Eprint
  {https://arxiv.org/abs/2306.16377} {arXiv:2306.16377 [hep-ph]} \BibitemShut
  {NoStop}%
\bibitem [{\citenamefont {Ellis}\ and\ \citenamefont {van
  Haasteren}(2017)}]{justin_ellis_2017_1037579}%
  \BibitemOpen
  \bibfield  {author} {\bibinfo {author} {\bibfnamefont {J.}~\bibnamefont
  {Ellis}}\ and\ \bibinfo {author} {\bibfnamefont {R.}~\bibnamefont {van
  Haasteren}},\ }\href {https://doi.org/10.5281/zenodo.1037579} {\bibinfo
  {title} {jellis18/ptmcmcsampler: Official release}} (\bibinfo {year}
  {2017})\BibitemShut {NoStop}%
\bibitem [{\citenamefont {{R Core Team}}(2018)}]{R}%
  \BibitemOpen
  \bibfield  {author} {\bibinfo {author} {\bibnamefont {{R Core Team}}},\
  }\href {https://www.R-project.org/} {\emph {\bibinfo {title} {R: A Language
  and Environment for Statistical Computing}}},\ \bibinfo {organization} {R
  Foundation for Statistical Computing},\ \bibinfo {address} {Vienna, Austria}
  (\bibinfo {year} {2018})\BibitemShut {NoStop}%
\bibitem [{\citenamefont {Odland}(2018)}]{tommy}%
  \BibitemOpen
  \bibfield  {author} {\bibinfo {author} {\bibfnamefont {T.}~\bibnamefont
  {Odland}},\ }\href {https://doi.org/10.5281/zenodo.2392268} {\bibinfo {title}
  {tommyod/kdepy: Kernel density estimation in python}} (\bibinfo {year}
  {2018})\BibitemShut {NoStop}%
\bibitem [{\citenamefont {{Hinton}}(2016)}]{Hinton2016}%
  \BibitemOpen
  \bibfield  {author} {\bibinfo {author} {\bibfnamefont {S.~R.}\ \bibnamefont
  {{Hinton}}},\ }\bibfield  {title} {\bibinfo {title} {{ChainConsumer}},\
  }\href {https://doi.org/10.21105/joss.00045} {\bibfield  {journal} {\bibinfo
  {journal} {The Journal of Open Source Software}\ }\textbf {\bibinfo {volume}
  {1}},\ \bibinfo {eid} {00045} (\bibinfo {year} {2016})}\BibitemShut {NoStop}%
\bibitem [{\citenamefont {{Vallisneri}}(2020)}]{2020ascl.soft02017V}%
  \BibitemOpen
  \bibfield  {author} {\bibinfo {author} {\bibfnamefont {M.}~\bibnamefont
  {{Vallisneri}}},\ }\href@noop {} {\bibinfo {title} {{libstempo: Python
  wrapper for Tempo2}}},\ \bibinfo {howpublished} {Astrophysics Source Code
  Library, record ascl:2002.017} (\bibinfo {year} {2020}),\ \Eprint
  {https://arxiv.org/abs/2002.017} {ascl:2002.017} \BibitemShut {NoStop}%
\end{thebibliography}%
\appendix
\section{Kernel Density Estimators}\label{appendix:kde}
    \begin{figure}
        \centering
        \includegraphics[width=\columnwidth]{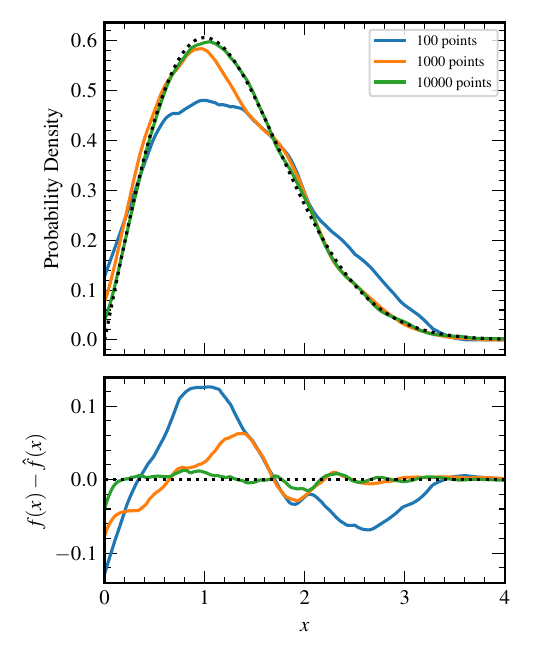}
        \caption{A demonstration on the importance of good sampling to construct an accurate KDE. We randomly drew 100, 1000, and 10000 points from a Rayleigh distribution $f(x)$, and created the KDE $\hat{f}(x)$ with the Epanechnikov kernel \citep{epanechnikov} and a bandwidth selected by the Sheather-Jones method \citep{sj}. The bottom panel shows the absolute difference between the actual distribution and its reconstruction.}
        \label{fig:gauss_toy}
    \end{figure}
Selecting the optimal kernel $K$ and bandwidth $h$ typically focuses on minimizing the asymptotic mean integrable squared error (AMISE) between the underlying distribution $f$ and the reconstructed estimator $\hat{f}$ given $N$ samples:
    \begin{equation} \label{eq:AMISE}
        \mathrm{AMISE}(h) = \frac{R(K)}{Nh} + \frac{1}{4}\sigma_K h^4 R(f'').
    \end{equation}
    Here, $R(g)=\int g(x)^2\mathrm{d}x$ for any function $g$, and $\sigma_K^2=\int x^2 K(x)\mathrm{d}x > 0$ is the second moment of the kernel, at a given point $x$. The second derivative $f''$ is with respect to $x$. Note that if the kernel is normal with standard deviation $\sigma$, $\sigma_K^2=\sigma^2$.
    
    The optimal bandwidth $h^*$ is found by minimizing \autoref{eq:AMISE} with respect to $h$ such that
    \begin{equation} \label{eq:dAMISEdh}
        h^* = \left(\frac{R(K)}{\sigma_K^4 R(f'')}\right)^\frac{1}{5} N^{-\frac{1}{5}}.
    \end{equation}
    If the kernel is normal with standard deviation $\sigma_K=1$, and the underlying distribution $f$ is known to be normal with standard deviation $\sigma$, bandwidth selection is trivial: $h^*=1.06\,\hat{\sigma}\,N^{-1/5}$, where $\hat{\sigma}$ is the standard deviation of the samples.

    However, $f$ is not always known and a method is required to reduce the AMISE without prior knowledge of $f$. One such method is the Sheather-Jones plug-in selector \citep{sj}. It computes the optimal bandwidth $h^*$ by estimating $R(f'')$ and iteratively solving \autoref{eq:dAMISEdh} with the Newton-Raphson method. This is a fast and effective bandwidth selector which we use in our KDE reconstructions.
    
    After the optimal bandwidth is selected, substituting \autoref{eq:dAMISEdh} into \autoref{eq:AMISE} finds the following relation between the AMISE and the kernel:
    \begin{equation} \label{eq:kAMISE}
        \mathrm{AMISE}(h^*) \propto \left[\sigma_K R(K)\right]^\frac{4}{5}.
    \end{equation}
    The optimal kernel is the kernel which minimizes this relation. This is the Epanechnikov kernel \citep{epanechnikov} which has the form
    \begin{equation} \label{eq:epanechnikov}
        K(x) = \frac{3}{4}(1-x^2), \ x \in [-1, 1].
    \end{equation}
    We expect to collect samples at the lower boundary of the free-spectrum prior. To ensure accurate KDE representation of the samples at the boundary, we mirror the data at the boundary point and fit the KDE to the mirrored data. This reduces the bias induced at the boundary. We then compute probability densities along a grid of $\log_{10}\rho$ within the prior boundaries that is finer than the bandwidth size. 

    \autoref{fig:gauss_toy} shows a toy model of using KDEs to recreate a distribution. We randomly drew 100, 1000, and 10000 points from a Rayleigh distribution, $f(x)=x\exp{\left(-x^2/2\right)}$, and recreate the distribution from those random samples using a KDE with the aforementioned optimizations. The reconstruction improves as the number of random draws in the training sample increases. Constructing a KDE with 10000 random samples from the distribution more accurately estimates the original distribution than using less number of samples. Therefore, the more data points we draw from the original distribution, the smaller the absolute difference between the distribution and its reconstruction.

\section{The Hellinger Distance}\label{appendix:hellinger}

\begin{table}
       \begin{tabular}{c|| c c c c c c c c}
            \hline
            \hline
            $n$ & 0.25 & 0.50 & 0.75 & 1.0 & 1.5 & 2.0 & 3.0 & 4.0 \\
            \hline
            $H$ & 0.09 & 0.18 & 0.26 & 0.34 & 0.50 & 0.63 & 0.82 & 0.93 \\
            \hline
        \end{tabular}
        \caption{The Hellinger distance, $H$, between two univariate normal distributions with equal standard deviations, yet with means offset by a certain number of standard deviations, $n$.}
        \label{table:hellinger_normal_offset}
    \end{table}

Given probability density functions $f(\vec{x})$ and $g(\vec{x})$ in $N$-dimensional parameter space, the Hellinger distance $H$ is defined as
    \begin{align*}
        H^2(f, g) &= \frac{1}{2}\int \left(\sqrt{f(\vec{x})}-\sqrt{g(\vec{x})}\right)^2\mathrm{d}^N x \\
        &= 1 - \int\sqrt{f(\vec{x})g(\vec{x})}\mathrm{d}^N x.
    \end{align*}

We choose the Hellinger distance as a metric for refitting accuracy over other distance measures---such as Jensen-Shannon---as it is bounded $0 \leq H \leq 1$, and valid for multivariate distributions. A value of $H=0$ implies that distributions are identical, while $H=1$ implies that they do not have any overlap and are completely different distributions. 

Our goal in building rapid and accurate refitting techniques is to ensure the Hellinger distance with respect to the posterior derived from the full PTA likelihood is sufficiently small. The interpretation of what \textit{sufficiently small} means is problem-specific, but some guiding intuition can be gleaned from simple analytic examples. One can show that the Hellinger distance between two univariate normal distributions, $f \sim \mathcal{N}(\mu_1,\sigma_1)$ and $g \sim  \mathcal{N}(\mu_2,\sigma_2)$, is
\begin{equation}
    H = \left\{1 - \sqrt{\frac{2\sigma_1\sigma_2}{\sigma_1^2 + \sigma_2^2}}\exp\left[ -\frac{1}{4}\frac{(\mu_1-\mu_2)^2}{\sigma_1^2 + \sigma_2^2}\right] \right\}^{1/2},
\end{equation}
where $\mu$ and $\sigma$ are the mean and standard deviation of the respective distributions. For distributions of equal standard deviation, but with their means offset from one another by a certain number, $n$, of these standard deviations, the Hellinger distance is
\begin{equation}
    H = \left[1 - \exp\left( -\frac{n^2}{8}\right) \right]^{1/2}.
\end{equation}
A $1$-$\sigma$ offset between these normal distributions may not typically be regarded as a significant disparity, and corresponds to a Hellinger distance of $0.34$. Values for other $n$ are given in \autoref{table:hellinger_normal_offset}.

\begin{figure*}
    \centering
    \includegraphics[width=\textwidth]{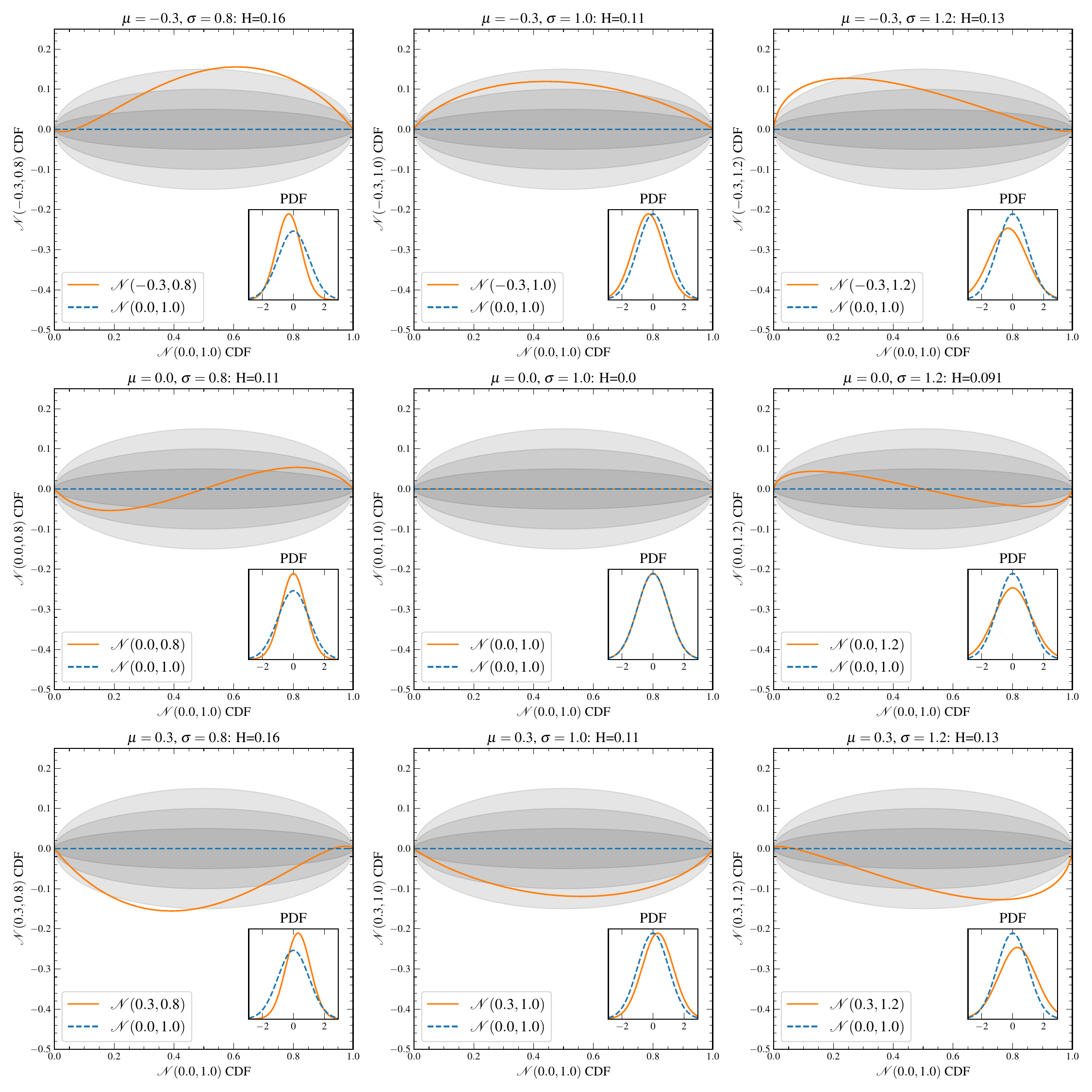}
    \caption{Examples of what $p$-$p$ plots look like for distributions that are approximately, but not entirely, equal to the true posterior (here the standard normal distribution $\mathcal{N}(0,1)$ -- the blue curves in the insets), if we assume that we create $n=100$ realizations of data. The orange curves in the insets show a modified normal distribution $\mathcal{N}(\mu,\sigma)$: our approximated posterior. The large figures show the corresponding $p$-$p$ plots. At the top we have indicated the associated Hellinger distance between the two posteriors.}
    \label{fig:pp-explanation}
\end{figure*}

In \autoref{fig:pp-explanation} we show some examples of univariate normal distributions with different means and standard deviations. Assuming we generate $n=100$ realizations from these distributions, we show what the associated $p$-$p$ plots would be, and the Hellinger distance between the distributions.

\end{document}